# A short account of thermoelectric film characterization techniques

Nagaraj Nandihalli*

Ames National Laboratory and Department of Materials Science and Engineering, Iowa State University, 2408 Pammel Drive, Ames, IA 50011-1015, USA

* Nagaraj Nandihalli ✉ nnandiha@uwaterloo.ca | nagaraj.nandi001@umb.edu

*Abstract*

Thermoelectric films and periodic structures have particularly intriguing electrical and thermal transport features due to their low dimensionality. As a result, they have piqued the attention of researchers from across the spectrum of disciplines. Their applications span from cooling fast CPUs to providing energy for wearable devices. The progress in the techniques for synthesizing TE materials and fabricating thin films has facilitated the emergence of a flourishing research domain in the field of electrical and thermal transport at the nanoscale. Further, a significant proportion of contemporary electronic, opto-electronic, and solar energy components are composed of materials featuring numerous interfaces and nanoscale contacts. Consequently, it is imperative to explore the thermal energy transfer and charge carrier transport across the films with thickness dimensions in the nanometer range. The development of cutting-edge approaches for grasping complicated processes at the nanoscale is critical. This review provides a concise overview of the prevalent methodologies developed over the last two decades and employed for the characterization of the Seebeck coefficient and electrical and thermal conductivity of thermoelectric films.

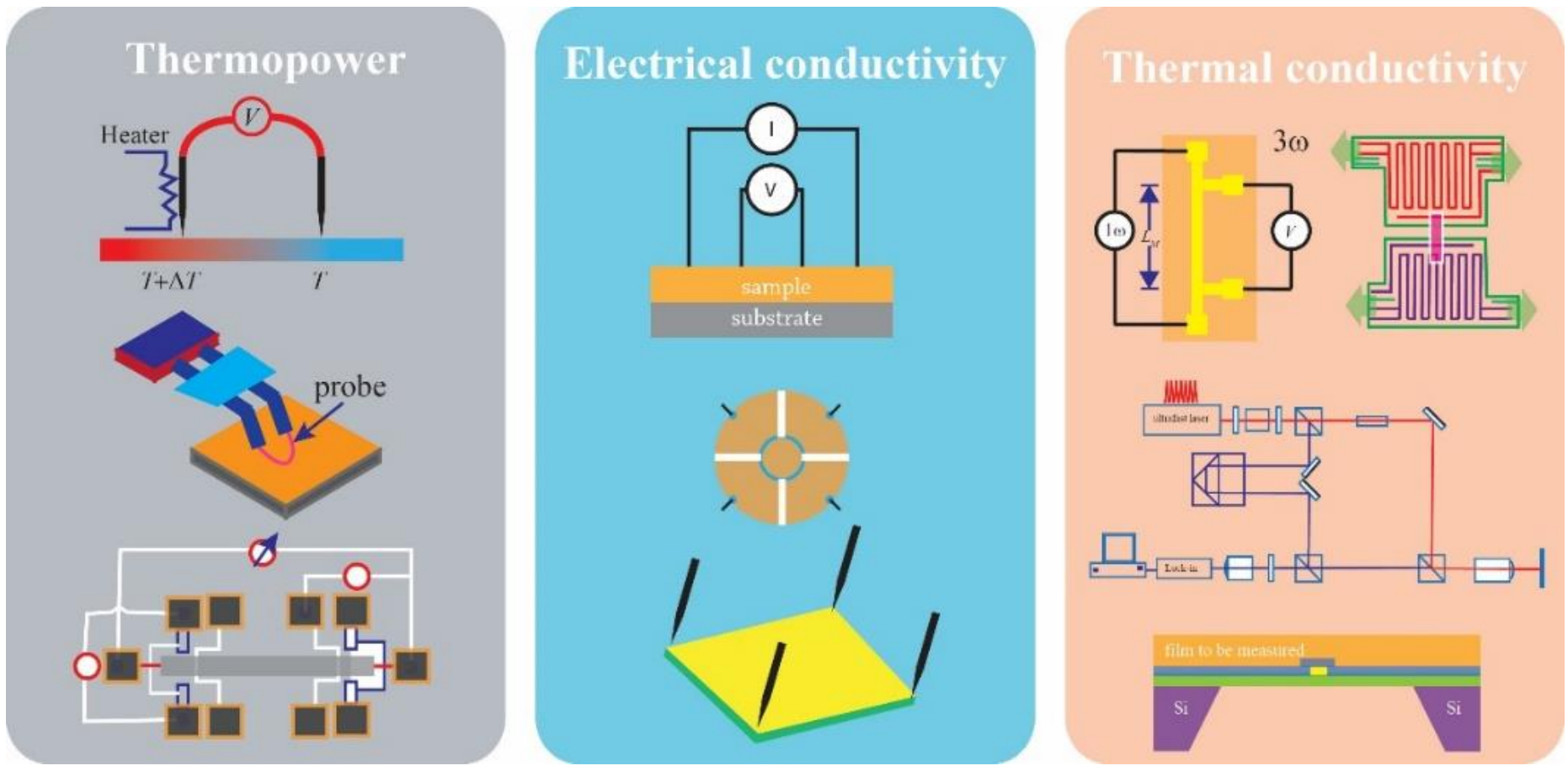

## 1 Introduction

Thermoelectric (TE) energy conversion technology is gaining popularity due to its superior properties. TE devices with no moving components find applications in power generating, cooling, and wearable devices.[1-5] The evaluation of thermoelectric (TE) materials is predominantly based on the dimensionless *figure-of-merit*, $zT = S^2\sigma T/\kappa$, where $S$ is the Seebeck coefficient, $\sigma$ is the electrical conductivity, $T$ is the absolute temperature, and $\kappa$ is the total thermal conductivity. To achieve a desirable TE performance, it is necessary for the material to exhibit a significant $S$, a high $\sigma$, and a low $\kappa$. The power factor ($PF = S^2\sigma$) is a metric that characterizes the capability of a TE material to convert electron energy. Because $ZT$ is related to $S^2$, $S$ is the most important parameter for TE property enhancement.

With reducing IC size and increasing processor frequency in microelectronics and optoelectronics, the amount of heat released locally (hot spots) by components is enormous (>1000 W/cm$^2$), negatively affecting their dependability and lifetime unless the dissipated heat is expunged. Localized, on-chip, solid-state thermoelectric cooling utilizing TE films is expected to facilitate the development of microelectronic chips with efficient heat management. A thermoelectric generator (TEG) can be used to generate electrical current from body heat to power wearable electronics and health sensors.[3, 6]

Many TE materials exhibit higher $zT$ values when they are in film or superlattice (SL) form, mainly owing to their reduced dimensionality.[6] In recent years, there has been a significant amount of research conducted on various TE materials.[3, 7-11] The TE properties of thin films are substantially influenced by the technique of material processing. The TE film fabrication involves the deposition of film on suitable substrates having desired thermophysical properties to achieve desired crystal structure orientation, crystal size, and strain-related effects. Complications emerge when attempting to characterize the TE parameters because most thin films are formed on semiconductor substrates whose TE effects may outweigh those of the test films and interfere with accurate measurements of parameters during the measurement. Thin-film thermophysical characteristics diverge significantly from bulk-form values for a variety of reasons. Readers are encouraged to refer to the author's article for more information.[6] To mention briefly, the TE periodic structures offer the freedom to customize the electrical and thermal transport properties. In these structures, the interaction of phonons with interfaces and grain boundaries significantly changes phonon transport. Furthermore, thin-film TE materials can alter the DOS by inducing a quantum effect, which boosts the $S$. Thermoelectric transports in SLs and films are very anisotropic and must be characterized individually in the in-plane (along the film plane) and cross-plane (perpendicular to the film plane) directions. Electrons and phonons may encounter substantially higher interface scattering in the cross-plane direction of SLs than in the in-plane direction.[6] As a result, to comprehend TE transports in the cross-plane direction, one must measure the $\kappa$, the $S$, and the $\sigma$ in the same direction, because device performance is dependent on the $zT$. The electronic contribution to $\kappa$ that results from an increase in $\sigma$ is often undesirable in TE materials/films. As a result, efforts should be made to comprehend the process that relates electrical and thermal transport properties by conducting nanoscale heat transfer studies. Thermoelectric attributes have recently been discovered in organic semiconductors as well as organic/inorganic composites, which are frequently used in the form of thin solid films. Thermal processes, on the other hand, have an impact on the performance of any polymer-based device. Understanding and manipulating heat and charge carrier transport at shorter length scales is crucial for improving the performance of materials designed for TE energy harnessing and TE miniatured thermoelectric cooling (TEC) devices. Aside from thermoelectrics, precise assessment of $\sigma$ and $\kappa$ of films of specific materials is critical in a number of vital technologies.[12, 13]

The electrical and thermal resistances that exist between the temperature sensor and the sample are of significant importance in thermal and thermoelectric measurements of thin films and nanostructures, particularly in cases where the samples have high $\kappa$. Several considerations must be taken into account to ensure precise measurement in TE films. Firstly, it is important to note that TE films typically require a supporting substrate and are not free-standing, which may impact temperature distribution. Secondly, finite-sized metal electrodes are necessary for monitoring the thermovoltage in low dimensional materials.[14] Thirdly, the anisotropic nature of TE properties of films necessitates measurements in both in-plane and out-of-plane directions to ensure a comprehensive analysis. For example, the extremely anisotropic $\kappa$ of 2D layered transition-metal dichalcogenides (TMDCs) thin films (13-81 W/m-K and 0.05-0.3 W/m-K for in-plane and cross-plane directions, respectively)[15-17] underscores the necessity of examining out-of-plane $S$ for TMDC thin films.

A large number of articles on TE films describe the use of custom-made setups to analyze electrical and thermal properties (Table 1, Table *2*, Table 3 and Table S1 in Supplementary information). As a result, numerous measuring tools for $\sigma$, $S$, and $\kappa$ were developed. There is no single consistent measurement method for all types of TE thin films. This review examines several prevalent methodologies.

## 1.1 Thin film configurations for TE applications

Contingent upon the direction of heat flow, two distinct thin film geometries can be delineated for thermoelectric (TE) applications. The planar architecture (Figure 1a) allows us to achieve a high temperature gradient ($\nabla T$) (assuming the substrate and film sample do not have particularly high $\kappa$) while displaying a small current section (film width × film thickness) and delivering a high voltage in an open circuit ($E_{oc}$) but with low power output. The fundamental disadvantage of this geometry is its high electrical resistance, which lowers the electrical power generation. The cross-plane geometry (Figure 1b) has a modest $\nabla T$ (along a 100 nm step), which results in a low $E_{oc}$ (i.e., electrical power output).[18] To attain high powers, the resistance must be kept as low as possible (through optimal film thickness, nanostructuring or composition) and the $\nabla T$ as large as possible. However, under the present TE technological landscape, having both is difficult. As a result, TE thin films (modules) are unsuitable for high-power applications, but they can be used in micro TE generators (μTEG) or temperature sensors.

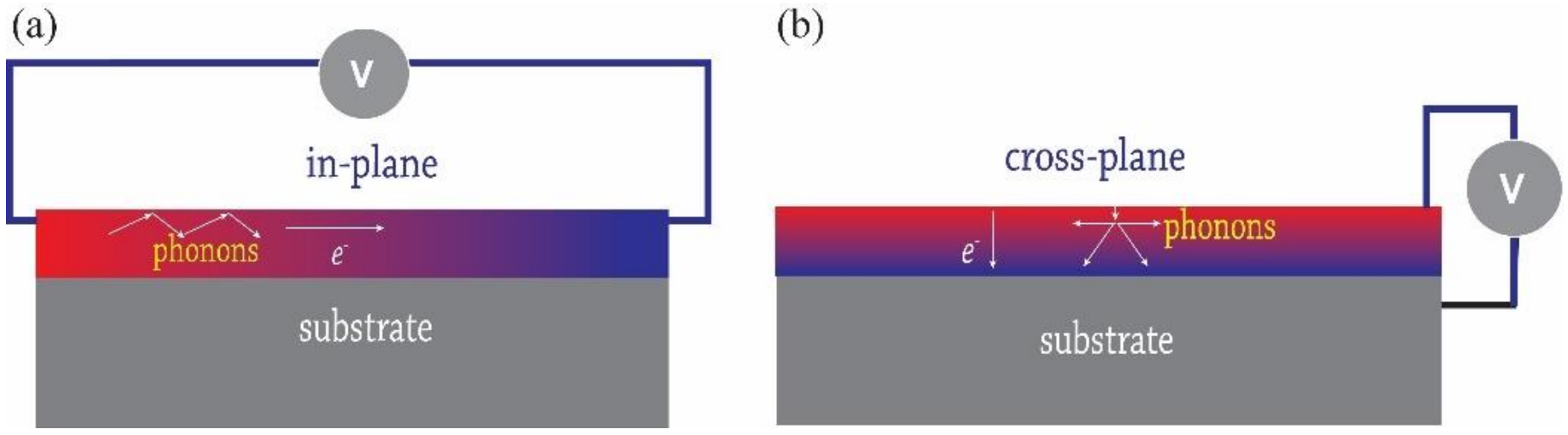

Figure 1: Thin film geometries based on the heat flow direction. (a) in-plane; (b) cross-plane.

Working TEG and TEC devices consist of segments of *p*-type and *n*-type–doped semiconductor materials connected by shunts (made of good electrical conductor, such as Cu) to form an electric circuit. In power generation mode (Figure 2b), the temperature difference triggers charge carrier movement, constantly supplying electrical energy to a load if the temperature difference is maintained. To achieve a large power output, the segments are cascaded as shown in Figure 2b. In cooling mode, the supplied electrical energy induces a cooling effect at one side of the module (Figure 2d). TE films are typically obtained using an appropriate process first, their TE properties are measured to determine their suitability, and the films are then utilized to construct TEG or TEC; the power generation or cooling performance is then quantified.

## 2 Seebeck measurement methods and instrumentation

The conventional approach for determining the Seebeck coefficient ($S$) of TE materials involves establishing a temperature gradient across the specimen and subsequently measuring the voltage and temperature differentials (Figure 2a). It should be noted that thermal gradient, $\nabla T$ = $\Delta T$/*distance between measuring points*. This approach is applicable for both inorganic bulk materials and TE films. The $S$ is defined as the ratio of an electric potential gradient ($\Delta V$) to an applied temperature gradient ($\Delta T$). The parameter appears simple, but its accurate measurement is far from a simple task. A fluctuation in the observed $S$ of < 5% at room temperature (RT) is normally acceptable. The sign of $S$ is also significant because it indicates the type of dominant charge carriers in the sample (i.e., +ve for *p*-type and -ve for *n*-type material). In general, the $S$ measuring procedure is divided into three parts: (1) Creating a temperature gradient. Most custom-built apparatuses drive direct current (DC) to ceramic heaters to generate a temperature gradient while also setting the required temperature range with temperature controllers. (2) Obtaining $\Delta V$ and $\Delta T$. This is an important step in evaluating the precision of the results for $S$, which is determined directly from the ratio of $\Delta V$ and $\Delta T$. A precise measurement of the temperature gradient can be obtained using any one of several different thermocouple (TC)s, which can be chosen according to the conditions of the test (i.e., K-, E-, T-, R-, and N-type). Besides, accurate voltage acquisition requires sensitive nano-voltmeters. (3) Data analysis. To control the entire apparatus in measuring and data collecting, a commercial or custom written program is used. It is demonstrated that a vacuum environment improves measurement precision.[19]The $S$ is often measured using one of two methods: the differential method or the integral approach.

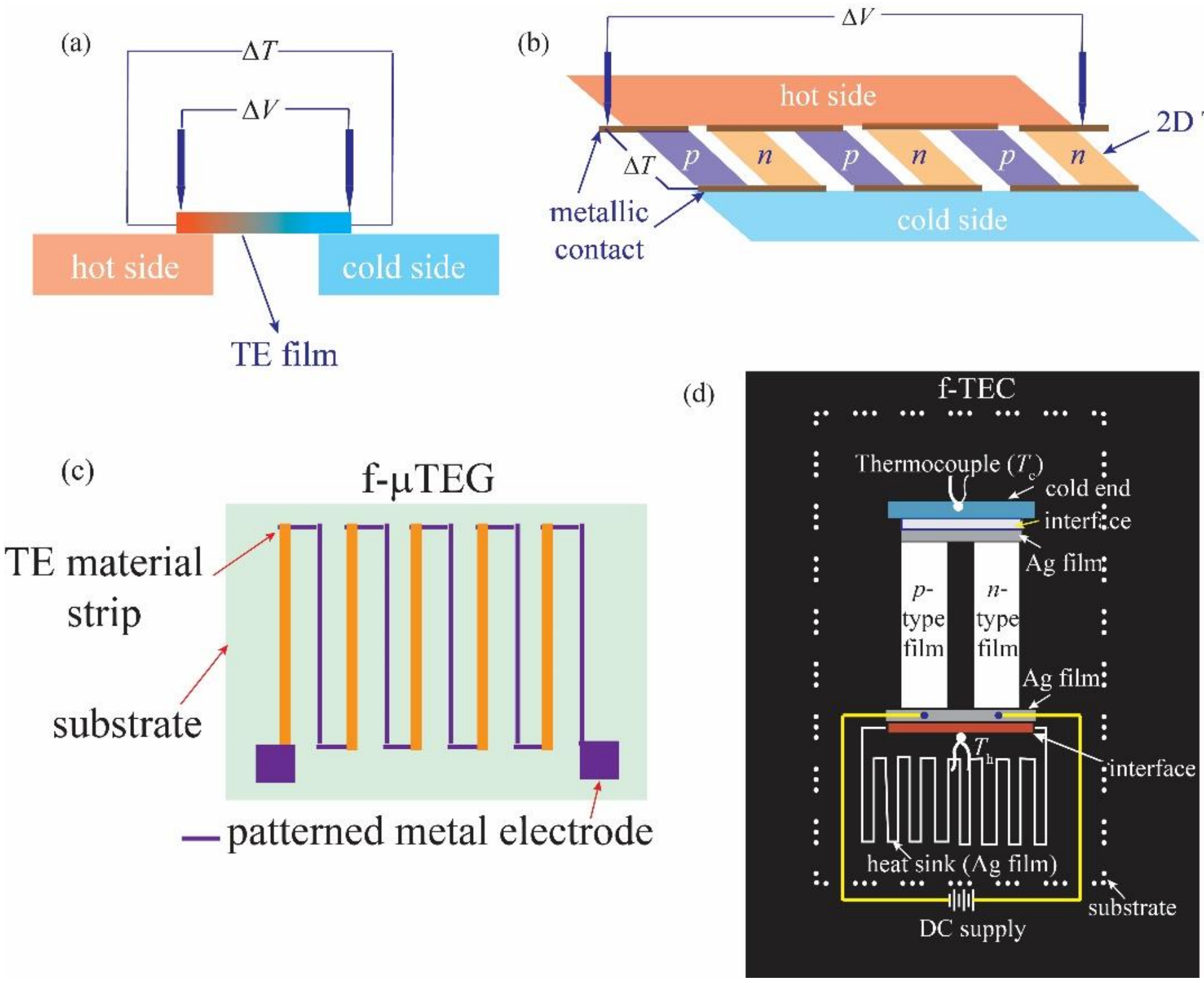

Figure 2: $S$ measurement scheme for TE films; power generation and cooling performance evaluation of TE generator (TEG) and TE cooler (TEC). (a) film having $\Delta T$ (side view); (b) TEG (top view). Here, *p*- and *n*-type elements are patterned in a cascaded fashion and connected using metal electrodes on a certain substrate to achieve maximum power output; (c) a flexible TEG (f-TEG, mono-mode) on a flexible substrate; (d) In operation, flexible TEC (f-TEC) on a suitable substrate. Only a pair of *p*- and *n*-type segments are shown.

It is important to note that the establishment of a temperature differential is applicable in all scenarios: (1) while measuring the $S$ of a standalone $n$- or $p$-type films (as depicted in Figure 2a); (2) while measuring the power generation capability of a TEG (as illustrated in Figure 2b); (3) while measuring the power generation capability a TEG operating in mono mode or single-mode;[3, 20] Table 1 lists the reports related to the case 1, and Table 2 displays the latter two cases. This review extensively covers the case 1. Figure 2d shows the TE cooling module in operation, where electrical energy is supplied to cool one side of the module. Table 3 presents the techniques employed for characterizing the TE parameters of diverse films, along with the cooling performance of miniature thermoelectric cooling (TEC) devices constructed from them.

## 2.1 The differential method and its instrumentation

In the differential method, a small $\Delta T$ (4-20 K) is set up for the sample at an average temperature of interest $T_0 = (T_1 + T_2)/2$, where $T_1 = T_0 - \Delta T/2$ and $T_2 = T_0 + \Delta T/2$. Under the condition that the applied temperature gradient $\Delta T = T_2 - T_1$ is << average temperature $T_0$, i.e., $\Delta T/ T_0 << 1$, the $S$ is calculated by dividing the electric potential $\Delta V$ by the temperature difference $\Delta T$ at a given temperature $T_0$ (i.e., $S = -\Delta V/\Delta T$).[3] The differential technique is divided into two categories: steady-state and quasi-steady-state. Under steady-state conditions, a specified $\Delta T$ is stabilized before taking the measurement in the related device,[21] and the $S$ is derived from a single measurement or an averaged value from a series of data at a constant $\Delta T$. However, compared to a quasi-steady-state, the stabilizing of each increment takes far longer. In quasi-steady-state settings, measurements are taken continually as the temperature differential gradually increases.[22, 23] Then, from the derivation of $\Delta V$ to $\Delta T$, $S$ is obtained. In either case, in order to improve the precision of the $S$ measurement, it is necessary to adhere to the following three fundamental requirements: (a) simultaneous measurements of temperature and voltage taken at the same location and time; (b) good thermal and electrical contact is maintained between the probes and the sample; (c) the effectiveness of the experimental setting to obtain microvolts or perhaps nanovolts.[24]

$S$ can be extracted from the $\Delta T$-$\Delta V$ plot in two ways: slope method and single point measurement. The slope approach generally yields $S$ by linear fitting multiple ($\Delta T$, $\Delta V$) data points. In this scenario, an offset voltage (the fitted line offsets the voltage axis) of several hundred µ$V$s could be achieved. There are various factors that may give rise to the offset voltage, such as disparities in TC wires, non-uniformity of materials, reactive specimens, and the cold finger phenomenon. The latter refers to the situation where the TC draws heat away from the specimen, leading to a temperature decrease between the specimen and the TC tip due to thermal contact resistance. If the offset voltage remains constant during the measurement, the slope method can eliminate the offset voltage, which is reflected in the distance from the fitting line diverging from the origin point, and extract only the thermoelectric voltage. In contrast, the single point method calculates the $S$ at $T_0$ by dividing $\Delta V$ by $\Delta T$, without isolating the contribution of offset voltage to the real $S$ from $\Delta V$.

Guan et al. developed an equipment that increased the lower limit of the $S$ measuring temperature range to 100 K.[25] This method involved the acquisition of the in-plane $S$ ($S_{\|}$) through a quasi-steady temperature difference technique, whereby two ceramic heaters were utilized to alternativelt heat the sample. During the measurement, one heater was heated to the desired temperature (this heating time was H1) and then turned off until the samples were equilibrated (this cooling phase was C1). The same procedure was followed with the other heater (the corresponding runs were H2 and C2). The retrieved $\Delta T$ and $\Delta V$ from the heating-up and cooling-down phases were combined and utilized for linear fitting. In the demonstration, the slopes of the $\Delta T$-$\Delta V$ curves were nearly identical and were treated as the same value of $S_{slop}$ (Figure 3a-c). The absolute $S$ was calculated using the following equation: $S = S_{slop} - S_{ref}$, where $S_{ref}$ was the $S$ of the TC leads. In this device, the resultant $S$ had an accuracy better than 1 µV/K on Pt sample.

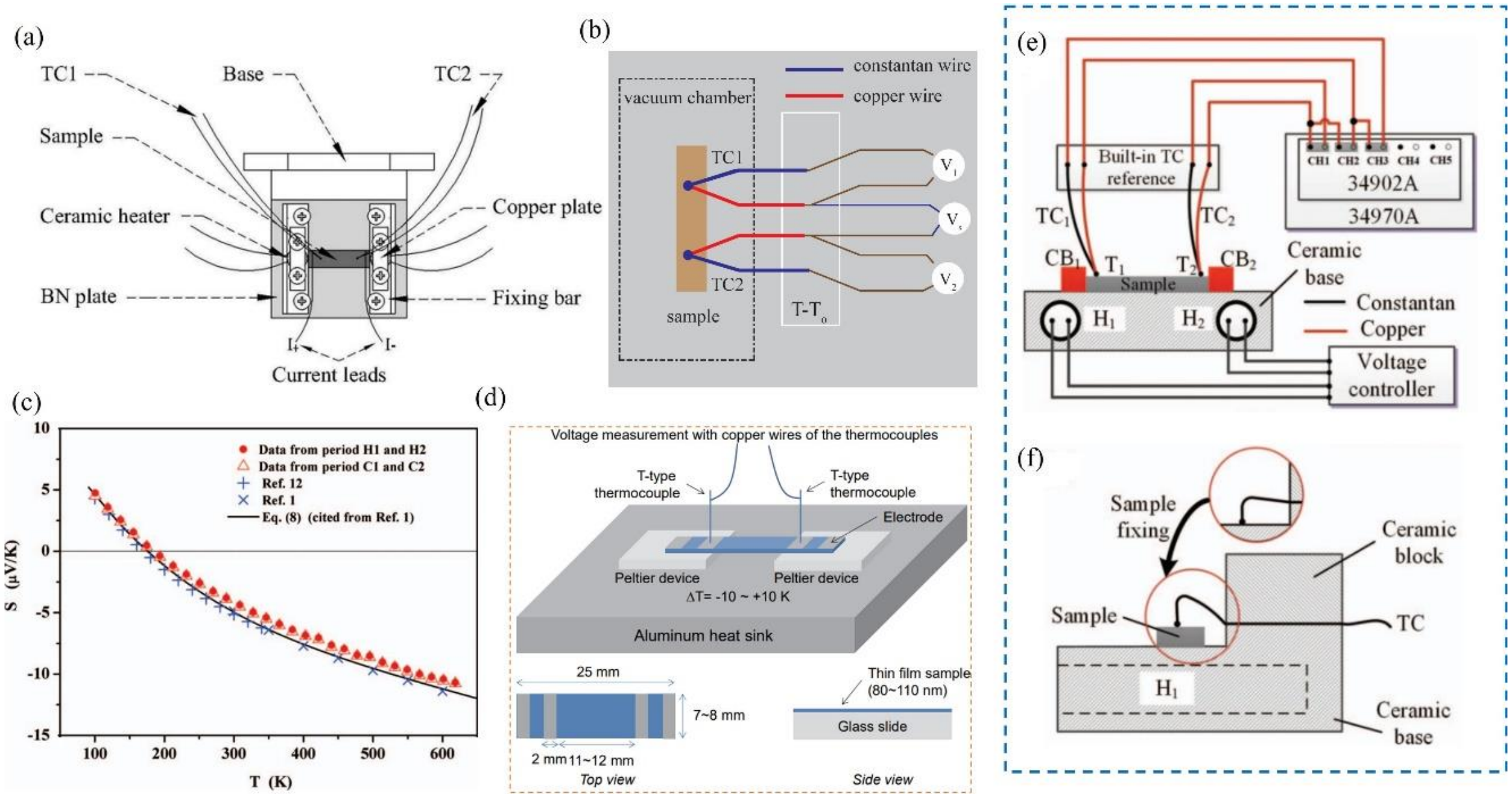

Figure 3: (a-c) quasi-steady temperature differential approach. (a) Top view of the sample holder. TC1 and TC2 are T-type TCs; (b) Schematic for Measuring $S$. Multimeters V1 and V2 record the voltage signals sensed by TC1 and TC2, respectively; multimeter $V$s measures thermoelectric voltage $\Delta U$. $T_0$ is ambient temperature; (c) The temperature dependent $S$ of platinum. Ref. 12 and 1 refer to refences[26] and [27] respectively.[25] (d) PEDOT-Tos)/CNT composite. A setup for measuring $\sigma$ and $S$.[28] (e-f): $LaCo_{1-x}Cu_xO_3$ ($x$ = 0.1 or 0.15) $S$ measurement setup. (e) customized setup; (f) Contacts between the sample and the TCs are made.[29]

A modified steady-state technique was used to determine the in-plane $S$ (and $\sigma$) of hybrid poly(3,4-ethylenedioxythiophene)-tosylate (PEDOT-Tos)/carbon nanotubes (CNTs) films.[28] Two Peltier devices were utilized to establish a temperature difference along the samples, as illustrated in Figure 3d. When one Peltier device heats the side that faces the sample, the other Peltier device cools the other side that faces the sample. Each Peltier device's temperature varied between +4 K and -4 K to achieve a maximum temperature difference of 6-8 K. T-type TC made of Cu and constantan wires were used to measure the temperatures of the hot and cold sides of the sample. To reduce electrical contact resistance, silver paint was used to create metal electrodes on the sample. Linear current-voltage relationships were observed, confirming the presence of ohmic contacts between the TC and the samples. To determine the thermopower of the samples, a series of 6 to 8 distinct temperature differentials were generated and the corresponding thermoelectric voltages were measured. The *S* of the specimens was determined by computing the linear gradient of the thermoelectric voltage as a function of the $\Delta T$. A similar technique was reported for measuring the *S* of $Cu_2Se$ thin films.[30] (see Table 1).

Zhou and coworkers claim a fast technique for measuring the $S_{||}$ based on the dynamic method, where the sample temperature rises continually with a predetermined and suitable temperature difference throughout the sample.[29] In this manner, a corresponding *S* can be obtained for the sample at each carefully determined temperature for the sample. This method can be considered a modified steady-state method because the temperature differential does not change throughout the process. From the acquired data, a relationship between the *S* and the sample temperature *T* could be well established by *S*-*T* curves. During the measurement, the heater H1 was turned on first (Figure *3*e) and after ~5 s, heater H2 was turned on. The data acquisition unit was instantly turned on to capture the temperatures *T*1 and *T*2 as well as the Seebeck voltage $\Delta V$. When the sample's average temperature ($T_{Avg} = (T1 + T2)/2$) exceeded 473 K, the two heaters and the data gathering unit were turned off. The raw data of *T*1, *T*2, and $\Delta V$ were analyzed to obtain the sample's *S*. Test measurements were performed on two samples of $LaCo_{1-x}Cu_xO_3$ ($x = 0.1$ or 0.15). Both samples were measured twice using voltage combinations of the two heaters of (70 V, 60 V) and (90 V, 60 V). The customized device's results deviated from data acquired by a commercial instrument by 8.4%.

Khottummee et al.[31] reported the measurement of $S_{||}$ of Copper-Zinc-Tin-Sulfide ($Cu_2ZnSnS_4$; CZTS; obtained by sol-gel method) films (thicknesses: 5.92 µm, 7.67 µm, and 3.08 µm) via custom-built device based on steady state method. Dip-coating was used to deposit the CZTS film on a glass slide. A $\Delta T$ was established at both ends (heater and cooler) of the film. To monitor the temperature and voltage differences, two TCs and voltmeters were directly affixed to the surface of the film (Figure 4a). The *S* was determined from; $S = \Delta V/\Delta T$, where $\Delta V$ is electrical voltage and $\Delta T$ is temperature differences. The peak *S* of CZTS film was $65 \times 10^{-6}$ mV/K. Germanium-Antimony-Tellurium (Ge-Sb-Te) thin films were deposited on $SiO_2$/Si wafers by pulsed dc magnetron sputtering and their RT Seebeck coefficients were measured by an in-house-built device based on steady state method (Figure 4b).[32]The GST thin-films were mounted between the Cu blocks sitting on an alumina base. To generate a $\Delta T$, these Cu blocks were kept individually within the heater controller. The $\Delta T$s were measured on the surface of the film samples using K-type TCs to determine hot ($T_{hot}$) and cold ($T_{cold}$) temperatures, whilst voltage differences ($\Delta V$) were measured using electrodes. Temperature ($\Delta T = T_{hot} - T_{cold}$) and voltage differences were used to calculate the *S* as heat flowed through the samples from the hot to the cool side. The temperature controller was constantly changed during the measurements to induce temperature variations (0-10 K range). A digital multimeter was used to measure $T_{hot}$, $T_{cold}$, and voltage. The *S*s were determined from: $S = \Delta V/\Delta T$; where $\Delta V$ and $\Delta T$ were the thermoelectric voltage and temperature differences, respectively. The GeSbTe film (397 nm-thick) annealed at 673 K yielded a *S* of 71.07 μV/K.

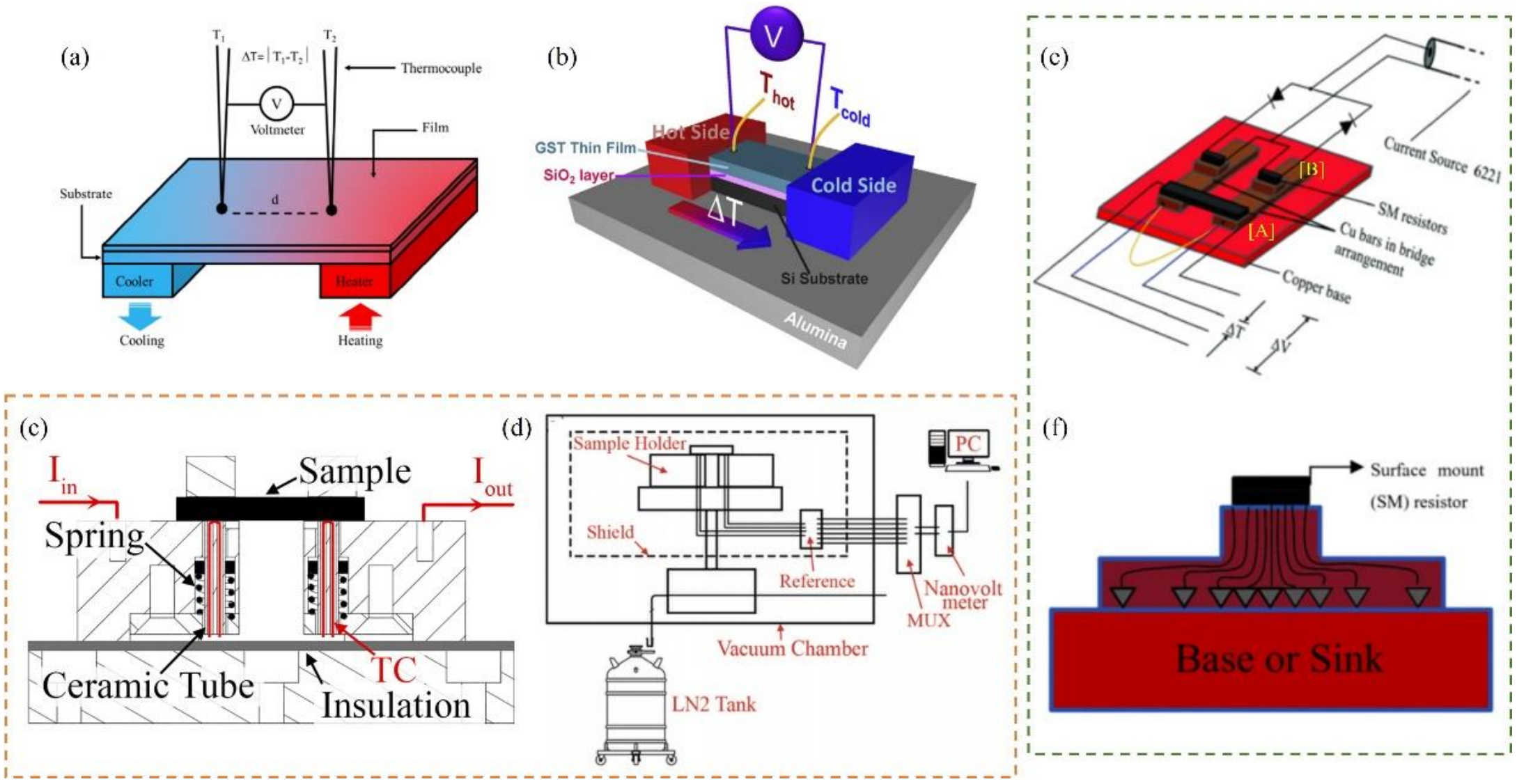

Figure 4: (a) $Cu_2ZnSnS_4$ film characterization via steady state method;[31] (b) Ge-Sb-Te thin films characterization via steady state method.[32] (c-d) PEDOT:PSS characterization via quasi-steady-state technique. (c) The sample holder's technical diagram. To ensure proper electrical and thermal contact, the TC wires are threaded through two alumina tubes that are pushed against the sample by two springs; (d) Measuring system outline. The system is a cryostat with a specific sample holder mounted on it. The TCs used in the measurements are connected to the reference, which has a PT100 monitoring the cold junction temperature. The reference is attached to the shield base. From the reference, only copper wires are utilized to link the electrical instrumentation.[33] (e-f) differential method. (e) Diagram of the setup; (f) Diagram of equilibrium heat flow when either SM resistor is heated.[34]

The measurement of the *S* of organic thin films presents a variety of problems that typical setups do not address. Because many highly conductive polymers have low *S*, measurement accuracy and precision are critical to enable material screening, particularly when exploring transport mechanisms. Beretta et al. developed a highly reliable, home-built instrument that operates in vacuum (Figure 4c-d).[33] The measurement employed a quasi-steady-state technique and covered a temperature range of 260 K to 460 K. This methodology eliminated any spurious voltages generated by the steady-state methods' requirement for $\Delta T$ stabilization. When compared to the steady-state method, faster measurements can be obtained by recording a set of data from continuously increasing $\Delta T$ at a specific sample temperature. Furthermore, the $S_{\|}$ was calculated as the derivative of the thermovoltage $\Delta V$ with respect to the temperature gradient $\Delta T$, which entirely cancels out spurious voltages. In addition, the apparatus was operated at a variety of sample temperatures to determine the relationship between the *S* and the sample temperature (this method is analogous to the dynamic method described before). The system was evaluated using a bulk sample of pure Ni as well as a thin film of commercial PEDOT:PSS that was formed on glass via spin coating. It was determined that the instrument has a reproducibility of 1.5% at RT and has the capability to characterize the soft materials.

Tripathi and colleagues devised an experimental platform utilizing the differential method to simultaneously measure the thermoelectric power (TEP) or $S_{\parallel}$ of two samples within the temperature range of 77 K to 500 K, employing a liquid nitrogen cryostat to speed up the testing process.[34] TEP had a maximum uncertainty of less than 0.5 $\mu$V/K and an accuracy of within 3%. It was comprised of two small surface mount (SM) chip resistors, a pair of diodes, and a differential geometry TC, making it economical. The utilization of TCs in differential configurations resulted in an improvement in the measurement accuracy of $\Delta T$ and, consequently, TEP measurement. Using only one current source, the diode setup was created for alternate heating of SM resistors. All the voltages (thermocouple voltage and Seebeck voltage) are measured with a single nanovoltmeter and a switching mechanism that alternately flips the inputs for thermal voltage ($\Delta V$) readings from the two samples. For calibration, TEP measurements on high purity Pt wire of thickness 50 μm and thin (80 μm) type K TC wires Chromel and Alumel were performed from 77 K to 500 K with respect to Cu lead wires. The setup can also be used to calibrate an unknown material against a known absolute TEP material with a repeatability of less than 2%. The symmetric position design of two miniature surface mount resistors (SM resistors), as shown in Figure 4e-f, offers the identical temperature differential across both samples positioned at the end of two bridges (at positions [A] and [B])); sample located at [A] position is shown only. During the process of measurement, by applying positive (+) and negative (-) currents on the same heater, it is possible to acquire the $\Delta T$ for $S$ calculation from the following generated average of $\Delta T$ (+) and $\Delta T$ (−). The $\Delta V$ acquisition was also computed by taking the average of $\Delta V$ (+) and $\Delta V$ (−). The computation of $S$ was contingent upon the mean values of $\Delta V$ and $\Delta T$ produced by both positive and negative currents. The maximum uncertainty in this setup was estimated to be 0.5 μV/K.

In specific contexts, the thermoelectric output is generated by a temperature gradient that is oriented in the out-of-plane (cross-plane) direction of the films. Consequently, it is important to perform cross-plane Seebeck coefficient ($S_{\perp}$) measurements in the direction perpendicular to the plane, which requires a temperature gradient along the cross-plane direction. $S_{\perp}$ measurement in thin films is challenging due to a small temperature differential across the film due to the films' extremely small thickness, resulting in a low signal-to-noise ratio of the Seebeck voltage. Soni and Okram have developed a device that is easily controllable for the purpose of out-of-plane measurement (Figure 5a-c).[19] Each measurement was conducted at a rate of 1 K/min, and lasted approximately 5 hours. This rate of operation was observed to be faster than previous measurements, which required extended periods of temperature stabilization and data collection. The aforementioned enhancement was attained through the optimization of the vacuum environment and the implementation of a more precise supplementary control mechanism. The utilization of a sandwich-like structure appears to be a necessary component in conducting out-of-plane measurements. The experimental setup involved the placement of a pellet sample, measuring approximately 5 mm in diameter and 0.5-2 mm in thickness, between two Cu blocks that were highly conductive and free of oxygen. TCs were utilized to anchor the sample thermally, and the leads of the TCs were affixed to ensure optimal electrical contact, as depicted in Figure 5c In this particular arrangement, the out-of-plane direction was subjected to a temperature difference ($\Delta T$) induced by Cu blocks and subsequently measured using TCs. The measurement of $\Delta V$ was conducted concurrently through the utilization of a Keithley nanovoltmeter. Based on the collected sample data, an estimated error of 3.76% was determined. This particular configuration is exclusively appropriate for thin film samples without substrates.

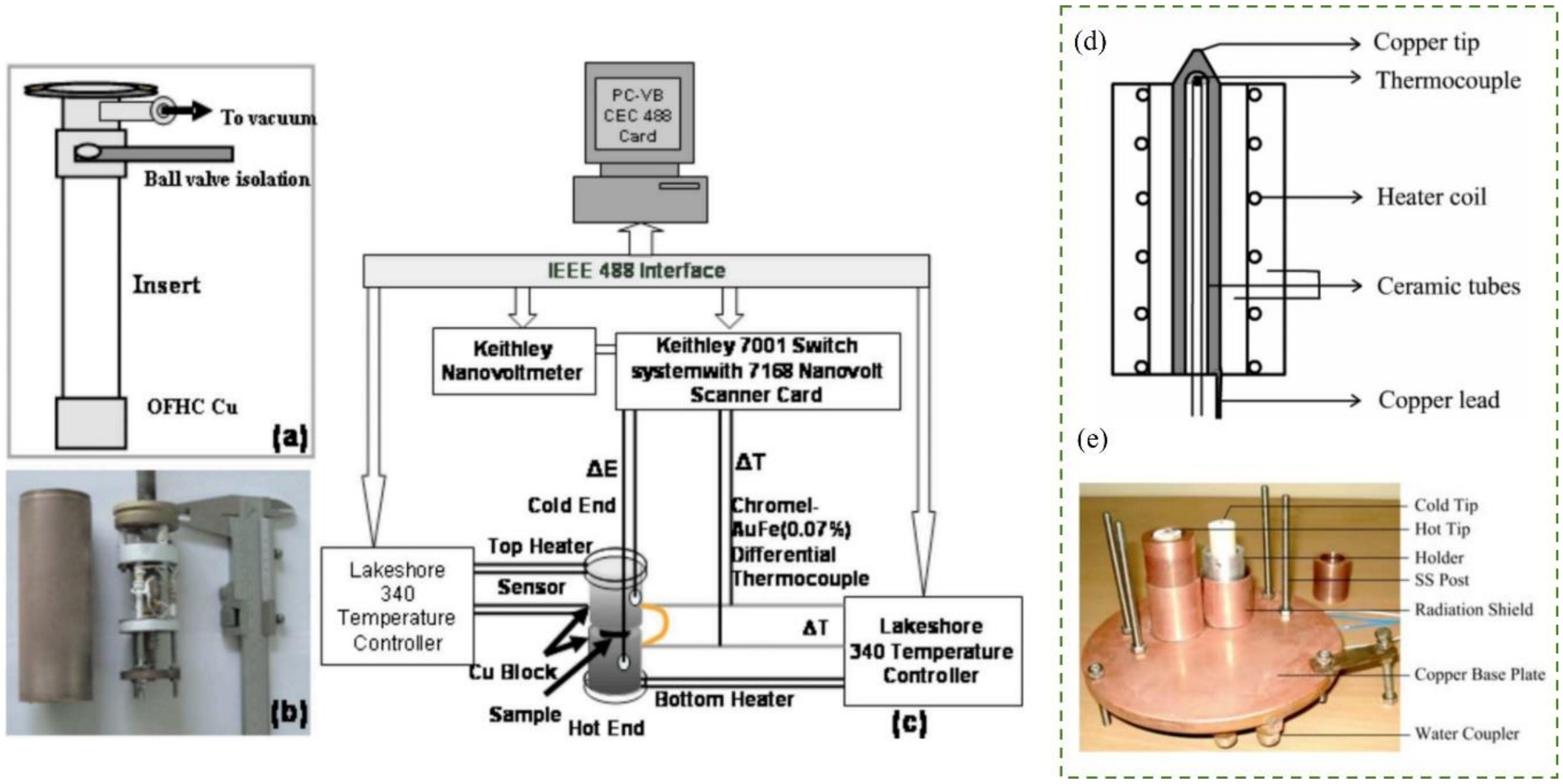

Figure 5: (a-c) out-of-plane $S$ measurement. (a) General-purpose vacuum insert schematic diagram; (b) Photograph shows the sample holder's bottom section and the radiation shield for thermoelectric power (TEP) measurements; (c) Diagram of a TE power measurement setup linked to a computer-interfaced system.[19](d-e) Integral method. (d) A graphic representation of the hot probe; (e) Photograph of the setup. The probes are installed on the water-cooled base plate inside the radiation shields. The stainless steel (SS) posts for installing the spring loaded sample holder are also visible.[35]

Pinisetty et al. [36] presented a similar sandwich device structure that was specifically developed to determine the $S$ of bulk materials, thin films, and nanowire (NW) composites within a temperature range of 120-350 K. The differential method, specifically the slope of $\Delta V$ versus $\Delta T$, was employed to conduct $S$ measurements. The design of the sample holder has been tailored to accommodate the physical property measurement system (PPMS), which is a commercial cryostat. The sample holder utilized in the PPMS was circular in shape and compact in size, designed to fit securely within the confines of the cryostat. The Seebeck sample holder has been constructed atop the aforementioned puck. The distinctiveness of this equipment lies in its capacity to facilitate the alteration of samples with relative ease, coupled with the highly precise temperature regulation provided by the PPMS mechanism. The $S$ of bulk samples of $Bi_2Te_3$, $Sb_2Te_3$, and Ni were measured during the apparatus testing process. To demonstrate the versatility of the apparatus, evaluative analyses were additionally conducted on thin films composed of Ni as well as composites of Ni NWs.

## 2.2 The integral method and its instrumentation

The integral method involves the fixation of one end of the specimen at a constant temperature T1, while the other end is subjected to a wide range of temperatures that are of interest. The extraction of the $S$ at a designated temperature is accomplished by determining the slope of the voltage and temperature (i.e., $S = -dV/dT$). The integral method is capable of effectively reducing the impact of voltage offsets that are commonly manifest in differential method, owing to the implementation of a significant thermal gradient. An additional benefit of utilizing the integral method is its ability to replicate the authentic operational condition of a TE device. However, maintaining temperature T1 at the fixed end of the specimen proves to be challenging due to the significant temperature gradient and elevated temperature. Hence, the essence of this methodology lies in minimizing the interface between the hot and cold ends and identifying appropriate fitting methodologies. Kumar and Kasiviswanathan[35] have developed an experimental setup utilizing the integral method, which incorporates the principle of localized heating through a "hot probe" that also serves as the measuring lead to measure Seebeck voltage from thin wires and thin films in the temperature range of 300–650 K (Figure 5d). In traditional configurations, precise temperature measurement is unattainable because of a limited separation between the TC and the point of contact. The problem at hand can be effectively addressed through the implementation of a TC that is appropriately isolated electrically and integrated into the probe. The diameter of the copper tip in direct contact with the samples was approximately 1 millimeter. This size led to the safeguarding of specimens against physical damage. The configuration of the cold probe was identical to that of the hot probe, and its layout is illustrated in Figure 5e. The findings indicated a 1% margin of error and demonstrated favorable reproducibility with established materials. Large $\Delta T$s are a characteristic inherent to the integral approach.

Resistive thermometry is a technique for measuring the temperature gradient generated by the power delivered to a heater.[37-40] A resistance thermometer measures the change in electrical resistance of pure metals, alloys, and semiconductors with temperature (i.e. $R = f(T)$). A microfabricated device with heaters and resistance thermometer can be used to determine the temperature difference between two locations on the surface of the device. By measuring the open-circuit voltage produced by a temperature gradient, these devices are commonly used to measure the $S$ of the material under test. Because resistance in a metal is proportional to temperature, it is feasible to determine the temperature beneath the thermometer after calibrating it by:[41] $R_2 = R_1[1 + \beta(T_2 - T_1)]$, where $\beta = (\Delta R/\Delta T)\times(1/R_0)$ is the temperature coefficient of resistance. Several researchers have provided similar testing systems that integrate heaters, thermometers, and metal contacts to calculate the in-plane $S$ of thin films (see Figure 6). [37-39]

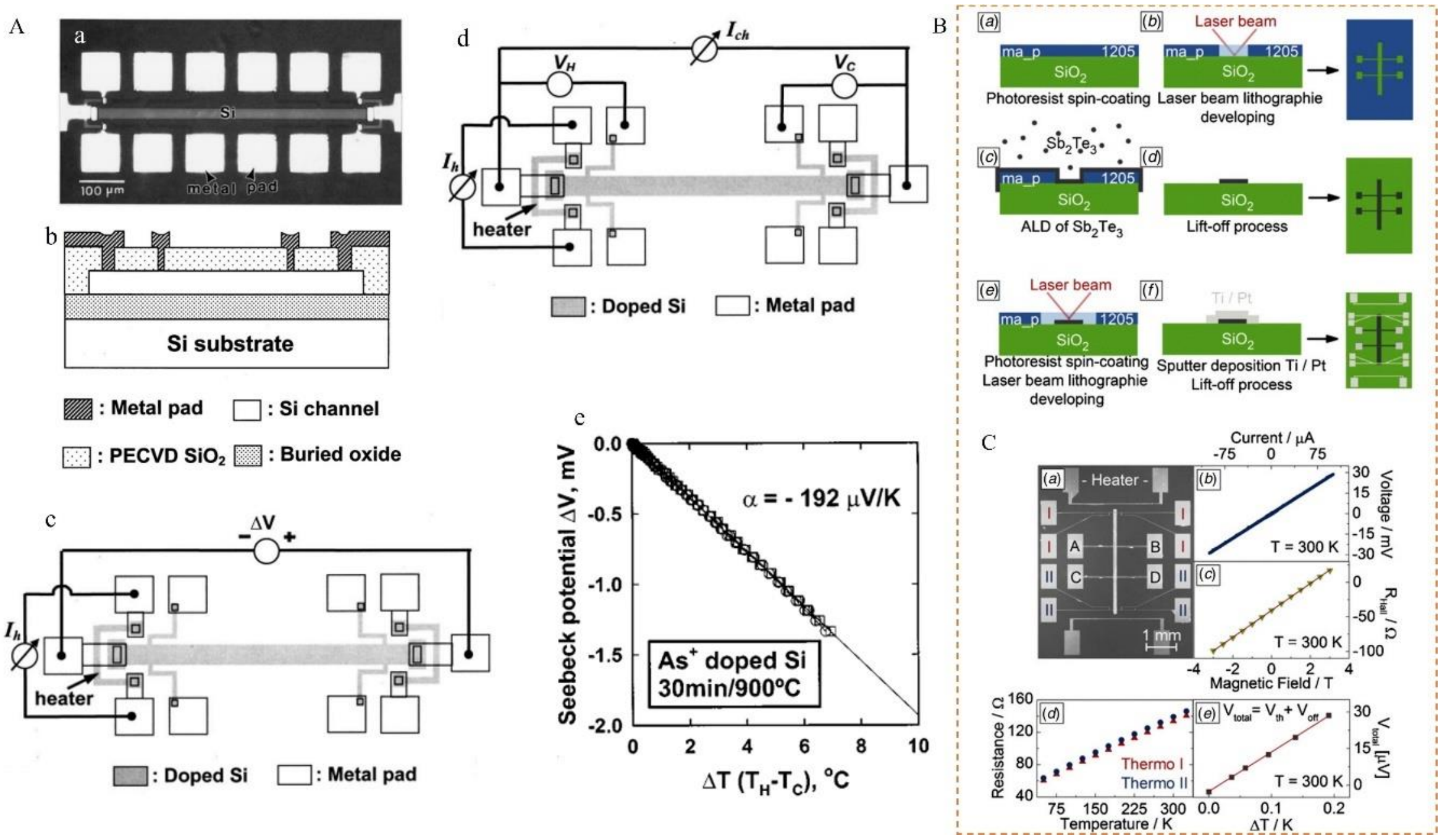

Figure 6: A: Si thin films. (a-b) plane view and cross-sectional view of the testing structure for *S* measurement; (c) Seebeck voltage measurement setup; (d) Si channel temperature measurement setup; (e) A plot of Seebeck potential vs. the temperature difference across the Si channel.[42] **Frame B**: $Sb_2Te_3$ thin films. (a) device fabrication steps, including two lithography patterning steps. First lithography step for pre-patterning of the Hall bar: the Si/$SiO_2$ substrate is spin-coated with maP-1205 photoresist; (b) The Hall-bar structure is patterned by laser beam lithography and the exposed photoresist is developed subsequently; (c) ALD deposition of $Sb_2Te_3$ on top of the whole substrate; (d) lift-off process in acetone. $Sb_2Te_3$ Hall-bar remains; (e) 2nd lithography step for the contact pattern on the Hall bar; (f) Sputter deposition of the metallic contact materials Ti/Pt and second lift-off process. **Frame C**: (a) Micrograph of the measurement device, showing a 128 nm thick $Sb_2Te_3$ Hall bar with inner contacts A, B, C, D, two micro-heaters, and two resistive thermometers I (red), II (blue); (b) *I-V* curve at 300 K; (c) Hall resistance $R_{Hall}$ vs. the magnetic field *B* at 300 K; (d) Temperature calibration of thermometers I (red) and II (dark blue); (e) Total measured voltage $V_{total}$ while stepwise tuning the $\Delta T$, where $V_{total} = V_{th} + V_{off}$ (with thermovoltage $V_{th}$ and offset voltage $V_{off}$).[39]

A method for measuring $S_{||}$ of thin film materials at the chip scale has been devised.[42] This technique was applied to explore the TE characteristics of Si thin films in SOI wafers. In this set up, there is a thin buried oxide layer between top Si thin film and SOI substrate. The buried oxide layer offers both electrical and partial thermal isolation. In the absence of any asymmetrical thermoelectric effects in Si thin films, the Si substrate, which is highly thermal conductive, tends to eliminate such effects. The testing structure utilized for measuring the *S* is depicted in Figure 6A(a & b). The channel of silicon with doping was measured to be 24 μm in width and 520 μm in length. To determine the *S* of Si thin films, it is necessary to measure the Seebeck potential that is generated within the Si channel when the two ends of channel are maintained at different temperatures (Figure 6Ac). By passing an electrical current through the Si heater, one side of the Si channel was subjected to a rise in temperature. The observed Seebeck potential ($\Delta V$), displays an increase in magnitude as the heater current is augmented. The temperatures at both ends of the channel were determined under the same heating setting by passing a modest continuous current through the main Si channel and measuring the voltage drop across a short section of the channel near both ends, as shown in Figure 6Ad. The determination of channel resistance, *RH* and *RC*, was executed through the application of Ohm's law. If the thermal coefficient of resistance (*TCR*) of Si is known, the change in Si channel resistance at both ends can be utilized to determine the temperatures at both ends. This technique offers the advantage of utilizing Si as an in-built temperature monitor, thereby obviating the need for additional temperature sensors. Temperatures, $T_H$ and $T_C$, were calculated from resistance changes, RH and RC. As a result, the *S* may be calculated using the plot of $\Delta V$ vs ($T_H$ - $T_C$) shown in Figure *6*Ae.

Zastrow et al. have designed a platform to characterize *S*, $\sigma$, and Hall coefficient $R_H$ on the same film (Figure *6* frames B and C).[39] The $Sb_2Te_3$ thin films (128 nm thick) were grown via ALD on Si wafers with a top layer of 300 nm thermal silicon oxide as a substrate. The integration of films was carried out on a measurement platform that was specifically designed for this purpose. This platform was comprised of a Hall-bar structure, resistive thermometers, and a microheater. Figure *6* frame B shows the details of the sample preparation. A typical device used in the experiment is shown in Figure *6*Ca. To determine the *S*, a direct current was administered to an on-chip heater (as depicted in Figure *6*Ca) near the film. This resulted in the generation of a $\Delta T$ along the Hall bar. The $\Delta T$ was ascertained by analyzing the resistivity change $dR/dT$ of two metallic thermometer lines, I (colored red) and II (colored dark blue), which were positioned at the ends of the Hall-bar. The experimental methodology involved the implementation of a four-terminal lock-in technique for each thermometer, whereby a sinusoidal excitation current of 1 μA was administered. Two thermometers were operated at different frequencies of 128.1 and 186.1 Hz to prevent any interference between them. These thermometers were calibrated based on the substrate temperature, which was determined by the base temperature of the sample stage in the cryostat. The observed calibration curves, as depicted in Figure 6Cd, exhibit metallic characteristics. For the calculation of the *S* using the formula $S = -V_{th}/\Delta T$, it is necessary to ensure that the assumption of a uniform *S* across the entire film and a negligible $\Delta T$ along the Hall bar is met. Hence, the selection of the current passing through the micro-heater was made in a manner that ensured the maximum temperature gradient remained below 0.5 K. The small temperature gradients had an additional impact whereby the cryostat temperature was maintained at the cold end of the film, within the four point thermometers' minimum resolution of 10 mK. This prevented an undue increase in the average temperature of the $Sb_2Te_3$ film. The thermovoltage $V_{th}$ was ascertained through linear regression analysis of the total voltage $V_{total}$, which was measured between the ends of the Hall-bar. This measurement was conducted by utilizing one contact of each thermometer as a probe while gradually adjusting the temperature gradient. The aforementioned methodology effectively isolated the offset voltage ($V_{off}$) from the threshold voltage ($V_{th}$), both of which contribute to the total voltage ($V_{total}$). The confirmation of the constancy of *S* during the Seebeck measurement and the negligible increase in the average temperature of the sample is evidenced by the linear dependence of $V_{total}$ on $\Delta T$ (Figure 6Ce). The measured values of the $S$ agreed quite well with other values that have been published in the literature.

Experimental designs based on microfabrication have been utilized for conducting $S_{\perp}$ measurements of thin films (Figure 7a-e and i). In these configurations, a resistive micro-heater is commonly deposited atop the thin film. To confine both thermal and electrical transport in the cross-plane direction, the film is etched down to form a mesa with a micro-heater on top. Thermal energy produced by the micro-heater traverses not only the thin film mesa but also the underlying substrate and additional layers. The total temperature difference between the two electrodes in such instances includes contributions from all device layers along the heat channel. The temperature difference and voltage across the thin film can be extracted to estimate the $S_{\perp}$ by measuring a reference sample that is similar to the test sample but placed on the substrate. Figure 7a-b shows a schematic of a sample having both a test device with a thin film mesa and a reference device without a thin film. The voltage electrode is in direct contact with the thin film's top surface, while the micro-heater is electrically separated from both the voltage electrode and the thin film. As a result, an insulating layer must be formed between the voltage electrode and the micro-heater. With this arrangement, both direct current (DC) and 3$\omega$ alternating current (AC) measurements are possible. The DC approach involves applying a DC bias to the micro-heaters to establish a steady-state temperature gradient across the film and measuring the resulting DC Seebeck voltage. In DC measurements, the signal-to-noise ratio of the Seebeck voltage is low. In the 3$\omega$ method, however, a lock-in procedure is used to reject the noise.

When a current with a frequency of 1$\omega$ is given to the micro-heater situated on the top of the thin film mesa, a thermal wave at a frequency of 2$\omega$ will be produced across the entire sample (right in Figure 7a). The temperature change on top of the thin film can then be obtained by measuring the 3$\omega$ component of voltage across the micro-heater line. The temperature change on the substrate surface can be obtained by doing the same experiment without the thin film mesa (left side in Figure 7a). The temperature difference across the thin film thickness caused by the applied current is the difference between these two temperature changes, i.e., $\Delta T_{2\omega,f} = \Delta T_{2\omega,f+sub} - \Delta T_{2\omega,sub}$. Because of the temperature oscillation, a Seebeck voltage (at the same frequency) is generated between the voltage contact and the GND contact. The Seebeck voltage across the thin film can be calculated by subtracting the corresponding value measured from the device without the thin film mesa, i.e., $V_{2\omega,f} = V_{2\omega,f+sub} - V_{2\omega,sub}$. Thus, the $S_{\perp}$ of the thin film is obtained by dividing this Seebeck voltage by the temperature difference:

$$S_{\perp} = -\frac{V_{2\omega,f}}{\Delta T_{2\omega,f}} = -\frac{V_{2\omega,f+sub} - V_{2\omega,sub}}{\Delta T_{2\omega,f+sub} - \Delta T_{2\omega,sub}}$$

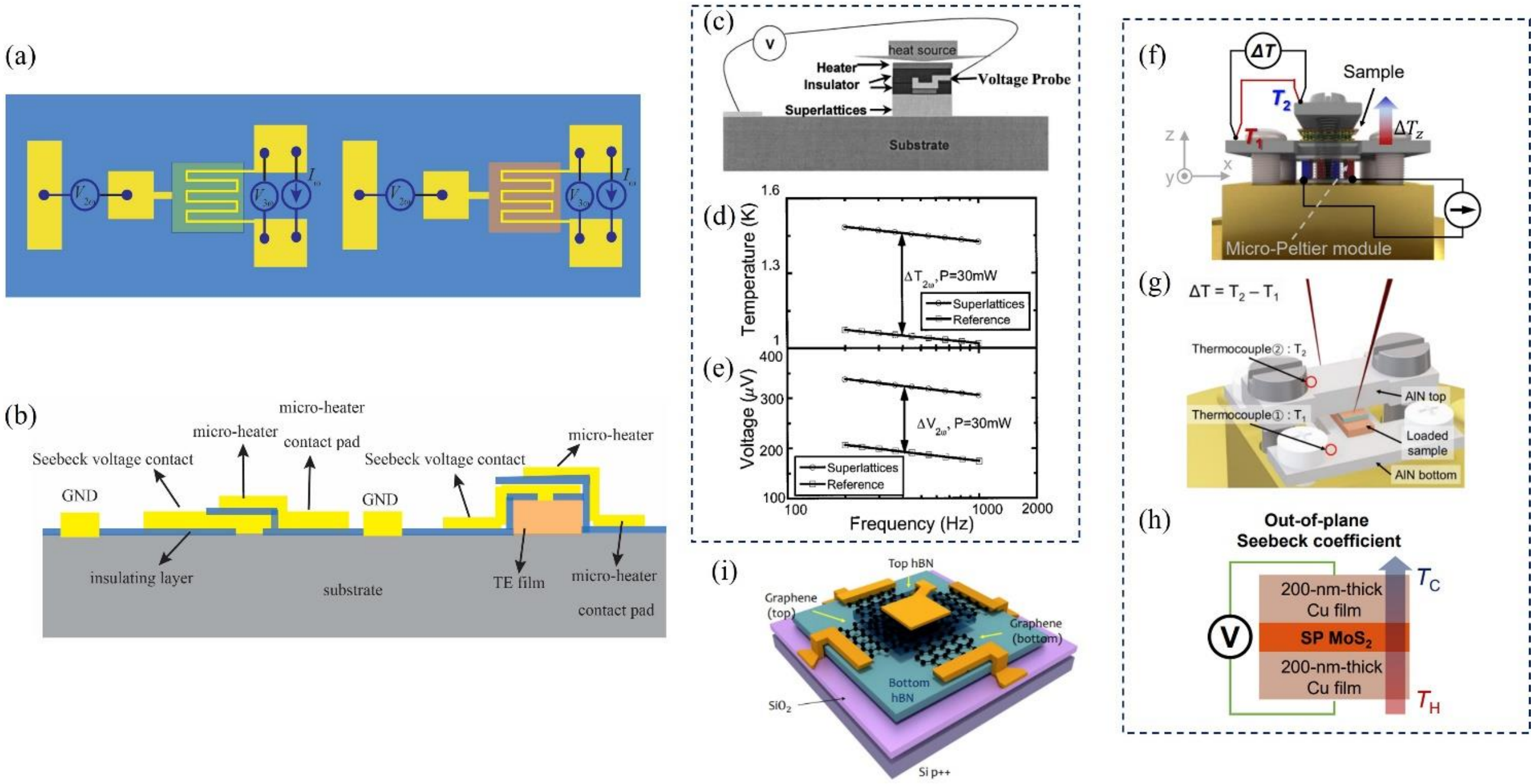

Figure 7: $S_\perp$ measurement of thin films. (a-b) Microfabricated test structure. (a) top view and (b) cross-sectional view of a sample test structure. Two devices are constructed, one without and one with a thin film mesa. A sinusoidal current with $1\omega$ frequency is injected into the micro-heater for an AC measurement, and the $3\omega$ component voltage is measured across the heater line to determine the $\Delta T$ across the device. Separately, the Seebeck voltage of $2\omega$ component is measured between the voltage pad and the GND contact. (c-e) $S_\perp$ measurement of *n*-type Si (75 Å)/Ge(15 Å) QD SL.(c) Schematic depicting a cross section of the SL sample equipped with a heater, temperature sensor, and voltage probe; (d) Measured amplitudes of the temperature and (e) the Seebeck voltage oscillations across the SL sample (circles) and the reference sample (squares) as a function of frequency at 320 K with a heat dissipation of 30 mW. $\Delta T_{2\omega}$ and $\Delta V_{2\omega}$ denote the $2\omega$ temperature and voltage drop across the Si/Ge SL film, respectively.[43] (f-h) $S_\perp$ of single phase and multi-phase $MoS_2$ thin films. (f) side view of test structure. $\Delta T$ was generated between the ends of the films along the *z*-direction using the micro-Peltier as a heater; (g) Schematic illustration of the measurement system used for the $S_\perp$ of the Cu/multilayer $MoS_2$ (~7 nm-thick)/Cu structure; (h) Seebeck voltage measurement contacts during $S_\perp$ measurement.[44] (i) Schematic of the hBN-encapsulated twisted-bilayer Gr device geometry and the contact pads used for the cross-plane TE measurement.[45]

A $3\omega$ method utilizing microfabrication techniques has been devised for the concurrent determination of the $S_\perp$ and $\kappa_{f,\perp}$ in an *n*-type Si(75 Å)/Ge(15 Å) quantum-dot SL.[43] The study revealed that the measured $S_\perp$ of Si/Ge SL exhibits a minor variation in comparison to the bulk values, unlike the $\kappa_{f,\perp}$. Using a similar technique, the $S_\perp$ and $\kappa_{f,\perp}$ of 3 μm thick ErAs:InGaAs/InGaAlAs SL films deposited on InP substrates were determined.[46] Using the DC method, the $S_\perp$ (from 50 K to 300 K) of *n*-type InGaAs (5 nm)/InAlAs (3 nm) SLs with varying doping densities has been measured.[47]

Despite the sub nanometer separation of the van der Waals gap (~0.5 nm), the coupling of the two graphene layers in twisted bilayer graphene (Gr) varies strongly with temperature ($T$) and the twist or misorientation angle between the hexagonal lattices of participating Gr layers.[48-50] It is anticipated that the inter-layer phonon drag will also play a crucial role in governing thermal and TE transport across the interface.[51, 52] Mahapatra et al. have measured the cross-plane TE characteristics of a hexagonal boron nitride (hBN)-encapsulated twisted-bilayer Gr structure prepared via a layer-by-layer mechanical transfer approach (common in van der Waals epitaxy).[45] The orientation of the two Gr layers was arranged in a cross configuration, which was fully enclosed by two layers of hBN (Figure 7i) To measure the Seebeck effect across the van der Waals junction, one of the Gr layers was local Joule heated (using the in-plane resistance of the top Gr layer, which serves as both heater and thermometer), establishing an interlayer temperature difference $\Delta T$ and measuring the resulting thermal voltage generated between the layers (Figure 7i). For a sinusoidal heating current $I_h(\omega)$, the thermal component was obtained from the second harmonic ($V_{2\omega}$) of the cross-plane voltage.[53, 54] The interlayer temperature gradient $\Delta T$ was wholly determined by $I_h$, with $\Delta T \propto$ to $I_h{}^2{}_{h,rms}$. It was determined that cross-plane TE transport is regulated by electron scattering and interlayer layer breathing phonon modes, resulting in a unique phonon drag effect over atomic distances.

The $S_\perp$ of CVD-grown $MoS_2$ (2D layered metal dichalcogenides (TMDCs)) film has been measured using a steady state method in order to steady the variation of $S_\perp$ in mixed-phase $MoS_2$ thin films.[44] Using conventional poly(methyl methacrylate) (PMMA)-assisted wet chemical etching, $MoS_2$ thin films were deposited onto a $Cu/SiO_2/Si$ substrate.[55] Finally, RF sputtering was used to form a Cu thin layer (200-nm thick) on the $MoS_2/Cu/SiO_2/Si$ substrate (i.e. $Cu/MoS_2/Cu/SiO_2/Si$ structure), and this was placed between AlN holders that acted as heat baths (Figure 7f-g). A micro-Peltier module was used as the heating source to provide an out-of-plane temperature difference ($\Delta T_\perp$) across the $Cu/MoS_2/Cu$ (~400 nm) sample (Figure 7h), and the two AlN holders were thermally connected by Mo screws to maintain a stable $\Delta T_\perp$ (6 K). Under steady-state conditions, $\Delta T_\perp$ was measured between the two AlN holders using two T-type TCs. The out-of-plane potential difference ($\Delta V_\perp$) was measured simultaneously between the upper and lower Cu thin films using two shielded tungsten needles. The $S_\perp$ for $MoS_2$ film was calculated by linearly fitting $S_\perp$ to $\Delta V_\perp/\Delta T_\perp$ at a given temperature. The measured $S_\perp$ for single- (SP) and mixed-phase (MP) $MoS_2$ films were ≈140 and 294 µV/K at RT. Table 1 list the $S_\parallel$ or $S_\perp$ of films of different materials and the techniques used to measure them.

### 2.3 Seebeck coefficient screening tools

Considering the latest progress in the field of TE material development, there have been new measurement techniques that have been devised to analyze the surface distribution of the $S$. These techniques are aimed at studying chemical and functional inhomogeneities that may arise due to various factors such as synthesis methods, defects, and combinatorial TE materials. The implementation of specific synthesis methods may result in non-uniformity within the material due to variances in composition and doping across the specimen, leading to localized fluctuations in TE characteristics. Certain variations exhibit dimensions within the sub-millimeter range or beyond, albeit their detection through conventional materials characterization methods, such as x-ray diffraction or scanning electron microscopy, may pose a challenge, particularly when the chemical variation is minimal. However, the doping variation, which significantly influences TE performance, may be substantial. Consequently, it is critical to conduct a systematic examination of the local fluctuations in the *S* pertaining to both bulk and thin-film specimens. The subsequent segment provides a succinct overview of such methodologies.

Yan et al. described a TE screening tool capable of measuring the *S* and electrical resistivity ($1/\sigma$) at temperatures ranging from 300 K to 800 K (Figure *8*a-b).[56] The measurement principle is comparable to the "hot probe" technique. The present technique involves the utilization of a TC probe that is subjected to heating by a miniature heater and subsequently placed in contact with the surface of the sample. This action leads to the creation of a thermal gradient in the immediate vicinity of the heated probe. Additionally, a second TC, referred to as the cold probe, is placed in contact with the sample surface at a distance from the hot probe. The *S* is calculated by dividing the thermoelectric voltage by the temperature difference, which may both be measured by the two TCs. In order to mitigate the impact of offset voltage interference in the S measurement, it is possible to modify the temperature of the hot probe. Subsequently, the *S* value can be determined by computing the slope of a linear regression of a sequence of thermoelectric voltages and temperature differentials. The two-probe technique involves the utilization of two TCs, which are positioned approximately 3 mm apart from each other, and are designated as hot and cold probes, respectively. During the scanning procedure, a pair of probes will move in unison. In the one-probe method, the cold probe is locked in place (in this case, the film edge on a 76.2 mm quartz wafer), and only the hot probe moves during the scanning process. The findings suggest that there was a high degree of similarity in the outcomes yielded by both techniques, regardless of whether the cold probe was stationary or co-located with the hot probe. Furthermore, the spatial resolution is constrained not by the separation distance of the two probes, but rather by the surface contact area between the heated probe and the sample. The *S* contour plots for a ternary $CoSb_3$-$LaFe_4Sb_{12}$-$CeFe_4Sb_{12}$ film were depicted in Figure *8*b, where two-probe method was employed for the measurement. It was reported that both methods measured similar values for the *S*, differing from each other by less than 5 μV/K.

Zhang et al., have presented a microprobe methodology that enables the concurrent measurement of both $\kappa$ and *S* of thin films.[57] The microprobe (V-shaped, Figure *8*c), which acts as a heater, a thermometer, and a voltage electrode, is heated by an alternate current with a high enough frequency that the amplitude of its temperature oscillations is negligible. The dc part of the electrical power creates a dc temperature rise above ambient $\Delta T_{sample}$ and a dc Seebeck voltage $V_{Seebeck}$ across the sample. From the observed electrical resistance of the probe and its calibrated *TCR*, the average rise in temperature of the probe wire along its length can be ascertained. When the microprobe is brought into touch with the surface of a sample, the $\Delta T_{probe}$ increases linearly with the amount of electrical power *P*, and the slope ($d\Delta T_{probe}/dP$) gives the effective thermal resistance ($R_{eff}$) of the probe. Underneath the film was a thin Au strip that in conjunction with the microprobe acquires the $V_{Seebeck}$.

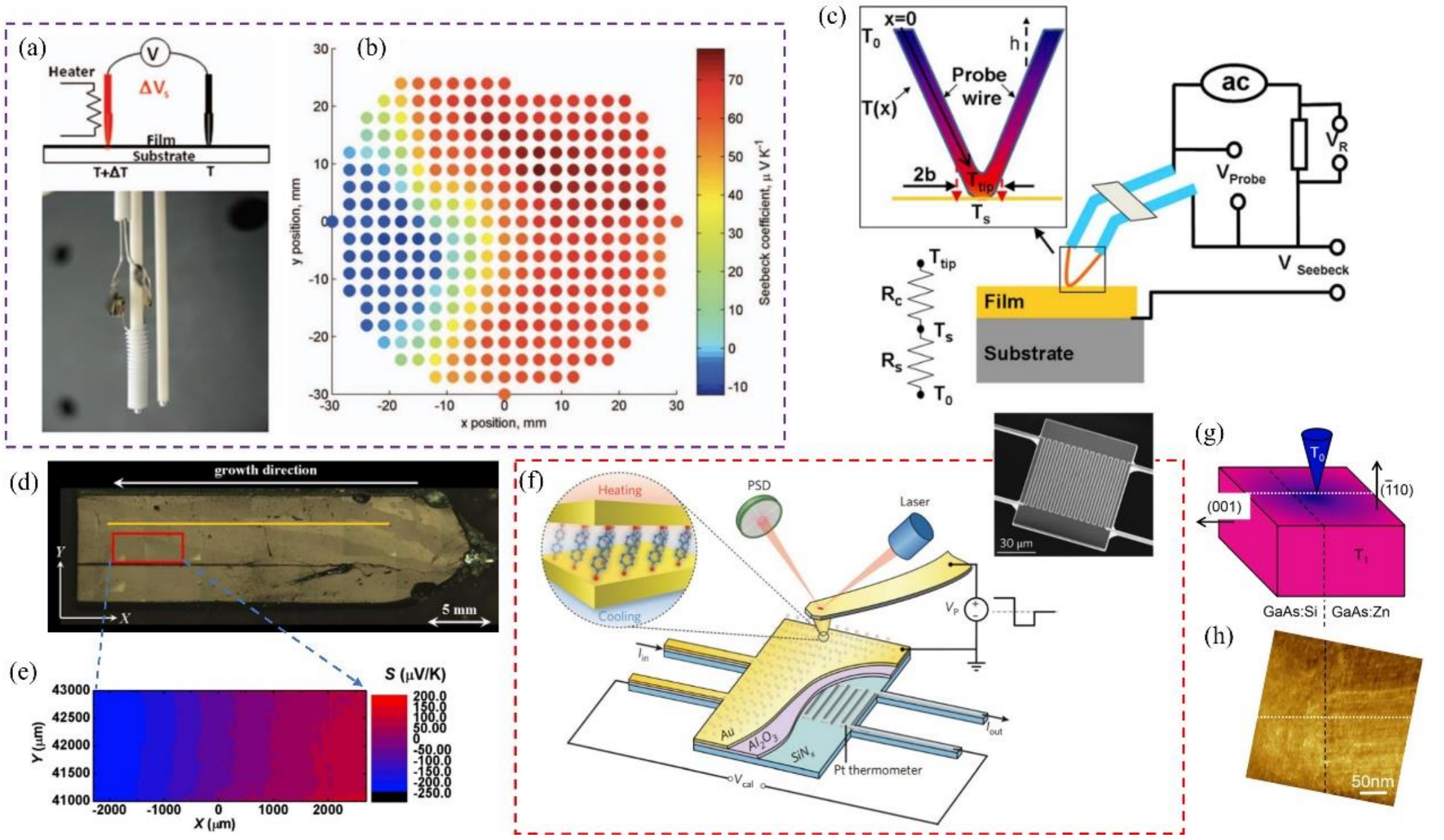

Figure 8: (a-b) hot probe technique for $S$ screening. (a) high temperature TE factor screening tool and the $S$ probe; (b) $S$ contour plots for a ternary $CoSb_3$-$LaFe_4Sb_{12}$-$CeFe_4Sb_{12}$ film deposited on a quartz wafer: measured by two-probe method.[56] (c) sketch of the microprobe measurement set up along with the parameters of the analytical model used to describe tip-sample heat transfer. $T_{tip}$, $T_s$, and $T_0$ are the temperatures of the tip, the surface of the sample, and the ambient, respectively. $R_c$ & $R_s$ are the thermal resistances of the tip-sample contact and the sample, respectively.[57] (d-e) micro-probe assisted spatial distribution analysis of the $S$ for the bismuth-telluride (BT) systems. (d) the cross section cut parallel to the crystal growth direction for the BT system; (e) details of the spread of the $S$ in the BT system measured for a rectangular area near the sign inversion boundary surrounded by red lines in Fig. d with measurement steps of $X$ and $Y$ = 50 μm.[58](f) $S$ at molecular junctions. Experimental platform. SEM image of the custom fabricated microdevice. (g-h) SThEM assisted $S$ screening. (g) SThEM setup with RT probe tip in contact with heated sample, consisting of the cleaved surface of a GaAs $p$-$n$ junction; (h) cross-sectional scanning tunneling microscopy of a $p$-$n$ junction, with junction location defined by the black dashed line and SThEM tip trajectory defined by the white dotted line.[59]

Nakamoto and Nakabayashi have devised a high spatial resolution $S$ measurement technique utilizing a micro-probe with a contact area of 10 μm in diameter.[58] The entire system was enclosed within a confined space to maintain a consistent temperature and minimize high-frequency interference during the measurement. At the base of the sample, a thermal bath was employed to maintain a constant temperature, while a microprobe located at the top was utilized to provide diverse temperature conditions. Moreover, utilizing the $XY$ stage with a spatial resolution of 1 μm enabled the acquisition of the $S$ to be limited to a specific region, thereby providing an opportunity to examine the isotropic characteristics of materials. Both configurations are evidently appropriate for conducting out-of-plane measurements. The Bismuth-Telluride (BT) crystal's cross-sectional view, cut parallel to its growth direction, is depicted in Figure 8d. Using the micro-probe technique a measurement was conducted in the vicinity of the boundary region spanning 2000 × 5000 $mm^2$, with a 50 µm increment, demarcated by red lines in Figure *8*d and shown in Figure *8*e. The phenomenon of sign inversion, specifically from *p*-type to *n*-type, is more pronounced. The observed sign inversion from *p* -type to *n*-type along the direction of crystal growth in this system can be attributed to the segregation of excess Te at the upper portion of the ingot. During crystal growth, the surplus Te acting as a donor is eliminated at the upper section of the ingot. Consequently, the lower region demonstrates *p*-type conductivity, while the upper region, which contains a greater quantity of Te, exhibits *n*-type conductivity.

The investigation of the *S* at molecular junctions has garnered interest in the field of interface engineering, particularly with regards to probing on a smaller scale. Cui et al. have constructed a novel experimental platform utilizing an atomic force microscope (AFM) to investigate the thermoelectric attributes of molecular junctions. In this set up, to produce the Au-molecule-Au molecular junctions, a self-assembled monolayer of organic molecules was sandwiched between an Au-coated AFM probe and an Au surface (Figure *8*f). A $\Delta T$ between the two Au contacts would be induced by applying a voltage bias ($V_P$) through the Au-coated probe due to the Peltier effect. A platinum (Pt) meandering line was inserted beneath the Au layer and separated by an insulating $Al_2O_3$. The Pt thermometer is provided with a constant electric current ($I_{in}$), and the voltage drop ($V_{Cal}$) across the Pt thermometer is regularly monitored. The $\Delta T$ induced by Peltier cooling or Joule heating within the molecular junction of the micro-device may result in a change of the resistance of the Pt thermometer. Consequently, $\Delta T$ is estimated by $\Delta T = V_{Cal}/I_{in}R_{Pt}\alpha$, where $R_{Pt}$ and $\alpha$ are the resistance and the temperature coefficient of resistance of the Pt thermometer, respectively. To determine the *S* of the molecular junctions, a sinusoidal electric current (0.5 Hz frequency) was applied to the Pt resistor. This resulted in a sinusoidal perturbation of the temperature of the suspended micro-device with an amplitude of $\Delta T_{2f}$ and a frequency of $2f$. Following this, a lock-in amplifier is utilized to gauge the thermoelectric voltage ($\Delta V_{2f}$) at $2f = 1$ Hz that results from the applied $\Delta T_{2f}$. The *S* of the junction was determined using the formula $S = S_{Au} - \Delta V_{2f}/\Delta T_{2f}$, where $S_{Au}$ represents the *S* of the Au thin film that was deposited on the micro-device. Walrath et al. reported a direct method for converting thermoelectric voltage profiles, $V(x)$, to *S* profiles, $S(x)$, by employing a quasi-3D conversion matrix that takes into account both the sample shape and the temperature profile.[59] Because of this deconvolution method, scanning thermoelectric microscopy (SThEM) was able to directly determine the *S* across a GaAs *p-n* junction contact (Figure *8*h). This combination computational-experimental technique should enable nanoscale *S* measurements across a wide range of heterostructure interfaces.

Ramrez-Garcia and Arnache-Olmos demonstrated a 2D *S* measurement device capable of scanning thermoelectric materials at many places over their surface.[60] The measuring setup was tested in a temperature range of ambient temperature to 35 °C using a homogeneous constantan (Cu55Ni45 alloy) foil. The results indicated good agreement with the data reported in the literature, with an error of less than 7%. The system employs a heater and a heat sink to reduce the heat flux through the scanning probe, hence lowering the measurement error. The *S* scanning method provides automated measurements in a maximum region of 15 mm × 5 mm with 1-mm steps. The statistical examination of the $S$ using up to 75 recorded data points can help identify inhomogeneities and defective spots along the sample. Once the sample is mounted, one of the heaters is turned on first, followed by a control system waiting for a user-defined target temperature to be reached; once the temperature stops rising, a set of pairs ($V_{chr}$, $V_{al}$) is measured until the temperature difference ($T_A$-$T_B$) reaches a user-defined value (Figure 9b). The Seebeck coefficient $S(T)$ is then estimated from the measured pairs ($V_{chr}$, $V_{al}$) using the equation: $S(T) = (S_{TC}(T)/ (1-\partial V_{chr}/\partial V_{al})) + S_{al}(T)$, where $S_{TC}(T)$ and $S_{al}(T)$ are the $S$s of the K-type TC and its Alumel wire, respectively. $\partial V_{chr}/\partial V_{al}$ is calculated as the slope of the linear fit of ($V_{chr}$, $V_{al}$) and $T = (T_A + T_B)/2$. Figure 9c depicts a plot of the average $S$ at each $X$ position, with error bars calculated using the standard deviation of the two $S$ values measured by two successive heating and cooling cycles.

A probe and system for measuring the $S$ were developed and implemented with the aim of achieving a high-throughput assessment of TE properties.[61] The probe is equipped with a pair of chromel-alumel TCs, with one of them being capable of regulating its own temperature to maintain a temperature differential between the two TCs. The probe was designed with the purpose of achieving uniform contact with any surface that exhibits heterogeneity. Figure 9e depicts a schematic picture of the two-point probe used to assess the $S$ using the Seebeck tester. Two K-type TC make up the two-point probe. A heating element (nichrome wire) covers one TC, which is temperature controlled. These two TCs were linked to the WE7000 PC-based measurement device. The TCs were attached to WE7000 channels CH1 and CH2, respectively (Figure 9d). The temperature of TCs was measured using channels CH1 and CH2. The temperature difference was computed by subtracting the value of CH1 from the value of CH2. Then, to CH3 and CH4, two alumel wires and two chromel wires were attached. CH3 and CH4 can measure the thermoelectric power of the alumel- and chromel-wires via these linkages. Using these interconnections, the $S$ can be computed using

$$S = \frac{V_{CH3}}{T_{CH1} - T_{CH2}} + [-0.0159 \times (T_{CH1} - T_{CH2}) - 5.9785]$$

The term, $[-0.0159 \times (T_{CH1} - T_{CH2}) - 5.9785]$ shows the $S$ of alumel-wire.

In this study, the probe was used to perform a verification test on $n$-type La-doped $Ca_{1-x}La_xMnO_3$ produced through the electrospray deposition (ESD) approach. Because the contact resistance between the two-point probe and the sample was larger than that of the commonly used $S$ measuring devices, the computed $S$ differed from previously published values. The variation of the $S$ by La substitution, on the other hand, showed the same tendency as the results reported by Lan et al.[62]

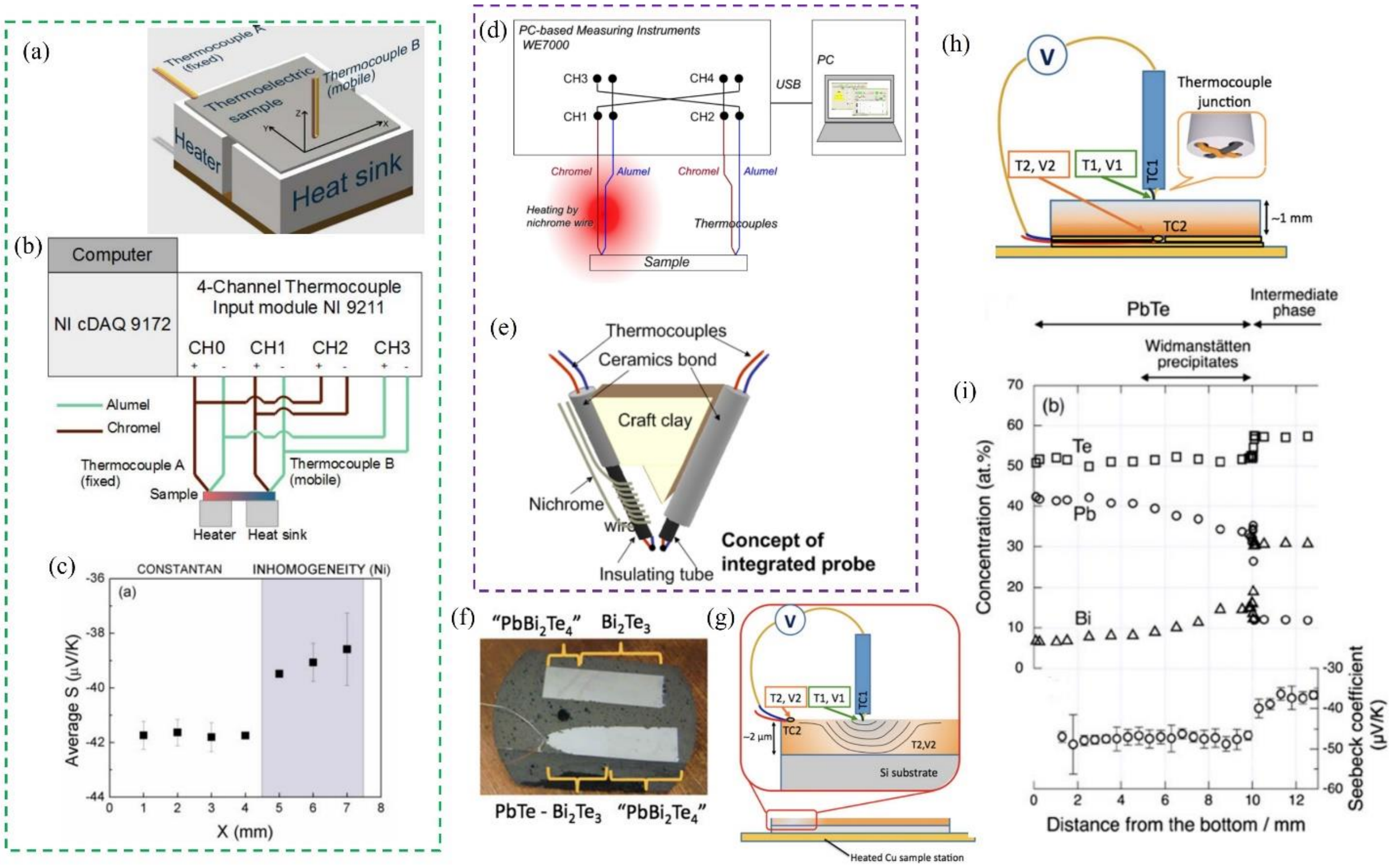

Figure 9 (a-c): 2D scanning *S* mapping system. (a) Photo of scanning system. An *XYZ* translation stage controlled with an Arduino board allows the TC B to be positioned at several points over a region of ~0.75 $cm^2$; (b) The experimental setup to conduct local *S* measurements using the scanning measuring setup. The transfer of thermal energy occurs from the heater to the heat sink via the sample. Two TCs are positioned over the sample, one fixed (A) and the other scanning (B); the wiring scheme illustrates the connections between TCs and the four-channel data gathering system. Channels CH0 and CH1 monitor the temperatures of TCs A and B, respectively, whereas Channels CH2 and CH3 measure the voltages between the chromel and alumel wires, respectively; (c) Avg. *S* of the two values along the modified constantan sample at each *X* point. The inhomogeneity induced by a nickel-based ink begins at $X = 5$ mm. The standard deviation is represented by the error bars.[60] (d-e) A probe and system for measuring *S*. (d) *S* measurement concept; (e) schematics of two-point probe.[61] (f-i) *S* measurement with cold scanning. Schematic of the setup for (g) thin-film sample (h) bulk sample; (f) Compositionally graded sample; (i) Correlation of Te, Pb, and Bi compositional change with *S* at various locations throughout the length of the sample.[63]

Iwanaga et al. demonstrated a simple scanning *S* measurement setup with a cold scanning tip on bulk and thin-film samples (Figure 9f-i).[63] The primary benefit of this system lies in its ability to evaluate the homogeneity of the film sample that has been deposited on an insulating substrate or a multilayer film that is insulated from a conducting substrate. This approach circumvents the need for substrate removal prior to the evaluation of the *S* of a film. Additionally, it mitigates the risk of material deterioration during subsequent processing procedures that are typically mandatory for Seebeck measurements using alternative techniques. The *S* of a bulk, compositionally graded material was measured using the apparatus, as illustrated in Figure 9f. The alloy has an overall composition of $Pb_{14}Bi_{28.8}Te_{57.2}$ and is produced via unidirectional solidification using the Bridgman process (Figure 9f).[64] Figure 9h-i depicts the changes in material composition as well as the measured $S$, which varies with location.

In the process of $S$ measurement, particularly when using the differential method, it is essential to consider the impact of offset voltages that may arise from non-ideal contact or material inhomogeneity. These factors must be minimized to ensure precise and accurate measurements. The total voltage $\Delta V$ obtained for $S$ calculation is in fact composed of three parts: voltage generated by the samples ($\Delta V_s$); voltage generated by the electrode materials ($\Delta V_e$), which can be removed from the final calculation because $S_e$ ($S$ of electrode material) is already known; voltage generated between the samples and the electrode wires ($\Delta V_c$). In other circumstances, the contact voltages $\Delta V_c$ that cause the offset are temperature invariant and $\Delta V_c << \Delta V$, therefore the effect of this component can be ignored. The resulting $S$ can be estimated as $S = -\Delta V/\Delta T + S_e$. Otherwise, it is necessary to consider such an effect. To solve this problem, inversion method can be invoked. In this procedure, two total voltages ($\Delta V_1 = -S\cdot\Delta T + \Delta V_e + \Delta V_c$, $\Delta V_2 = -S - \Delta T + \Delta V_e + \Delta V_c$) are obtained by inverting the temperature gradient ($\Delta T$, $-\Delta T$) and the $S$ is extracted as $S = (\Delta V_1 - \Delta V_2)/2\Delta T$. Inaccuracies in the determination of the temperature gradient across the samples pose a potential problem for the precision of the measurement. In general, thermal energy can propagate through radiation into the surrounding environment, leading to significant changes in the profile of temperature distribution. This phenomenon has a significant impact on the integral technique, particularly because the hot side with a higher temperature will radiate a large quantity of heat into the surrounding environment, causing the cold side to remain at a constant temperature. To remedy this problem, encasing the heat source to reduce heat loss or providing a constant temperature for the cold source may be an option. Furthermore, because there is a $\Delta T$ between the TC and the detecting point on the sample, if the TC is used as both a temperature sensor and a heat source, the $\Delta T$ across the sample may be overstated. Since most of the TE films are deposited on certain substrates, the transfer of heat into the substrates can have an impact on the proper measurement of the $S$. When using microprobe methods to determine the distribution of the local $S$ (which shows how uniform the sample surface is), the flow of heat into the substrate changes the acquisition of the $S$s.

## 3 Thin film $\sigma$ characterization techniques

The primary challenge in the adaptation of traditional techniques for the determination of $\sigma$ in semiconductors pertains to the level of precision they offer. In the context of semiconductor materials, it is observed that a significant contact resistance exists between the metal wire and the semiconductor within the measurement module. This resistance is known to surpass the bulk resistance of the semiconductor. The presence of contact resistance throughout the measurement circuit results in a significant contribution of contact resistance to the measured resistivity. Furthermore, the voltmeter can quantify the electrical potential difference of the conductive pathways, concomitant with the electrical current traversing them, thereby leading to augmented resistance measurement and diminished conductance. Hence, the conventional approach is inadequate for precise determination of the electrical resistance of semiconductor materials. The prevalent techniques employed for the practical determination of resistivity comprise the four-probe method, bridge method, and Van der Pauw method. The subsequent section will provide a brief overview of each measurement method.

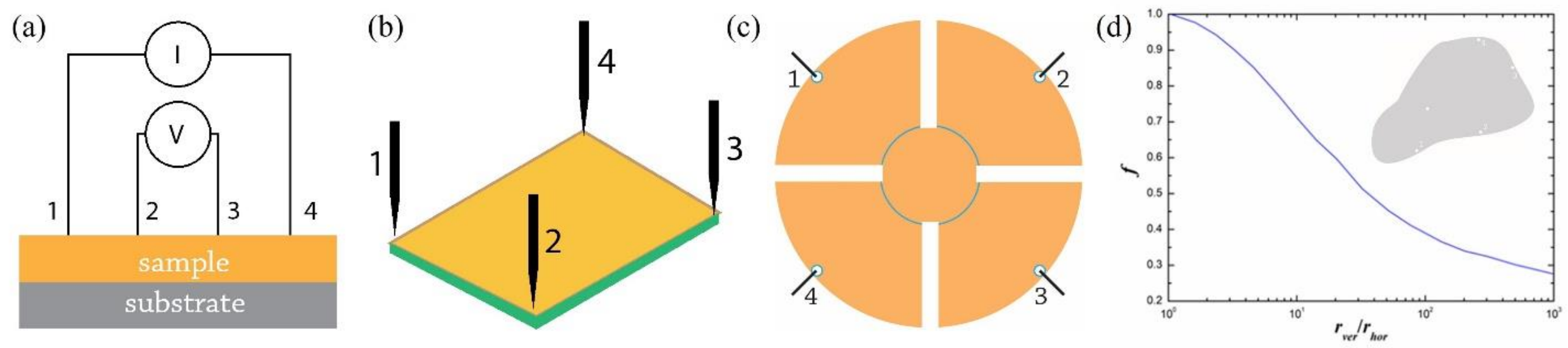

Figure 10: (a) Typical collinear four probe configurations for measuring $\sigma$ of thin film; (b) conventional probing geometry for the van der Pauw method; (c) circular cloverleaf geometry; (d) $r$ vs. $f$ in Equation 1. Inset shows sample of arbitrary shape.

The $\sigma$ of thin films can be determined through utilization of a four-point probe technique. As illustrated in (Figure 10a), the sample surface is marked with four contact points that are collinear and equidistant. A current, $I$, is passed via the outer contacts (1 and 4), and the voltage drop $V$ across the inner electrodes (3 and 4) is measured. When the sample size is significantly larger than the probe spacing and the thickness of the film ($d$) is less than half of the contact spacing, the $\sigma$ of the film can be calculated using the formula $\sigma = \frac{ln2}{\pi d}\frac{I}{V}$. When the thickness of the film is exceedingly thin, this method can reduce any contact resistance during the measurement. To carry in-plane $\sigma$ and $S$ measurements, it is necessary for the substrate to possess non-conducting properties. On the other hand, cross-plane measurements necessitate the utilization of specially fabricated devices to confine the current and minimize contact resistance, thereby reducing the film thickness. This method is not suited for integrated semiconductor layers. The placement of four contacts on a conducting layer can prove to be a challenging task due to the limited surface area available. The van der Pauw method (Figure 10b) is a simple four-probe methodology for measuring the $\sigma$ of thin film samples of arbitrary shape.[65] There are five prerequisites for using this technique:[66] 1. The sample must be flat and uniform in thickness; 2. The sample must not include any isolated holes; 3. The sample must be homogeneous and isotropic; 4. All four ohmic contacts must be located at the sample's edges; and 5. Individual ohmic contacts must be infinitely small, with an area of contact at least an order of magnitude less than the total sample area. Moreover, it is imperative that the thickness of the sample is less than both its width and length. The method entails obtaining two resistance measurements, namely $r_{ver}$ and $r_{hor}$, to determine a singular measurement of $\sigma$. For one direction of current, example, from contact 1 to 2 (Figure 10b), when the voltage is measured between 3 and 4, the first resistance is obtained: $r_{12,34} = V_{34}/I_{12}$. For accuracy, the contact points are swapped in terms of applied $I$ and measured $V$ to obtain $r_{34,12} = V_{12}/I_{34}$. $r_{ver}$ is the average of $r_{12,34}$ and $r_{34,12}$. The $r_{hor}$ is obtained in a similar fashion. A current is passed from contact 1 to the adjacent contact 4, and the voltage between 2 and 3 is measured to get $r_{14,23} = V_{23}/I_{14}$. And again $r_{23,14} =$ $V_{14}/I_{23}$ is also obtained. $r_{hor}$ is the average of $r_{14,23}$ and $r_{23,14}$. To remove parasitic voltages that may be present during the measurement, it is often necessary to reverse the current polarity and test various voltages created by varied magnitudes of current. Then according to Van der Pauw's theorem for an isotropic conductor[65]:

$$exp^{\left(\frac{\pi d}{\rho} r_{hor}\right)} + exp^{\left(\frac{\pi d}{\rho} r_{ver}\right)} = 1 \qquad \textit{Equation 1}$$

Where $d$ and $\rho$ are the sample thickness and resistivity, respectively. If $d$, $r_{hor}$, and $r_{ver}$ are known, $\rho$ can be obtained via Equation 1. For a rotationally symmetric sample, $r_{hor} \approx r_{ver}$, in that case $\rho = \pi d(r_{hor}+r_{ver})/2\cdot\ln2$. A circular cloverleaf sample geometry is preferred to minimize measurement error (where the diameter of the inner area is ¼ of the whole sample, Figure 10c).

However, in the general case of arbitrary shape, Equation 1 is equivalent to *Equation 2*.

$$\rho = \frac{\pi d}{2ln2}\frac{r_{ver}+r_{hor}}{2}f(r) \qquad \textit{Equation 2}$$

$$cosh\left(\frac{r-1}{r+1}\frac{ln2}{f}\right) = \frac{1}{2}\exp\left(\frac{ln2}{f}\right) \qquad \textit{Equation 3}$$

Where $f$ is called geometric factor, which can be determined numerically or graphically.[67-70] $f$ is the only function of $r = \frac{r_{ver}}{r_{hor}}$, which satisfies the *Equation 3* and graphically represented in Figure 10d. Thus, to determine $\rho$, we need to have $r$ value first, then get the corresponding $f$ value from the graph and calculate $\rho$ from Equation 2. Following the introduction of the vdP method, several alternative approaches have been proposed for the determination of $f$. de Vries and Vieck presented results for $f(\frac{r_{ver}}{r_{hor}})$ for polyethylene samples with parallelepiped geometries and aspect ratios (length/width) ranging from 0.6 to 2.6.[68] Furthermore, several reports have considered non-ideal conditions in vdP measurements, taking into account contact, sample thickness, and sample inhomogeneity corrections.[71-76]

Table 1: Film, periodic structure, growth method (& steps), electrical conductivity ($\sigma$) etc, Seebeck coefficient ($S$), power factor ($PF = S^2\sigma$), thermal conductivity ($\kappa$ or $\kappa_l$), and $zT$ of different material films reported in the literature. Units of charge carrier concentration, $n$ is $cm^{-3}$; Units of mobility, $\mu$ is $cm^2/V\text{-}s$. Abbreviations: EChem = Electrochemical; MS = Magnetron Sputtering; RFMS = Radio Frequency MS; DCMS = Direct Current MS; ther. co-evap = thermal co-evaporation; $T_{sub}$ = substrate temperature; $T_{an}$ = annealing temperature; $t_f$ = film thickness; RT = room temperature, $\kappa_{f,\perp}$ = cross-plane $\kappa$ of the thin film; $\kappa_{f,\parallel}$ = in-plane $\kappa$ of thin film; $\sigma_{f,\parallel}$ = in-plane $\sigma$ of thin film; vdP = van der Pauw.

| Film, periodic structure | Growth method/substrate | $\sigma$ ( 1/Ω-cm) &/or $\mu$, $n$/test method | $S$ μV/K//test method | $PF$ $\mu W/cm\text{-}K^2$ | $\kappa$ W/m-K//test method | $zT$ | Ref. |
|---|---|---|---|---|---|---|---|
| $Ge/Si_{1-x}Ge_x$ *p*-$Ge/Si_{0.25}Ge_{0.75}$ QW width: (65 Å) | LEPECVD/SOI (001) | $\sigma_{f,\parallel}$ 279 4-point DC | 255 ($S_{\|\|}$); microfabricated device | 18.2 | 4.0; SThAFM | 0.135 at RT | [77] |
| *n*-type strained Si(5 Å)/Ge(7 Å) SL; $t_f$ = 1200 Å; 100 periods | MBE/Si | 20; Micro-probe | -500; differential | | $\kappa_{f,\perp}$ 1-2; 2$\omega$ | 0.11 at 300 K | [78] |
| $Si_{1-x}Ge_xAu_y$ SL $t_f$ = ~200 nm | MBE/sapphire | 1000; 4-probe | 502; steady-state | 40 | $\kappa_{f,\perp}$ 0.72; TDTR | expected: 1 | [79] |
| Bi-doped *n*-type PbTe/ $PbSe_{0.20}Te_{0.80}$ SLs ($t_f$ = 1.8 nm) & *p*-type PbTe SL ($t_f$ < 1 nm) | MBE/$BaF_2$(111); different number of periods (*P*s) | *n*-type: $\mu$ = 1087-1218 (*P*s of 155-660); *p*-type: $\mu$ = 604-672 (*P*s of 54-460) | | *n*-type SLs: ~26.5 for *P*s of 660-155; *p*-type: 25 at 300 K (*P*s of 54-460) | $\kappa_{f,\parallel}$ *n*-type: 1.7,1.93, 2.3; *p*-type: 2.32 for *P*s = 460; bridge method[80] | *n*-type: 0.45 (*P*s = 155); *p*-type: 0.33 (*P*s = 460) | [81] |
| *p*-$Bi_{0.5}Sb_{1.5}Te_3$ *n*-$Bi_2Se_{0.3}Se_{2.7}$ & $t_f$ = 2 μm | magnetron co-sputtering/$SiO_2$ | for *p*-: 590; *n* = 5.1 × $10^{19}$; $\mu$ = 73; for n-:650; *n* = 6.8 × $10^{19}$; $\mu$ = 62; 4-probe | 207 -196; differential | | 0.96 (*p*-) 0.91 (*n*-) laser flash method | 0.79 (*p*-) 0.82 (*n*-) at RT for both | [82] |
| *p*-type $Bi_{0.5}Sb_{1.5}Te_3$ $t_f$ = 420 nm | MS/PI (Bi$_{0.5}$Sb$_{1.5}$Te$_3$ and Te as targets) | 9.2 × $10^2$; *n* = 2.73 × $10^{19}$; $\mu$ = 157.8; 4-probe | | ~23.2 at RT; peak *PF*: ~25 at 360 K | ~0.8 at RT; TDTR | | [83] |
| $Bi_2Te_3$/CNT scaffold $t_f$ = 20 nm | MS/CNT scaffold; Te & $Bi_2Te_3$ targets | $\sigma_{f,\parallel}$ 73-86 at RT; 4-probe | -142 to -147 at RT; differential | 1.6 | $\kappa_{f,\parallel}$ 0.19 at RT; Suspd. membrane | 0.25 at RT | [84] |
| $Bi_2Te_3$ $t_f$ = ~1 μm | co-evaporation/ PI | $\sigma_{f,\parallel}$ 1000; *n*= 3×$10^{19}$ to 20×$10^{19}$ & $\mu$ = 80-170; 4-probe vdP | 250 | 48.7 | 1.3 at RT; differential method**[85]** | expected *zT* of ~1 at 300 K | [86] |
| $Bi_{1.1}Te_{3.0}/Sb_2Te_3$ multilayer system $t_f$ = 10 nm each layer | two electron gun evaporators/Si & AlN cross-plane meas. | | $S_\perp$:127.2 for 5 BT/6 ST layers; 125.5 | | 0.62 & 0.51 for 11 &29 layers resp.; | | [87] |
| | | | for 19 BT/ 20 ST layers; integral | | $3\omega$ | | |
| $Bi_{0.5}Sb_{1.5}Te_3$ $t_f$ = 500 nm | DC MS sputtering/Cu | $\sigma_{f,\parallel}$ 400; 4-probe | 150 at RT; differential | 12 at 200 °C | $\kappa_{f,\perp}$ 1 at RT; $3\omega$ | 0.73 | [88] |
| *p*-type $Sb_2Te_3$ (ST) and *n*-type $Bi_2Te_3$ (BT) $t_f$ = 1 μm both | RFMS/glass; different $T_{an}$s | $\sigma_{f,\parallel}$ at RT 1077 (ST, $T_{an}$ = 400 °C); 519 (BT, $T_{an}$ = 300 °C) 4-probe | at RT ST and BT: 140 and -140 resp. both at $T_{an}$ = 300 °C; integral | ST: 12.7; BT: 10.2, $T_{an}$ = 300 °C | | ST: 0.48 BT: 0.60 both at RT | [89] |
| *n*-$Bi_2(Te,Se)_3$ nanowire array $t_f$ = 1 μm | co-evaporation/ $SiO_2$ | $\sigma_{f,\parallel}$ 490; $n = 3.2 \times 10^{19}$; $\mu$ = 97; 4-probe | -189; differential | | $\kappa_{f,\parallel}$; 0.52; Laser PIT | 1.01 at RT | [90] |
| $Bi_{1.5}Sb_{0.5}Te_3$ 3D freestanding film $t_f$ = 10 μm | EChem/epoxy template | 615 at RT; 4-point probe | -145 at RT K; integral | 13 at RT | 1.14 to 0.89 Microfabricated device and TDTR | ~ 0.56 at 400 K | [91] |
| $\beta$-$Cu_2Se$ (Cu/Se = 1 to 9) $t_f$ = 600-850 nm | pulsed hybrid reactive magn. sputtering (PHRMS)/Kapton | $\sigma_{f,\parallel}$; 1000; 4-terminal method | ~120; differential | 11 (in-plane) (Cu/Se = 2) | $\kappa_{f,\perp}$ $\kappa$ ~0.8; SThM-$3\omega$ | 0.4 at RT (Cu/Se = 2) | [92] |
| $Cu_{1.83}Ag_{0.009}Se_{0.77}S_{0.23}$ $t_f$ = 60-90 nm | precursor spin coating & heating /quartz or Si; $T_{an}$ = 350 °C | 1010 at RT; vdP method | 51 at RT; steady state slope | | 0.43 at RT; differential $3\omega$ | | [93] |
| $Cu_2Se$ $t_f$ = 100 nm | MS/glass | $4.55 \times 10^3$ at 310 K; dc 4-terminal | 33.51 at 374 K; steady state | 2.4 at 368 K | 2.14 at RT; TDTR | 0.073 at 374 K | [94] |
| $Cu_2Se$ $t_f$ = several μm | cosolvent method/spin coating/glass | ~250; 4-probe vdP | 200-250; differential | 6.53 | < 1 TDTR | 0.34 | [95] |
| $Cu_2Se$ $t_f$ = 50-100 nm | spin-coating/glass; $T_{an}$ = 400 °C | 1000 at RT 4-point vdP | 80 at RT; integral | | $\kappa_{f,\perp}$; 0.9 at RT; TDTR | 0.14 at RT | [96] |
| SnSe $t_f$ = 180 nm | reactive evaporation/glass | 0.11 to 0.15 in 4-300 K; | 7863; differential [19] | 7.2 at 42 K | 0.023-0.045 W/m-K; | 1.2 at 42 K | [97] |
| | | $n$ = 8.7 × $10^{16}$; $\mu_h$ = 10.8; DC 4-probe | | | steady state method[19] | | |
| SnSe $t_f$ = 730 nm, 300 nm for normal & glancing angle resp. | PLD/($SiO_2$/Si)-normal (N) & 80° glancing angle (G) | 21 at 300 K and 274 at 580 K (both for G); 4-point | 193.7 at 477 K (N); 498.5 at 426 K (G); differential | 18.5 at 478 K for G; | $\kappa_{f,\perp}$ max. ~0.189 at 340 K; 3$\omega$ | | [98] |
| $a$-axis oriented SnSe films $t_f$ = 400 nm | PLD/sapphire($r$-plane) | 28.3 at 800 K; 4-probe | 264 at 800 K; custom built | 1.96 at 800 K | 0.35 at 300 K; 3$\omega$ | 0.45 at 800 K | [99] |
| SnSe $t_f$ = 1 µm | thermal evaporation/glass | $\sigma_{f,\parallel}$; 4.7 at 600 K; 4-probe | > 600 at RT; 140 at 600 K; differential | 0.11 at 505 K | 0.08 (375-450 K); 3$\omega$ in Volklein geometry | 0.055 at 501 K | [100] |
| Yb filled $CoSb_3$ skutterudite $t_f$ = 130 nm | RT DCMS/oxidized Si; 1020 K heat treatment | at RT: 61; $\mu$ = 80; $n$ = 4.8 × $10^{18}$; vdP | at RT: -160; -270 at ~620 K differential | | $\kappa$ ~1.1 at 700 K; 3$\omega$ method | 0.48 at 700 K in $t_f$ = 130 nm | [101] |
| Indium filled $CoSb_3$ $t_f$ = 180 nm | co-sputtering/PI | 270 for 14% In filling 4-probe | gradient method | | $\kappa_l$: 1.10 - 0.05; hot-wire | 0.05-unfilled; 0.56-14% In filling | [102] |
| $n$-type $CoSb_3$ $t_f$ = 190 nm | DC sputtering-annealing at 200 °C/Si(100) & $Al_2O_3$; $T_{an}$ = 350 °C | at RT, 7.87 & 5.9 for Si & $Al_2O_3$ substrates resp. 4-probe | -250 at 550 K for both substrates; differential | . | $\kappa_l$ = 3 at RT; differential 3$\omega$ | | [103] |
| $n$-$SrTi_{0.8}Nb_{0.2}O_{2.75}$ $t_f$ = 300 ± 10 nm | PLD/ $LaAlO_3$ | 614; $n$ = 132 × $10^{19}$; 4-probe | -125; differential | ~9.6 | $\kappa_{f,\perp}$ 3$\omega$ | ~0.29 | [104] |
| $p$-type poly Si & $n$-type poly Si $t_f$ = 1.5 µm, both | PECVD/$SiO_2$/Si | $p$-type: 79 $n$-type: 1333; 4-probe | 460.2; -87.7 | | 30.5; 12.6; 3$\omega$ method | | [105] |
| $p$-CuI $t_f$ = 300 nm | reactive sputtering/ glass + PET | 143 $n$ = 11 × $10^{19}$; 4-Terminal | 162; steady state | 3.7 | $\kappa_{f,\parallel}$; 0.55; differential 3$\omega$ | 0.21 at 300 K | [106] |

## 4 Thermal conductivity ($\kappa$) characterization

Thermal conductivity is a conspicuous transport coefficient to measure consistently, especially when the sample specimen is very small in dimension. According to Fourier's heat conduction equation, $\kappa$ is expressed via: $\dot{Q} = -\kappa\ \Delta T/\Delta l$, where $\dot{Q}$ [W/m$^2$] is the thermal energy transfer rate passing through the specimen (heat flux); $\Delta l$, the sample length; and $\Delta T$, the temperature difference across the sample. The thermophysical experimental systems commonly exhibit a significant level of uncertainty in the measured heat flux values. This is primarily attributed to the parasitic heat transfer that occurs through radiative heat exchange with the surroundings via IR, as well as losses that arise from thermal resistance at interfaces. These factors are inherent and inevitable. Measurements of sample dimensions can contribute significantly to the uncertainty in $\kappa$ measurements.[107] Consequently, despite adhering to established measurement standards, the collective measurement uncertainty for bulk materials can potentially attain a maximum of 20% for measurement setups. The diminution of sample dimensions in thin films exacerbates the situation, necessitating meticulous attention at each stage of experimentation to maintain an acceptable level of overall uncertainty.

The methods for determining $\kappa$ can be classified into two primary groups based on the time dependence of the heating sources, namely steady-state methods and transient methods. The steady-state techniques involve the generation of a consistent temperature gradient across the specimen through the imposition of a heat flux that remains constant over time. The $\kappa$ can be determined by utilizing the information regarding the quantity of heat that is conducted through the sample, the resulting temperature differential, and the dimensions of the sample. Transient approaches impose a time-dependent heat flow on the sample and quantify the related thermal response. Transient approaches are further classified into two types: time-domain methods and frequency-domain methods. The utilization of picosecond transient thermoreflectance for thermal diffusivity measurements of metals has stimulated the development of both time-domain and frequency-domain reflectance measurements.[108] Time-domain techniques involve the acquisition and examination of the thermal response over time subsequent to the application of a heating pulse to the sample. In contrast to the time-domain approach, the frequency-domain approach typically applies cyclic thermal disturbance, resulting in a cyclic signal that is subsequently subjected to phase and amplitude analysis. Certain transient techniques, such as the laser flash method, encounter challenges in directly quantifying $\kappa$ due to the intricacies involved in precisely measuring the heat flow that is conducted through the sample. In such instances, it is frequently advantageous to quantify the thermal diffusivity $\alpha$ or thermal effusivity $e$ instead. The $e$ is related to $\kappa$ via $e = (\kappa c_p \rho_s)^{1/2}$, where $c_p$ is specific heat capacity, and $\rho_s$ is density. The $\kappa$ measurement techniques can also be classified as contact or noncontact, depending on whether the sample comes into direct contact with the heat source and temperature sensors. Noncontact approaches often include optical heating and sensing. Noncontact methods have the advantage of eliminating thermal contact resistance between the sample and the heat source/temperature sensors, allowing for high-accuracy observations.

## 4.1 Steady-State Methods

### 4.1.1 Cross-plane thermal conductivity ($\kappa_{f,\perp}$) measurement

To evaluate the $\kappa_{f,\perp}$ of a film, a temperature drop across a thin film sample must be generated and measured. When the thickness of the specimen is reduced to a range of a few nanometers to tens of microns, the task of generating and quantifying a decrease in temperature becomes exceedingly difficult. The schematic of two commonly used steady-state measurement configurations is depicted in Figure *11*. Thin films with thickness $d_f$ are grown or deposited over a substrate with high $\kappa$ and low surface roughness (e.g., polished silicon wafer) in both configurations. A metallic strip (length $L$ and width $2a$ with $L \gg 2a$) is then deposited onto the thin film whose $\kappa_{f,\perp}$ is to be measured. A metallic strip, such as Cr/Au film (with a high temperature coefficient of resistance), should be used. A direct current (DC) passing through the metallic strip heats it during the experiment. Thus, the metallic strip functions as an electrical heater as well as a sensor for measuring its own temperature $T_h$. The average heater temperature $T_h$ is assumed to be the same as the temperature at the top of the film $T_{f,1}$. The simplest method would be to use a different sensor to directly detect the temperature $T_{f,2}$ at the bottom side of the film (Figure *11*a). However, this method complicates sample preparation processes, which often include cleanroom microfabrication. The other method is to utilize another sensor located at a specified distance from the heater/sensor to detect the temperature rise of the substrate directly beneath it (Figure *11*b). The substrate temperature rise at the heater/sensor location is inferred by utilizing the measured substrate temperature rise at the sensor location through a 2D heat conduction model.

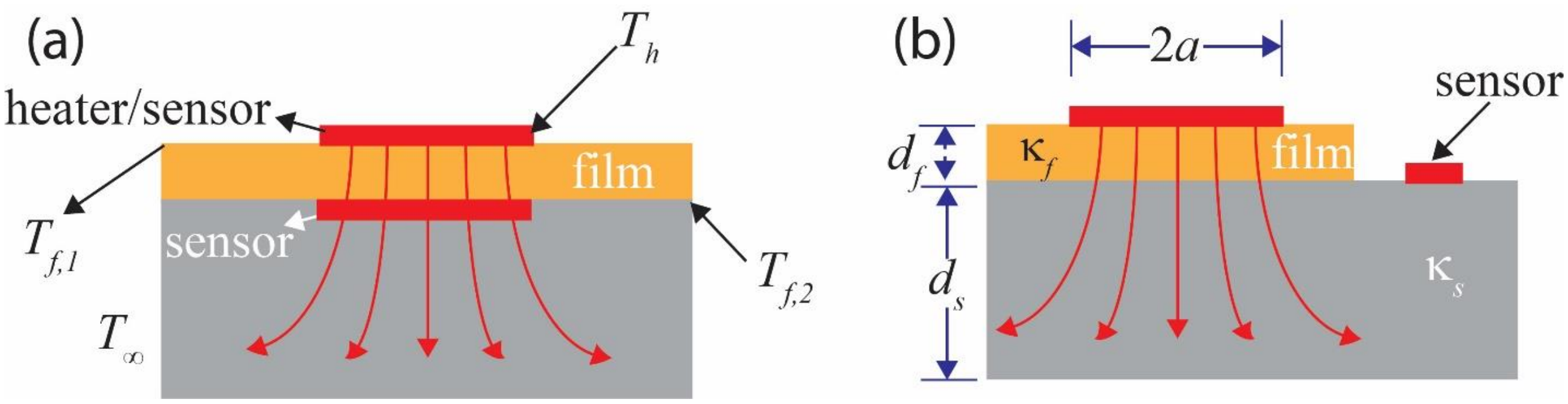

Figure 11: Measuring the $\kappa_{f,\perp}$ of thin films using steady-state methods. The metallic strip in both situations serves as an electrical heater as well as a temperature sensor, monitoring its own temperature rise. The temperature on the thin film's top side $T_{f,1}$ is generally accepted to be the same as the heater/sensor average temperature ($T_h$). $T_\infty$ is the ambient temperature. To measure the temperature $T_{f,2}$ at the thin film's bottom, either (a) a sensor is placed between the thin film and the substrate for direct measurement (figure a), or (b) a sensor is placed at a predefined distance away from the heater (figure b). The temperature at the bottom of the thin film is then calculated using a 2D heat conduction model.

Table 2: Different TE films and their power generation capabilities when TE generators (TEGs) are constructed from them and measurement methods; $t_f$ = film thickness; ⦸ = cross section; PI= polyimide; ∥ = in-plane; vdP = van der Pauw method; $S_{\parallel}$ = in-plane $S$; $\sigma_{\parallel}$ = in-plane $\sigma$

| *p*/*n* materials | Method | Substrate | Number of pairs | $\Delta T$ (K) | Output voltage (mV) | Output power (µW) | Power density µW/ $cm^2$ | $S$, $\sigma$ meas. method | Ref |
|---|---|---|---|---|---|---|---|---|---|
| $Bi_{0.5}Sb_{1.5}Te_3$/ $Bi_2Te_{2.7}Se_{0.3}$ | EChem | metallic | 160; 50 µm thick legs | 20 | 660 | 77 | 770 | differential and 4-probe | [109] |
| $Bi_{2+x}Te_{3-x}$ | EChem, photolithography and etching | mold Si→Su8; ; *p* and *n* leg-length = 100 – 300 µm | | 40 | | | 278 | | [110] |
| $Sb_2Te_3$/ $Bi_2Te_3$ $t_f$ = 500 nm | RF magnetron co-sputtering | kapton | 100 | 40 | 430 | 0.032 | | differential and vdP | [111] |
| $Sb_2Te_3$/$Bi_2Te_3$ $t_f$ = 20 µm | EChem | Si | 242 | 22.3 | 294 | 5.9 | | $S_{\parallel}$ -differential and 4-point probe | [112] |
| Bi-Te alloys *p* & *n* column height = 150 µm | pulsed EChem with EG in | $SiO_2$/Si | 1 ⦸ = 300 $\mu m^2$ | 10 | 1.4 | 0.024 | | temperature differential using Peltier modules | [113] |
| $Bi_2Te_3$-Cu, Cross-plane | Pulsed EChem | Si→Su8 mold | 71 pairs; 80–135 µm thick legs; ⦸ = 300 $\mu m^2$ | 38.64 | 215.5 | | 2434.4 | | [114] |
| $Bi_{0.5}Sb_{1.5}Te_3$/ $Bi_2Se_{0.3}Se_{2.7}$ $t_f$ = 2 µm | magnetron co-sputtering | $SiO_2$ | 98 | 4 | 120.5 | 145.2; $\Delta T$ = 14.6 K for 160 mA | | $S$ and $\sigma$- 4-probe; $\kappa$- laser PIT | [82] |
| $Sb_2Te_3/Bi_2Te_3$ $t_f$ = 10 µm | Pulsed EChem | Si | 127 | 52.5 | 405 | 2990 | 9200 | $S_{\parallel}$- differential and 4-probe | [115] |
| $Bi_2Te_3$-Cu and $Bi_2Te_3$-$Sb_2Te_3$ $t_f$ = 200 µm | EChem | $SiO_2$/Si | 24 | 2-4 | | | 1 & 4 resp. | $S_{\parallel}$-differential/ $\sigma$- 4-probe | [116] |
| $Bi_{0.68}Sb_{1.24}Te_{3.08}$ /$Bi_{1.92}Te_{3.08}$; RTG | EChem | PI | 32 | 57.8 | 170 | 0.247 | | $S$- $\Delta T$ using peltier modules; $\sigma$– Hall meas.syst. | [117] |
| $Sb_2Te_3/Bi_2Te_3$ $t_f$ = 200 µm | EChem | glass template/$SiO_2$/Si | 4 | 138 | 40.89 | 19.72 | | $S$- temperature differential using TE cooler | [118] |
| $Sb_2Te_3/Bi_2Te_3$ | lithography/ photoresist melting | Si | 127 | 123 | 18.5 | 3.14 | ~28.5 | integral method | [119] |
| $Bi_{0.4}Sb_{1.6}Te_3$ $Bi_2Te_{2.7}Se_{0.3}$ | flash evaporation | glass | 7 | 30 | ~84.0 | ~0.21 | ~0.07 | $S_{\parallel}$- temperature gradient; $\sigma$- 4-point probe | [120] |
| Bi-Te Π-structure | EChem | $SiO_2$/(100 nm Au/10 nm Cr); leg height = 20 µm | 880; leg ⦸ = 50 × 50 µm$^2$ | | 17.6 | 0.96 | | $S$- 2-probe technique in a steady state | [121] |
| CuI | reactive sputtering | PET | | 10.8 | 2.5 | 0.008 | | $S$ and $\sigma$; differential; $\kappa_{f,\parallel}$- diff. 3$\omega$ | [106] |
| $Bi_{0.5}Sb_{1.5}Te_3$/ $Bi_{0.5}Te_{2.7}Se_{0.3}$ solar- TEG | MS | PI | 12 | | 150 under 30 mW/cm$^2$ illumination | | | steady state | [122] |
| $Bi_{0.5}Sb_{1.5}Te_3$/ $Bi_2Te_{2.7}Se_{0.3}$ solar- TEG | MS | PI | 12 | 100 | 220 | ~80 | | *S* and *σ*: steady state; *κ*~ hot disk; IR thermal imager | [123] |
| $Sb_2Te_3/Bi_2Te_3$ $t_f$ = 1 µm | RFMS | glass | 11 | 28 | 32 | 0.15 | | $S_{\parallel}$-integral; $\sigma_{//}$ -4-point probe | [89] |
| $Sb_2Te_3/Bi_2Te_3$ | sputtering | PI | 13 | 24 | 48.9 | 0.6935 | | $S_{\parallel}$ and σ - differential | [124] |
| $Sb_2Te_3/Bi_2Te_3$ | Pulsed laser ablation | AlN | 200 | 88 | 500 | | 1040 | $S_{\parallel}$- steady state | [125] |
| Zn-Sb/Al-doped ZnO | DC magnetron co-sputtering | PI | 10 | 180 | | 246.3 | | temperature gradient method and 4-probe | [126] |
| *p*-type $Sb_2Te_3$ (680 nm)/*n*-type $Bi_2Te_3$ (550 nm) | magnetron sputtering | PI | 8 pairs | 60 K | 129 mV; 6.8 mV under $\Delta T$ = 5K when wore on body | 2000 nW | 1420 | differential and 4-probe method | [127] |
| $Sb_2Te_3/Bi_2Te_3$ (20 layers) $t_f$ = 1.5 nm each layer | e-beam evaporation | Si | 128 × 256 elements | 0.5 K/µm | 51 | 0.021 | | differential | [128] |
| $Bi_{0.5}Sb_{1.5}Te_3$/ $Bi_2Te_3$ $t_f$ = 300 nm each layer | RFMS; $T_{an}$ = 200 ºC | Si | 4 | 50 | ~3.6 | 0.0011 | | $S_{\parallel}$ -differential | [129] |
| $Sb_2Te_3/Bi_2Te_3$ $t_f$ = 20 µm | MS/EChem | $SiO_2$/Si | 3.8× 2.7× 0.8 mm | 25-100 | | 3-56 | 30-62 | | [130] |
| $Sb_2Te_3/Bi_2Te_3$ $t_f$ = 100 µm | Pulsed EChem | $SiO_2/Si$ | | 22 | 56 | | 3 | differential | [131] |
| $Bi_{0.48}Sb_{1.52}Te_3$/-$Bi_2Se_{0.3}Te_{2.7}$ $t_f$ = 1 & 0.4 µm | evaporation | PI | 11 | 100 | 50 | | | steady state | [132] |
| *p*-type $Ag_{0.005}Bi_{0.5}Sb_{1.5}Te_3$ | MS | PI | 4 | 60 | 31.2 | | 1400 | *S* and *σ* - steady state | [133] |
| *p*-$(Sb_2Te_3)(Te)_{1.5}$ (630-700 nm)/*n*-$Bi_2Te_3$ thin films | DC sputtering and RFMS (for Sb & Te resp.)/thermal diffusion reaction | PI | 18 legs | 20 K | 46.73 mV | 161.44 nW | >280 | differential and 4-probe method | [134] |
| $Ag_{1.8}Se$ | thermal evaporation | PI | 4 | 50 | | | 4680 | *S* and *σ*- steady state; vdP | [135] |
| $MoS_2$ (*p*-type) and $WS_2$ (*n*-type) planner structure $t_f$ = 700 nm | RFMS | glass | 7.5 cm × 3.6 cm | 240 | 0.7 | | | differential method and linear 4-probe; *κ*-TDTR | [136] |
| Ag/$Cu_2Se$ films(400 nm) | Co-sputtering | PI | 10 legs | 50 K | | 70 nW | 428 | differential and 4-probe method; *κ*-Van-der-Pauw test[137] | [138] |
| $Sb_2Te_3$ thin films/textured Ag-doped $Bi_2Te_3$ (446.6 nm) film | thermally induced close diffusion reaction | PI and glass | 40 pair | 64 K | | | 2100 | differential and 4-probe | [139] |

Table 3: Techniques used to characterize TE parameters of films and the cooling performance of miniature TEC devices. Abbreviations are listed in Table 2.

| *p*/*n* materials | Method | Device configuration; film $S$, $\sigma$, and $\kappa$ characterization techniques | Size (mm$^2$) | Pairs | Cooling performance, $\Delta T$ (K) | Cooling flux (mW/cm$^2$) | Ref |
|---|---|---|---|---|---|---|---|
| $Bi_2Te_3$ SL/$Bi_2Te_3$ SL | MOCVD | cross-plane; $S$, $\sigma$, and $\kappa$ measured simultaneously | | 2 | 55.0 | 128 | [140] |
| $Sb_2Te_3$/$Bi_2Te_3$ | evaporation | cross-plane; integral and 4-probe | 1& 2 stages | 1 & 3 | 13 & 19 | | [141] |
| poly-Si film/poly-Si film | E-beam evaporation/electroplating | in-plane; $\sigma$: 4-probe; $\kappa_{f,\perp}$: $3\omega$ | 100 | 62500 | 5.6 | | [105] |
| $Sb_2Te_3$/$Bi_2Te_3$ | EChem | cross-plane | 2.89 | 63 | 2.0 | | [142] |
| ($Bi_2Te_3$/$Sb_2Te_3$ SL)/ ($Bi_2Te_3$/$Bi_2Te_{2.83}Se_{0.17}$ SL) | MOCVD | cross-plane; integrated method | 3.5 × 3.5 | 49 | 14.9 | $1.3 \times 10^5$ | [143] |
| $Si_{0.89}Ge_{0.10}C_{0.01}$/Si SL $t_f = 2$ μm | MBE | cross-plane; $\kappa_{f,\perp}$: $3\omega$ method | 50 × 50 μm$^2$ | | 2.8 at 25 K; 6.9 at 100 K | | [144] |
| $Sb_2Te_3/Bi_2Te_3$ clustered multipillar (MP) structure | EChem; MEMS | column-type; cross-plane; $\sigma$: 4-probe; $\kappa_{f,\perp}$: $3\omega$ method | 6 × 6 | 100 ‖ ST; 100 ‖ BT MPs | 1.2 | | [105] |
| $Bi_{0.5}Sb_{1.5}Te_3$/ $Bi_2Te_{2.7}Se_{0.3}$ $t_f$ = 10 µm | DC MS sputtering/Cu | In-plane ($S$ & σ,): steady state; 4-probe; | 1 mm × 1 mm | 98 | 6; $\kappa$: modified $3\omega$ method | 1.38 × $10^5$ | [88] |
| $Sb_2Te_3/Bi_2Te_3$ $t_f$ = 700 nm both | co-evaporation/glass | $S$: Integral and VdP | | 1 | 15 | | [145] |
| $Bi_2Te_3/Sb_2Te_3$ SL/ $\delta$-doped $Bi_2Te_{3-x}Se_x$ | MOCVD | $S$, $\sigma$, and $\kappa$ measured simultaneously | | | 43.54 | 257600 | [146] |

### 4.1.2 In-plane thermal conductivity ($\kappa_{f,\parallel}$) measurement

### Volklein Method

The Volklein method is a chip-based approach for determining the $\kappa_{f,\parallel}$ of a thin film less than a micrometer thick.[147, 148]The heater is patterned in the center of a suspended sheet or membrane. Because the thermal sink is positioned at the structure's edge, the in-plane thermal flow is ensured, and thus the temperature rise is governed by $\kappa_{f,\parallel}$.[80] The chip incorporates a rectangular suspended dielectric (e.g., $SiN_x$) membrane with a thickness of 50 to 100 nm. This membrane serves as a substrate on which the film specimen can be deposited. (Figure *12*). Along the long axis of the membrane, a thin metal stripe (Cr/Au film) is created, which serves as both a heater and a thermometer. The Si ridge that surrounds the membrane functions as a heat sink. All measurements must be taken in a vacuum. First, the thermal conductivity of the bare membrane must be ascertained. In order to initiate this process, a direct current is administered to the metallic strip, whereby the resulting thermal energy is subsequently transferred through the membrane and ultimately dissipated into the heat sink. Simultaneously, there will be a loss of heat through radiation from the surfaces of the membrane and metal stripe. An apparent thermal conductance associated with the chip ($G$) can be estimated using the temperature rise and the input electrical power. With appropriate assumptions, it is possible to demonstrate that the thermal conductance of the membrane due to heat conduction and radiation, ($G_M$) ≈ $G$.[149] Because $G_M$ is a function of the unknown thermal conductivity $\kappa_{M,\parallel}$ and emissivity $\varepsilon$ of the membrane, two experiments with two different membrane geometries are required to acquire $\kappa_{M,\parallel}$ and $\varepsilon$ concurrently,[149] i.e., the difference in heater/sensor temperature rise of two measurements using two different thin film widths can then be used to calculate $\kappa_{f,\parallel}$ while keeping all other parameters constant.

Any vapor deposition processes (magnetron sputtering, thermal evaporation, or atomic layer deposition) can be used to deposit TE thin film specimens on the top side (top configuration) or bottom side (bottom configuration) of the membrane. In the event that the film is affixed to the upper surface of the membrane, it is advisable to apply a slender insulating coating onto the metal stripe to prevent any potential electrical shortcuts (Figure *12*b). The thermal conductivity of the composite layer (containing test film and supporting membrane) $\kappa_{C,\parallel}$, is measured in the same way as $\kappa_{M,\parallel}$, to yield the in-plane thermal conductivity of the deposited film ($\kappa_{f,\parallel}$).

The $\kappa_{f,\parallel}$ can be deduced from the following equation[149, 150]

$$\kappa_{f,\parallel} = \frac{\kappa_{C,\parallel} \cdot d_c - \kappa_{M,\parallel} \cdot d_M}{d_f}$$

where $d_f$, $d_M$, $d_C$ are the thicknesses of the thin film, the membrane, and the composite layer, respectively. Using this technique, the $\kappa_{f,\parallel}$ for a $Bi_{87}Sb_{13}$ film (110 nm thick) was measured (120–450 K).[150] The obtained results were in good agreement with the previously published report.[148]

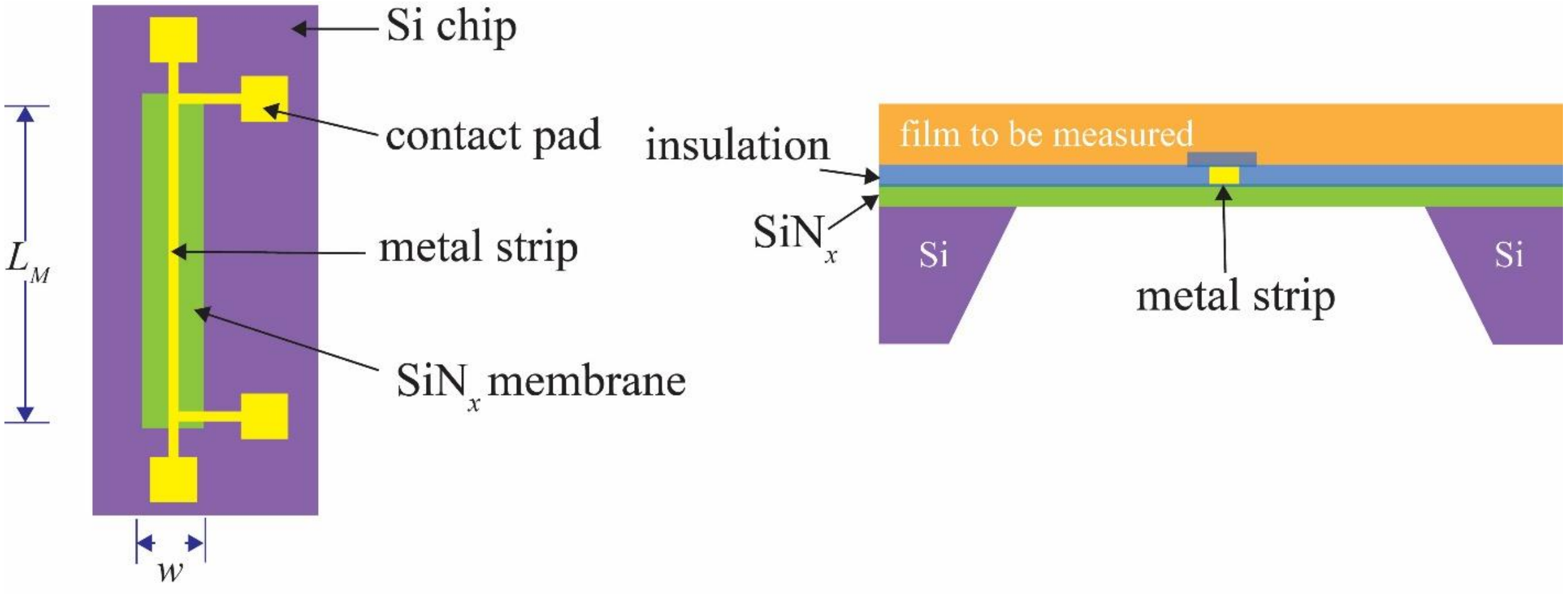

Figure 12: Volklein method. (a) The measurement chip used in the experiment; (b) The TE film is deposited on top of the suspended membrane.

The employment of an ultrathin suspended membrane presents a significant benefit in this approach, as it obviates the parasitic heat transfer from the sample to the substrate. This factor is crucial in ensuring reliable measurements of sub-micrometer thick thermoelectric films.[80, 149, 150] The top arrangement enables for the preparation of samples using various solution-based procedures: spin-coating, dip-coating, drop casting, ink-jet printing, and so on.[3] Moreover, the present experimental design has the potential to integrate with measurement techniques for other TE properties on a single chip. This enables the acquisition of all necessary parameters for the computation of the in-plane *zT*-value from the same setup as demonstrated in a study.[150] However, this approach is deemed inadequate for measuring the $\kappa_{f,\perp}$ of thin films. The undertaking of the measurement necessitates a vacuum environment. The precision of the measurement is contingent upon the comparative values of the thermal conductance linked with the coated film and the uncoated membrane. The thermal conductivity of the bare membrane should, if at all possible, be smaller relative to that of the deposited film ($\kappa_{M,\parallel}.d_M < \kappa_{f,\parallel}.d_f$).

### Microfabricated suspended device

A novel approach was devised to address the issue of parasitic heat transfer from the sample to the substrate. This involved the use of microfabricated suspended devices (MFSD) in a steady-state method to determine the $\kappa$ of nanowires, nanotubes (in the axial direction), or narrow thin film stripes (in the in-plane direction).[151, 152] This approach employs established techniques for film fabrication (spin-coating and thermal evaporation). The present methodology entails the acquisition of temperature data at both extremities of the specimen. The experimental measurements are conducted under vacuum conditions to mitigate the effects of convective heat dissipation. One configuration is schematically shown in Figure *13*a.[153] The test platform is made up of two neighboring suspended $SiN_x$ membranes that are held together by five identically long and thin $SiN_x$ beams. On each membrane, a platinum resistance thermometer (PRT) is fabricated to serve as the heater or thermometer. The specimen is transferred onto the device in order to establish a bridge between the two membranes. This is why, in some literature, this method is called the *suspended thermal bridge method*. One membrane serves as a heating membrane, allowing a dc current to heat the PRT, while the other serves as a sensing membrane. The rise in temperature on both membranes is thought to be the same, and heat loss from gas flow, convection, and radiation is not taken into account. The Joule heat generated by the heater ($I^2R_h$ where $R_h$ is the resistance of Pt resistance thermometer) and the two Pt current leads ($2I^2R_L$, where $R_L$ is the resistance of each Pt lead) on the heating membrane are $P_H$ and $2P_L$, respectively. A portion of the generated heat is delivered directly from the heating membrane to the environment via the five beams. The remaining heat passes through the sample into the sensing membrane and is then transferred to the environment via the five beams attached to the sensing membrane. The expression for the measured thermal conductance of the sample, $G_s$, can be demonstrated as follows:[153]

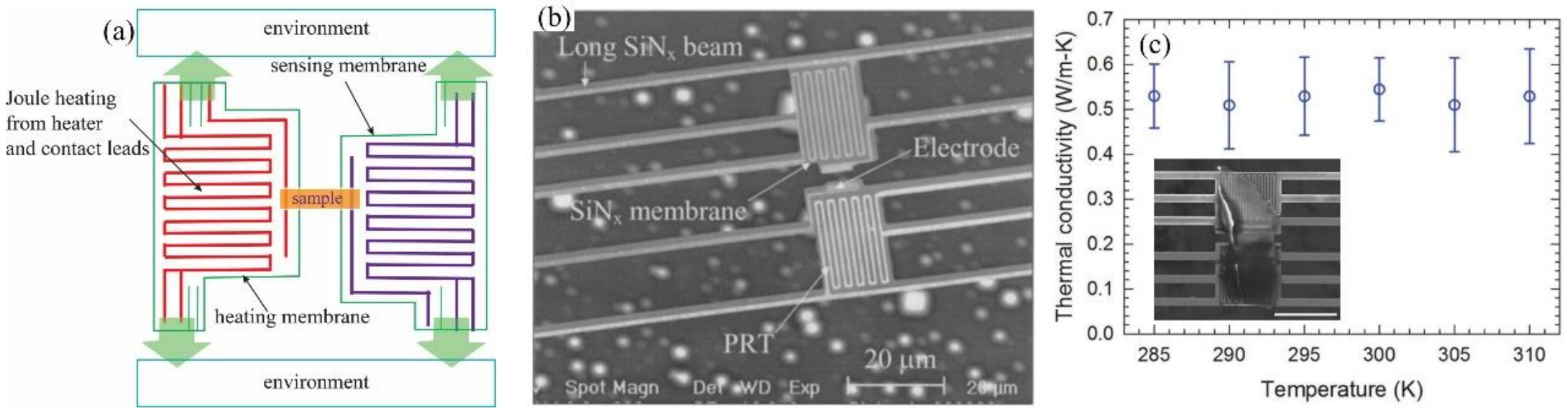

Figure 13: MFSD technique. (a) Schematic illustration of the proposed experimental configuration; (b) SEM micrograph of a microdevice for thermal property measurements of nanostructures.[153](c) $\kappa$ of PEDOT-Tos/CNT nanocomposite at 285-310 K. inset: PEDOT-Tos/CNT nanocomposite sample placed between two suspended membranes of a microfabricated device. Scale bar: 20 µm. Reproduced with the permission from Choi et al. [28] Copyright © 2018, ACS.

$$G_s = \frac{\Delta T_c (P_H + P_L)}{\Delta T_h^2 - \Delta T_c^2}$$

where $\Delta T_h$ and $\Delta T_c$ represent temperature rises at the heating and detecting membranes, respectively and both can be measured from the resistance change of the PRT. The measured thermal resistance (i.e., $G_s^{-1}$) consists of the intrinsic thermal resistance ($R_n$) of the sample and the thermal contact resistance between the sample and the two membranes ($R_{contact}$). $G_s^{-1} = R_n + R_{cont}$. Only if $R_{contact}$ is small enough can the sample's thermal conductivity be estimated directly from $G_s^{-1}$ by

$$\kappa = \frac{L_s}{R_n A_s} \approx \frac{L_s}{G_s^{-1} A_s}$$

where $L_s$ and $A_s$ are the length and cross-sectional area of the sample, respectively. As a result, one of the primary problems while using this strategy is how to limit $R_{contact}$. In practice, $R_{contact}$ can be lowered by deposition of metal or amorphous carbon at the points where the sample comes into contact with the membranes.[153] An alternative approach involves the determination of the thermal contact resistance and intrinsic thermal resistance of the sample through the utilization of the sample's thermoelectric voltage for the computation of temperature drops at the contacts.[154, 155]

Aside from thermal contact resistance, another critical issue related with the MFSD method is the device's measurement sensitivity, which determines the lowest limit of thermal conductance that can be detected for a sample.[156] The sensitivity of the measurement is constrained by various factors, such as the heat transfer that occurs through residual gas molecules, the radiation emitted from the membranes, and the fluctuations in temperature of the sample environment.[157]

MFSD method has been used to characterize the thermal conductivity of various materials: nanofilms[155, 158], 2D materials, such as graphene [159, 160], boron nitride[161], and TMDC materials[162]. Apart from these, MFSD technique has been widely employed to characterize the thermal properties of polymers and their composites.[163, 164] [165, 166] Choi et al. measured the $\kappa_{f,\parallel}$ of a PEDOT-Tos/CNT nanocomposite sample mounted between two suspended membranes of a microfabricated device (Figure *13*c inset).[28] The study concluded that the introduction of PEDOT-Tos in CNT junctions resulted in the suppression of thermal transport, reducing the $\kappa$ of the nanocomposites (Figure *13*).

Weathers et al. studied the $\kappa_{f,\parallel}$s of a series of PEDOT:PSS and PEDOT:Tos films, which revealed a clear relationship with the in-plane electrical conductivity, $\sigma_{\parallel}$. Thermal conductivity varied from about 0.5 W/m-K to as high as 1.8 W/m-K at 300 K as the $\sigma_{\parallel}$ increased from 20 to 500 S/cm.[165] The value of 0.5 W/m-K is comparable to the $\kappa_{f,\parallel}$ (0.58 W/m-K) of a low $\sigma_{\parallel}$-PEDOT:Tos film measured using a chip-based 3$\omega$ technique.[167] MFSD test platform can be designed to couple with the characterizations of the $S$ and the $\sigma$ of the same sample.[165] The method's principal downsides are that it necessitates substantial microfabrication work, and transferring as-prepared samples to microdevices may be difficult. Thermal contact resistance and measurement sensitivity are two more crucial factors that must be carefully evaluated.

### Steady-state infrared thermography

Steady-state infrared thermography is a noncontact absolute method that directly measures the $\kappa_{f,\parallel}$ of sub-micrometer thin films. In the experimental set up in a vacuum chamber, the thin film sample is suspended above a cavity in an opaque substrate that has a high $\kappa$ (Figure *14*a). The sample is irradiated uniformly from below by visible light with an absorbed power density $Q_A$. The wavelength of the light is precisely chosen to create great optical absorption and significant heating. The substrate functions as both a shadow mask for illumination and a heat sink for heat flow. An infrared camera records the temperature distribution. The temperature as a function of distance $r$ from the center of the cavity can be expressed as follows:[168]

$$T(r) = -\frac{Q_A}{4\kappa_{f,\parallel} d_f} r^2 + \frac{Q_A}{4\kappa_{f,\parallel} d_f} r_o^2 + T_o \qquad \textit{Equation 4}$$

where $r_0$ is the radius of the cavity. $\kappa_{f,\parallel}$ and $d_f$ are in-plane thermal conductivity and the thickness of the film, respectively. $T_0$ is the temperature of the heat sink. Because the last two terms in *Equation 4* are merely constants, $\kappa_{f,\parallel}$, may be calculated from the parabolic temperature distribution near the cavity's center, the $Q_A$, and the $d_f$. The measurements, however, necessitate rigorous calibrations of the light source and the IR camera. To apply the approach to organic hybrid thermoelectric materials, the samples must have a significant absorptance at the illumination wavelength and a comparably strong emissivity for IR detection.

The relation between the thermoelectric $zT$ and morphology was studied in TE thin composite films of poly(3,4-ethylenedioxythiophene):polystyrene sulfonate (PEDOT:PSS)/different amounts of Si nanoparticles (Si NPs).[169] In this study, the $\kappa_{f,\parallel}$ of free-standing composite films was investigated using steady-state IR thermography. The process provided by Greco et al. was adapted to prepare free-standing films (Figure *14*f).[170] The PEDOT:PSS/Si NPs films were applied to an opaque aluminum substrate. Holes in the substrate (0.3 to 0.8 mm in sizes) permitted a homogenous visible light source to illuminate a free-standing circular part of the film from below, effectively heating this section of the film. The homogeneous heating produced a parabolic temperature distribution in the film, which was determined by the size of the film and its thermal conductivity. Contact-free measurement of this temperature distribution was accomplished using IR microscopy using a DCG systems InSb 640 SM high speed IR camera (Figure *14*g), and thermal conductivity was determined by fitting the resulting temperature paraboloid. A circular Si photodiode was used for calibration.

The $\kappa_{f,\parallel}$s of thin films of PEDOT:PSS and of polyimide (PI) obtained via a steady-state IR thermography are reported to be 1.0 W/m-K and 0.4 W/m-K at RT, respectively.[168] The opaque substrate was utilized to suspend the films above a hole, which were subjected to heating through a visible light beam. The IR microscopy technique was utilized to capture the temperature distribution within the thin films. Subsequently, the obtained data was fitted to the analytical expression derived for the specific hole geometry, in order to determine the value of $\kappa_{f,\parallel}$.

The results for $\kappa_{f,\parallel}$ vs. the temperature differential between the center and the edge of the hole are presented in Figure *14*e. The study involved the characterization of films with varying hole diameters and illumination power densities. A larger hole diameter and stronger illumination output resulted in a greater temperature rise. For increases less than 4 K, the results for PEDOT:PSS demonstrate a consistent value of $\kappa_{f,\parallel}$ = 1.0 W/m-K for all hole diameters. The measured values of $\kappa_{f,\parallel}$, increase as the temperature rises. The obtained value for PEDOT:PSS were closely comparable to $\kappa_{f,\parallel}$ values obtained via TDTR method[171] and laser flash method.[172]

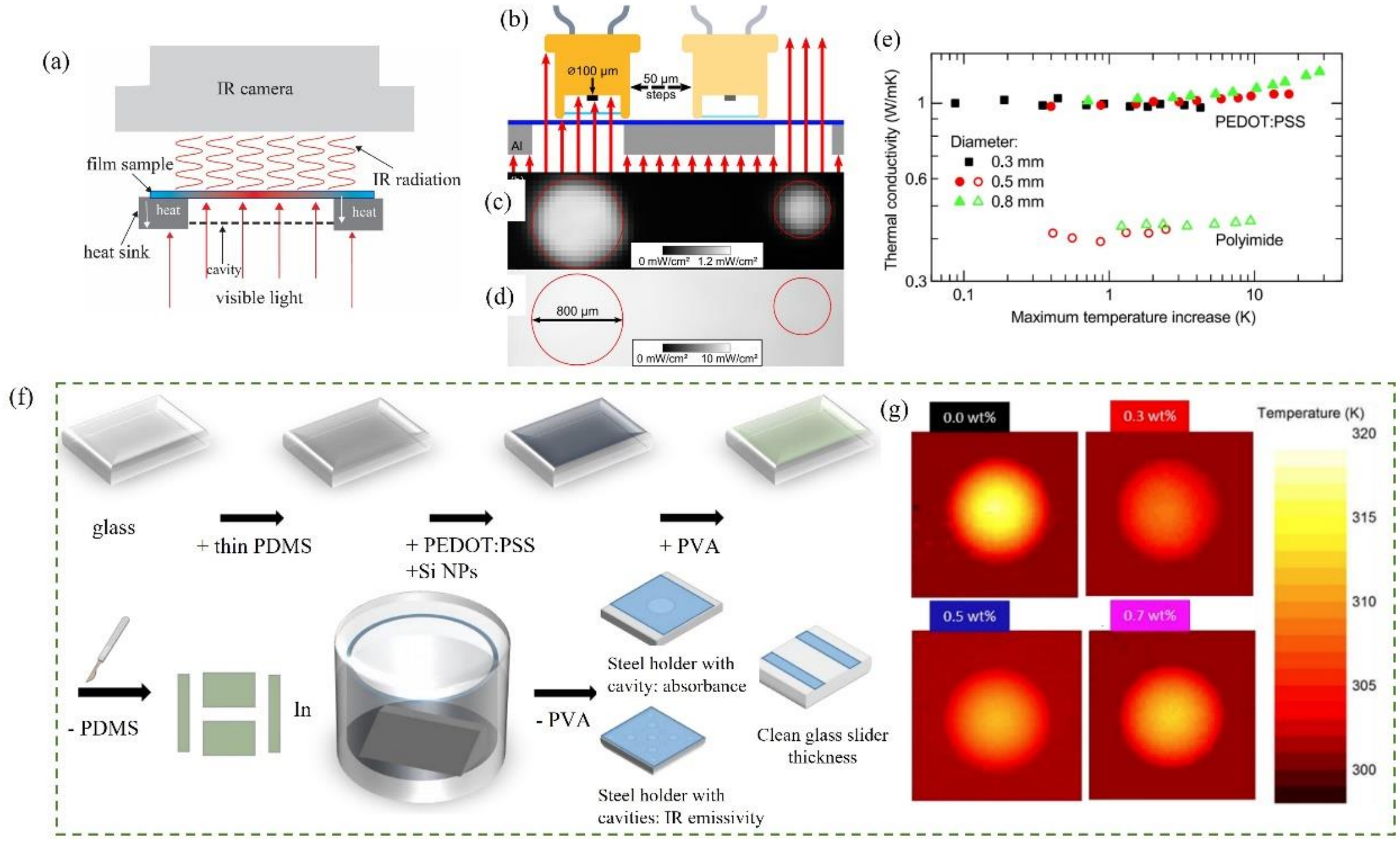

Figure 14: IR thermography technique. (a) A film that has been suspended over a hole in an opaque substrate. Visible illumination (red arrows) from below uniformly heats the free-standing film. Heat is transported through the film and into the substrate, which acts as a heat sink. In steady state, a maximum temperature of the free-standing film and a maximum thermal radiation are attained at the center of the hole; (b-e) $\kappa_{f,\parallel}$ measurement of PEDOT:PSS and PI films. (b) Scanning the sample with a calibrated photodiode (black) with a diameter of 100 µm by moving the chamber and a blue LED underneath the fixed diode by 50 µm steps; (c) & (d) show the measured power density with and without a PI sample on a supporting substrate, respectively. The red circles indicate the position of the holes. The difference in absolute intensity between (c) & (d) is due to the low transmittance of PI at the used illumination wavelength; (e) The $\kappa_{f,\parallel}$ of PEDOT:PSS films (250 nm-thick) and PI films (9 µm-thick) vs. the maximum temperature increase resulting from the LED illumination. The illumination power density varied between 2 $mW/cm^2$ and 100 $mW/cm^2$. (b-e) adapted from ref.[168] (f-g): PEDOT:PSS/Pt NP films. (f) Steps for preparing freestanding PEDOT:PSS films and then transferring them to various substrates for IR thermography and $\kappa$ measurements; (g) Images of IR thermography of PEDOT:PSS/Si NP composite films with varying Si NPs concentrations. (f-g) adapted from ref.[169]

Since steady-state infrared thermography requires no supplementary information for temperature calibration (except for calibration on the sample itself), this renders the method suitable for application on newly developed materials, including but not limited to highly porous SiGe films, composite materials, and organic-inorganic hybrid materials.[173]

## 4.2 Time-domain thermoreflectance method (TDTR)

The TDTR technique is an optical pump-probe methodology that is employed to ascertain the $\kappa_{f,\perp}$ and interfacial thermal conductance of thin films and bulk materials.[108, 174] The proposed technique involves the utilization of a pump laser pulse with a constant wavelength to irradiate the surface of the sample. This results in the generation of a transient change in temperature, leading to a corresponding modification in surface reflectance. The latter can be detected by a probe laser. Generally, a slender metallic film (known as a transducer layer) is applied onto the surface of the specimen, which exhibits alterations in reflectance at the wavelength of the probe laser in response to temperature elevation. Figure 15c depicts the time evolution of the thermoreflectance signal. Upon exposure to a brief laser pulse, the surface temperature of the sample undergoes an initial increase, succeeded by a gradual cooling phase. By comparing the thermal characteristics of the sample with appropriate theoretical frameworks, its thermal properties can be determined. The TDTR techniques can be executed in different time domains, depending on the pulse width of the pump laser and the data acquisition time scale.[175-177] In the nanosecond domain, these methods can be utilized to evaluate the thermal properties of thin films. In the picosecond domain, they can be employed to measure the thermal properties of thin films and their interfacial properties.[108, 178-180] In the femtosecond regime, these techniques can be utilized to investigate electron and phonon processes.[177, 181] A continuous wave (CW) laser can be employed as the probe laser if the thermoreflectance technique is used in the nanosecond domain.[175-177] When operating in the picosecond or femtosecond domains, however, an ultrafast pulsed laser system should be used as the light source for pumping and probing.[182, 183] In this experimental configuration, to prevent the laser beam from bouncing back into the laser oscillator, a broadband Faraday optical isolator is fitted at the oscillator's output. A half-wave plate positioned before the isolator can be utilized to modify the laser power. A polarizing beam splitter (PBS) divides ultrafast pulsed laser light into the pump and probe beams. The temporal evolution of the surface temperature can be recorded by adjusting the time delay between the probe pulse and the pump pulse using a mechanical delay stage. TDTR measures thermoreflectance response as a function of the time delay between probe arrival and pump pulses at the sample surface. To improve the signal-to-noise ratio, an electric-optic modulator (EOM) modulates the train of pump pulses at frequency $f_{mod}$ so that the thermal response signals may be detected by a lock-in amplifier.[184] The modulation frequency $f_{mod}$ typically ranges from 1 MHz to 10 MHz, allowing thermal penetration depth to be limited to the sub-micrometer range.[182] Therefore, the TDTR technique has emerged as a compelling approach for evaluating the thermal conductivity of thin films with sub-micrometer thickness, as it enables the omission of substrate-related effects. Simultaneous measurement of $\kappa$ and $C_V$ can be achieved through the application of a frequency-dependent TDTR technique.[185, 186]

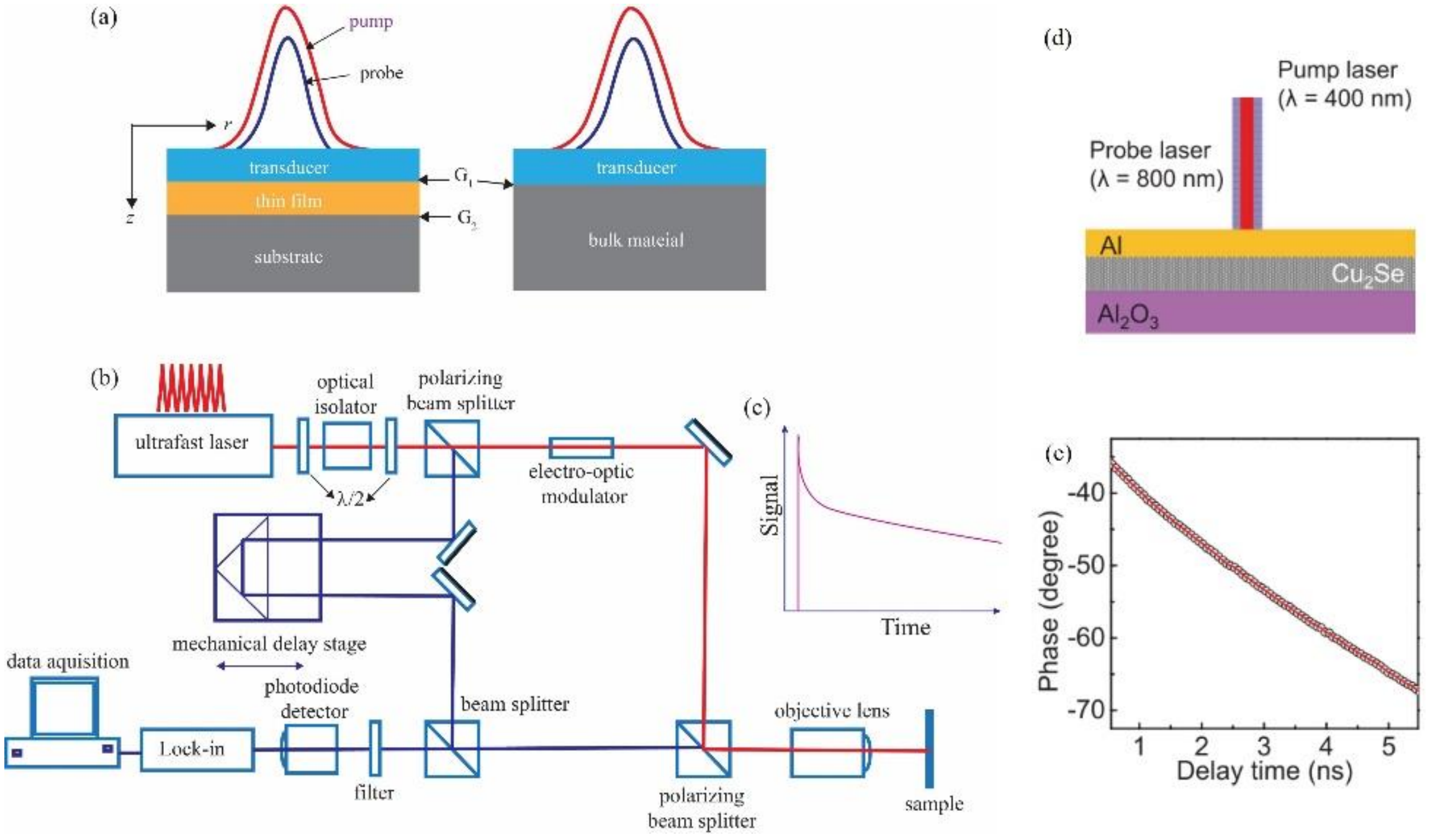

Figure 15: (a-c) TDTR technique. (a) Typical thin film and bulk material samples tested using the TDTR technique with concentric pump and probe beams. A thin metal transducer coating is often applied to samples. In the Figure, $G1$ is the interfacial thermal conductance between transducer and the sample. $G2$ is the interfacial thermal conductance between thin film and substrate; (b) The experimental setup for TDTR based on an ultrafast pulsed laser system; (c) time-dependent behavior of the thermoreflectance signal. (d-e) $Cu_2Se$ thin film characterization. (d) Schematic of the measurement setup, showing the Al transducer film on top, the pump laser heating a small area of the film, and the probe laser measuring the change in thermoreflectance; (e) The diffusive model fits the normalized TDTR phase shift versus time data. The best fitting from the multilayer thermal transport model (red line) and the experimental curves (black circles). (d-e) are adapted from ref.[187]

The TDTR method was originally developed to measure the $\kappa_{f,\perp}$. The TDTR approach, on the other hand, can be adapted to characterize $\kappa_{f,\parallel}$ by using a variable spot size configuration or a beam offset configuration.[184, 188, 189]The TDTR technique offers a prompt and non-invasive approach to acquire thermal conductivity of materials spanning from films with sub-micrometer thickness to bulk samples, with a notable lateral spatial resolution. TDTR method does not entail the intricate sample preparation procedures that are typically necessary for the conventional $3\omega$ technique. The method has received a lot of attention as a useful methodology for determining the $\kappa$ of various inorganic materials, as well as organic/hybrid TE materials: PEDOT:PSS films (100–400 nm thick,[190], 2.5 μm thick,[191] and 80 nm thick,[192]),[171], poly(3-hexylthiophene-2,5-diyl) (P3HT) thin films (77–200 nm thick),[193] graphene based organic nanocomposite (65 nm thick).[194] With TDTR method, Liu et al.[171] have observed that the $\kappa_{f,\perp}$ of the drop-cast DMSO-mixed PEDOT:PSS film is ~0.3 W/m-K, which is consistent with the data obtained by the conventional $3\omega$ technique.[195] In the same study, the authors were able to obtain $\kappa_{f,\parallel}$ (1.05 W/m-K) for the PEDOT:PSS films by properly orientating the sample (embedded in an epoxy matrix) with respect to the direction of the laser beam.[171]

Lin et al. have employed the TDTR method to characterize the thermal conductivity of various $Cu_2Se$ thin films prepared via a spin coating process and subsequent annealing.[187] In this experimental design, a femtosecond pulse laser is split into a pump and a probe beam. The pump laser ($\lambda$ = 400 nm) impulses cause an instantaneous temperature to rise on the sample surface, and the probe beam ($\lambda$ = 800 nm) monitors the time-dependent temperature decay (Figure 15d). A transient decline curve of this type is then fitted with a multilayer thermal model to determine thermal conductivity. Figure 15e depicts typical TDTR data and how it was fitted to the thermal model.

The $\kappa_{f,\perp}$ of thin films of $WSe_2$ was measured using the TDTR method using an Al (60-85 nm-thick) film as the transducer.[17] The $\kappa_{f,\perp}$ was calculated by comparing the time dependence of the ratio of the in-phase ($V_{in}$) and out-of-phase ($V_{out}$) signals from the radio-frequency lock-in amplifier to calculations from a thermal model.[196] Because the $WSe_2$ films have a very low thermal diffusivity, the sensitivity of the measurements was improved by using different pump beam modulation frequencies: a low frequency (580 kHz) for the 70 nm and 26 nm films; a high frequency (9.8 MHz) for the 360 nm films and single crystal sample. The $k_{f,\perp}$ of $WSe_2$ thin films at RT was 0.05 W/m-K, which was 30 times lower than the *c*-axis thermal conductivity of single-crystal $WSe_2$.

The TDTR technique exhibits a high degree of versatility and can be effectively employed for the thermal characterization of materials in both bulk and thin film forms. The TDTR measurement can provide an accurate and reproducible value of thermal diffusivity for films as thin as 100 nm, while separating the effect of substrate heating. The TDTR method's key drawbacks include lack of commercial tools, its complicated experimental setup, which necessitates advanced precision optics, as well as the surface coating needs. Furthermore, in order to avoid diffusive reflections, the surface of the sample must be smooth when performing the TDTR measurement.[189] The requirement may make sample preparation challenging, especially for some polymer materials with difficult to regulate surface roughness. Undulations or striations, for example, can arise spontaneously on the surface of spin-coated or cast polymer films.[197] Adding to these problems is the anisotropic $\kappa$ of organic materials, which demands additional requirements.

### 4.3 Traditional 3$\omega$ Method

Ever since it was first introduced in 1990, the 3$\omega$ method has been widely used to measure the thermal properties of both bulk materials and thin films.[198] The 3$\omega$ method was initially employed to conduct measurements on a substrate that possessed a significantly greater thickness than the thermal penetration depth of the thermal wave generated by the sinusoidal current. Figure *16* depicts a common 3$\omega$ measurement schematic. The desired thin layer is grown or coated on a substrate (e.g., silicon, sapphire). A metallic strip (e.g. Al, Au, Pt) is deposited on top of a substrate. The metallic strip's dimensions are typically half-width $a$ = 10-50 μm and length $L$ = 1000-10000 μm, which is considered as infinitely long in the mathematical model. The metallic strip functions as an electrical heater as well a temperature sensor.

An alternating current at frequency $\omega$ passing through the heater/sensor is expressed as:

$$I(t) = I_o \cos(\omega t) \quad \text{Equation 5}$$

Where $I_o$ is the and amplitude. The electrical resistance of the resistive heater/sensor causes Joule heating at a frequency of $2\omega$. A $2\omega$ heating causes a temperature change in the heater/sensor at a $2\omega$ frequency and is given as:

$$\Delta T(t) = \Delta T_o \cos(2\omega t + \varphi) \quad \text{Equation 6}$$

where $\Delta T_o$ is temperature change amplitude and $\varphi$ is phase. The temperature change modulates the heater/sensor's electrical resistance at $2\omega$:

$$R_e(t) = R_{e,o}(1 + \alpha_R \Delta T) = R_{e,o}(1 + \alpha_R \Delta T_o \cos(2\omega t + \varphi) \quad \text{Equation 7}$$

Where $\alpha_R$ is the temperature-coefficient of resistance of the heater/sensor. $R_{e,o}$ is heater/sensor's electrical resistance at the starting state. The amplitude $\Delta T_o$ and the phase $\varphi$ are directly related to the thermal conductivity of the sample and to the angular frequency $\omega$. When multiplied by the $1\omega$ driving current, a small voltage signal across the heater/sensor at $3\omega$ frequency can be detected:[199]

$$\begin{aligned} V(t) &= I(t)R(t) \\ &= R_{e,o} I_o \cos(\omega t) + \frac{1}{2} R_{e,o} I_o \alpha_R \Delta T_o \cos(\omega t + \varphi) \\ &+ \frac{1}{2} R_o I_o \alpha_R \Delta T_o \cos(3\omega t + \varphi) \end{aligned} \quad \text{Equation 8}$$

This change in voltage at $3\omega$ frequency (3rd term in *Equation 8* carries the information about the footprint of thermal transport inside the sample. However, because the $3\omega$ voltage signal (amplitude $\frac{1}{2} R_o I_o \alpha_R \Delta T_o$) is very faint and usually three orders of magnitude smaller than the amplitude of the applied $1\omega$ voltage, such a measuring technique is normally implemented using a lock-in amplifier.

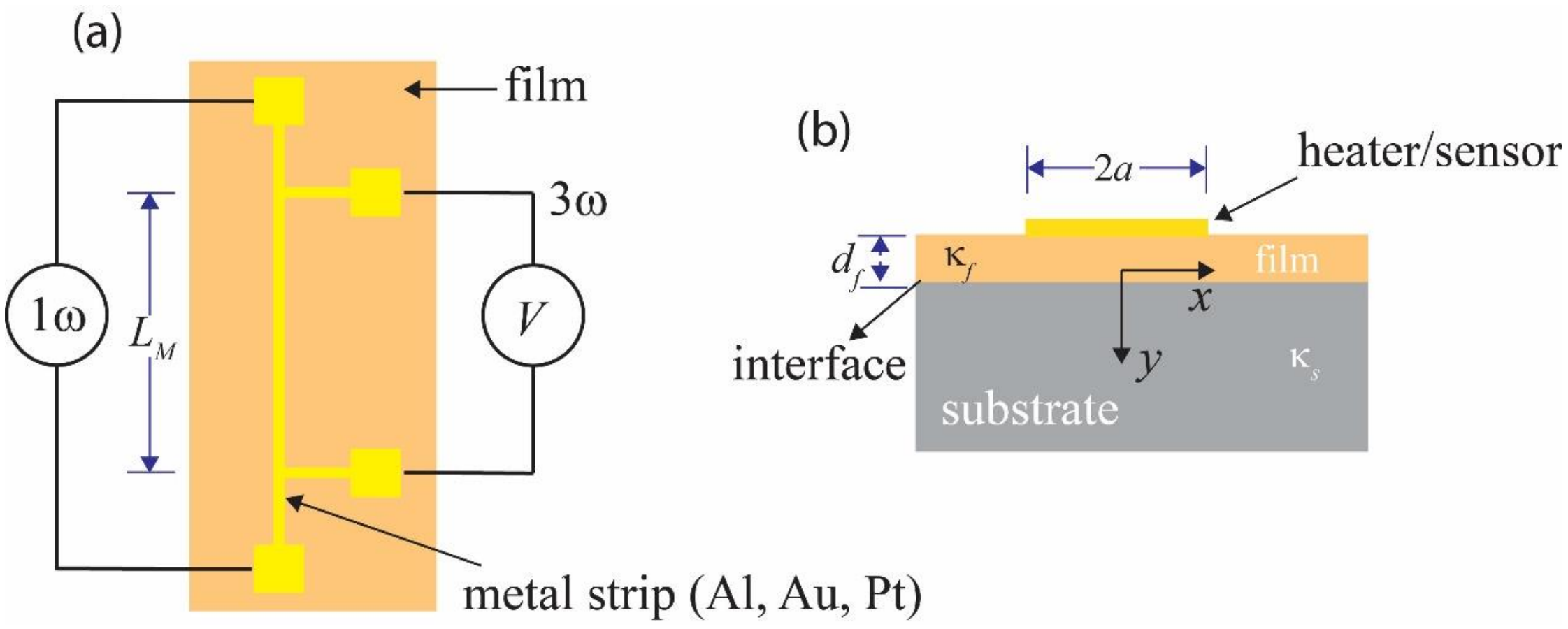

Figure 16: (a) Top view and (b) cross-sectional view of a typical microfabricated $3\omega$ test structure for thermal characterization of thin films. A metallic strip (i.e., heater/sensor) with width $2a$ is deposited on top of the thin film. This strip works as both heater and temperature sensor. An alternating current with frequency $1\omega$ heats the heater/sensor with frequency $2\omega$. The change in temperature of the heater/sensor changes the resistance, which results in a small change in the voltage at $3\omega$ frequency. An electrically insulating layer is required between the thin film specimen and the heater/sensor while characterizing the electrically conducting thin films.

To measure the $\kappa$ of an electrically conductive or semiconducting material, it is necessary to introduce an electrical insulation layer between the thin film and the electrical heater/sensor. Depending on the width of the heater, the $3\omega$ technique can be used to test both the $\kappa_{f,\perp}$ and $\kappa_{f,\parallel}$ of thin films. For determining the $\kappa_{f,\perp}$ and $\kappa_{f,\parallel}$, approximate analytical equations are typically used.[200] In the event that a metallic heater/sensor is placed on an isotropic substrate without a thin film, the heater/sensor can be treated as a line source under the condition that the thermal penetration depth $L_p = (\alpha_S/2\omega)^{1/2}$ is significantly greater than the heater/sensor half width $a$.

By selecting a suitable heating frequency for the heating current, it is possible to confine thermal penetration to a specific region within the substrate. Then the complex temperature rise of the heater/sensor can be approximated as:[201]

$$\Delta T_s = \frac{p}{\pi L \kappa_s}\left[0.5\ln\left(\frac{\alpha_s}{a^2}\right) - 0.5\ln(\omega) + \eta\right] - i\left(\frac{p}{4L\kappa_s}\right) = \frac{p}{\pi L \kappa_s} f_{linear}(ln\omega) \qquad \textit{Equation 9}$$

where subscript $S$ is stands with substrate, $\eta$ is a constant, $\kappa$ is thermal conductivity, $p/L$ is peak electrical power per unit length, and $f_{\text{linear}}$ is a linear function of $\ln\omega$. From *Equation 9* it is evident that the isotropic $\kappa_S$ can be obtained from the slope of the real part of the temperature amplitude as a linear function of the logarithm frequency $\ln(\omega)$. This method of substrate thermal conductivity measurement based on equation above is called as "slope method" and the slope of the real part of the temperature amplitude vs. log frequency is refer to as "slope". To determine the $\kappa_{f,\perp}$ of a thin film deposited on a substrate, it is necessary to estimate the temperature drop across the film (Figure *17*a). In practical applications, it is common to assume that the temperature at the upper surface of the film is equivalent to the temperature of the heater/sensor due to the negligible values of contact resistances.[202]. The prevalent approach for ascertaining the temperature at the bottom side of the film involves computing the experimental heat flux utilizing the substrate thermal conductivity $\kappa_s$. This value is typically established or can be measured via the $3\omega$ method, as demonstrated in *Equation 9*. Assuming 1D heat conduction across the thin film (Figure *16*a), the thermal conductivity of the thin film can be easily deduced from:

$$\Delta T_{s+f} = \Delta T_s + \frac{p d_f}{2aL\kappa_{f,\perp}} \qquad \textit{Equation 10}$$

Where subscript $f$ stands for thin film properties, subscript $S+f$ denotes thin film on substrate structure. Then $\kappa_{f,\perp}$ is obtained by fitting the experimentally measured temperature rise data under a variety of heating frequency $\omega$ to *Equation 10* (see Figure *17*c).

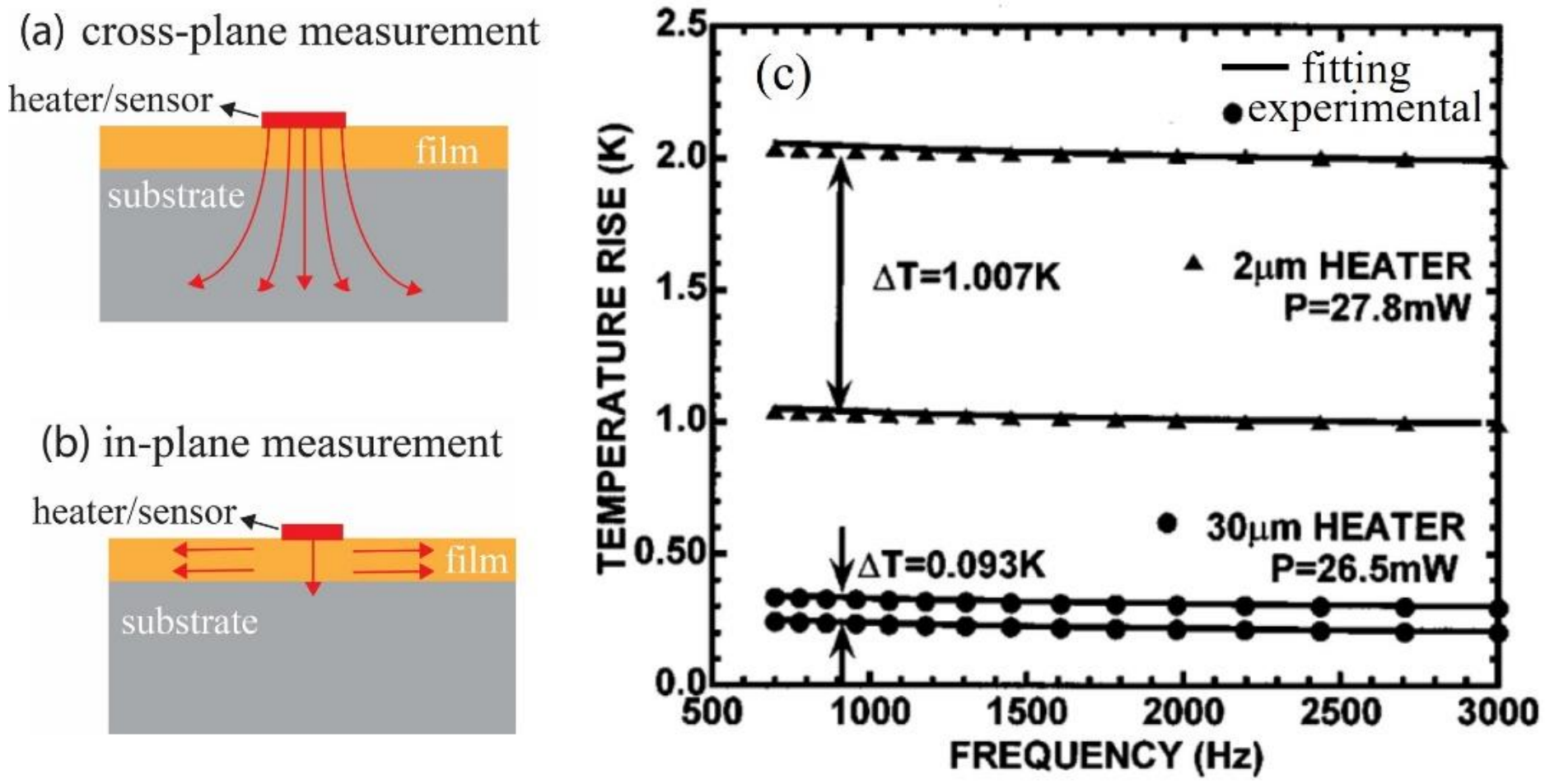

Figure 17: Thermal conductivity measurement configurations for the 3$\omega$ method. (a) cross-plane ($\kappa_{f,\perp}$); (b) in-plane ($\kappa_{f,\parallel}$). The heater width should be comparatively larger than the thickness of the thin film during $\kappa_{f,\perp}$ measurements in order to ensure that the heat conduction across the thin film is one-dimensional. A narrower-width heater is utilized for the $\kappa_{f,\parallel}$ measurement so that the $\kappa_{f,\parallel}$ can be determined by heat spreading in the thin film; (c) the temperature rise of the 30 µm and 2 µm width heaters/sensors deposited on the Ge-quantum-dot-SL and the reference samples as a function of frequency. The experimental signal is compared to the predictions for the temperature rise of the heater/sensor calculated based on Eqs. (24), (25) and (27) for the fitted values of $\kappa_{f,\parallel}$ and $\kappa_{f,\perp}$ of the film.

The 3$\omega$ technique has also been widely utilized to determine thin film $\kappa_{f,\parallel}$. As illustrated in Figure *17*b, a considerably narrower heater is employed in comparison to the cross-plane thermal conductivity test, so that the heat transfer mechanism within the film is sensitive to both $\kappa_{f,\parallel}$ and $\kappa_{f,\perp}$. In this case, the half width $a$ of the heater should be small enough to satisfy:[200]

$$\frac{a}{d_f}\left(\frac{\kappa_{f,\perp}}{\kappa_{f,\parallel}}\right)^{1/2} \leq 0.1 \qquad \textit{Equation 11}$$

Where $\kappa_{f,\perp}$, $\kappa_{f,\parallel}$ and $d_f$ are the cross-plane, in-plane thermal conductivities of the thin film, and film thickness respectively. The utilization of a 2D heat transfer model is necessary for data reduction due to the sensitivity of lateral heat spreading to in-plane thermal conductivity. The temperature drop across the thin film is expresses as:[201]

$$\Delta T_f = \frac{p}{\pi L}\left(\frac{1}{\kappa_{f,\perp}\cdot\kappa_{f,\parallel}}\right)^{1/2}\int_0^\infty \frac{sin^2\lambda}{\lambda^3} tanh\left[\lambda\left(\frac{d_f}{a}\right)\left(\frac{\kappa_{f,\parallel}}{\kappa_{f,\perp}}\right)^{1/2}\right] d\lambda \qquad \textit{Equation 12}$$

*Equation 12* gives the temperature drop of a thin film normalized to the value for purely 1D heat conduction through the film as a function of the $\kappa_{f,\parallel}$ and $\kappa_{f,\perp}$s and the heater/sensor half width $a$. In reality, $\kappa_{f,\perp}$ is frequently measured first with a much wider heater/sensor that is solely sensitive to cross-plane thermal conductivity. $\kappa_{f,\parallel}$ is then determined using a considerably smaller heater/sensor width (Figure *17*c). In Figure *17*c the temperature rise of the 30 µm and 2 µm width heaters/sensors deposited on the Ge-quantum-dot-SL and the reference samples as a function of frequency are represented by the experimental data points.

The $3\omega$ technique has the potential to measure thin films with dielectric, semiconducting, and electrically conducting properties. The $3\omega$ method has a considerable advantage over conventional steady-state methods because the error due to radiation heat loss is drastically reduced. Errors due to thermal radiation are shown to depend on the experimental geometry's characteristic length. Even at a high temperature of 1000 K, the calculated error of a $3\omega$ measurement caused by radiation is less than 2%. In the case of electrically conducting and semiconducting materials, it is necessary to ensure electrical isolation of the samples from the metallic heater/sensor through the use of an additional insulating layer.[203] However, this introduces an additional thermal resistance, which inevitably leads to a reduction in both sensitivity and measurement accuracy. An additional obstacle pertains to the implementation of the $3\omega$ technique, which necessitates microfabrication procedures for the metallic heater/sensor. In contrast, optical heating and sensing techniques, such as the transient thermoreflectance method, typically require only minimal sample preparation.

The primary factor contributing to uncertainty in the $3\omega$ $\kappa$ measurement arises from the process of determining the $V_{3\omega}$ signal at various frequencies and the resulting linear regression fit. Because we are solely concerned with the slope of the $V_{3\omega}$ vs. $\ln(2\omega)$ curve, the stability and linearity of the $V_{3\omega}$ values throughout measurement are more significant than the absolute accuracy. Therefore, the standard uncertainty is calculated as a slope uncertainty in the linear regression fit. This uncertainty is determined for each thermal-conductivity measurement at each temperature point individually.

Zhou et al. employed $3\omega$ methods to investigate the $\kappa$ of four samples of micrometer-thick $Bi_2Te_3$ TE films produced by pulsed electroplating.[204] The measurement devices were made with a dielectric layer of sputtered $SiO_2$ and heaters of Au lines (Figure *18*a). As a reference, an Au-coated Si substrate was used. The $\kappa_{f,\perp}$ of the films was determined using the differential method and the slope method separately. The moderate fluctuations in the $3\omega$ voltage with frequency and thickness, as well as the consistent measurement results using these two approaches, suggest that the characterization methods are viable and dependable. The electroplated film's $\kappa_{f,\perp}$ decreases from 1.8 W/m-K to 1.0 W/m-K as the pulse potential increases from -100 mV to 50 mV, which is attributable to the films' improved microstructure. In addition, the thermal conductivity anisotropy of the $Bi_2Te_3$ film was determined by employing a two-wire $3\omega$ technique. It was observed that the measured $3\omega$ voltages ($V_{3\omega}$) were linear functions of logarithmic frequency, and the calculated effective temperature drops across the TE film ($\Delta T_f$) were unchanged with frequency up to 1000 Hz. As observed in Figure 18b, the calculated $\kappa_\perp$ of $Bi_2Te_3$ films of different thicknesses were almost same (with a mean value of 1.48 W/m-K). As the slope method does not require a reference device, total thermal resistance of the sample $R_{th}$ of the four sample devices were determined and plotted as a function of thickness (Figure *18*c). From the slope of the fitted line, a $\kappa_{f,\perp}$ value of 1.48 W/m-K for the $Bi_2Te_3$ films was found. This value was identical to the result determined by the differential method.

To characterize the $\kappa_{f,\perp}$ of the boron doped $Si_{0.8}Ge_{0.2}$ thin film, a differential $3\omega$ method was employed.[205] A schematic of the microfabricated thermoelectric test structure is shown in Figure *18*d. On the sample, aluminum heaters 20 μm wide and 2 mm long were deposited. To electrically insulate the sample from the heater, 0.2 μm $SiO_2$ was utilized. To subtract the unknown thermal characteristics of the insulating layers, a reference sample of similar structure without the boron doped nanocrystalline $Si_{0.8}Ge_{0.2}$ thin film was used. The experimental results depicted in Figure *18*e illustrate the temperature amplitudes sensed by heaters with a width of 20 μm that were deposited onto both the reference and the boron-doped nanocrystalline $Si_{0.8}Ge_{0.2}$ thin film specimen.

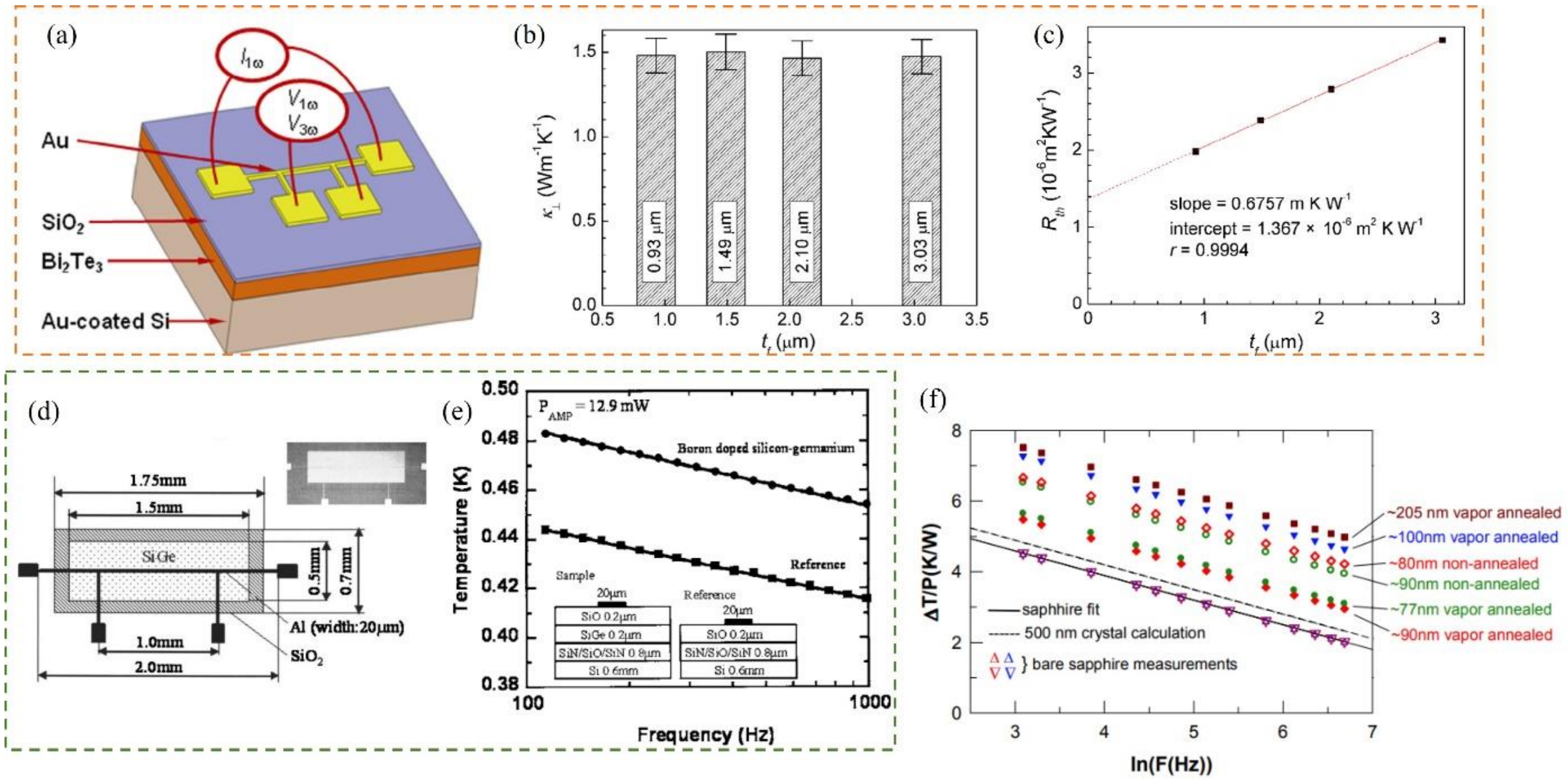

Figure 18: Differential and slope methods for $Bi_2Te_3$ TE films. (a) Schematic of the multilayer device structure for 3$\omega$ measurements; (b) calculated $\kappa_\perp$ of the $Bi_2Te_3$ films with different thicknesses deposited at pulse-on potential, $E_{on}$ = 0 V; (c) Thermal resistance ($R_{th}$) of the samples as a function of $Bi_2Te_3$ film thickness ($t_f$). The red line is the linear fit to obtain $\kappa_\perp$ via a slope method.[204]**(d-e)** $Si_{0.8}Ge_{0.2}$ thin films. (d) microfabricated test structure for thermal conductivity measurement; (e) The temperature amplitudes experienced by 20 μm wide heaters deposited onto the reference and the boron doped nanocrystalline $Si_{0.8}Ge_{0.2}$ thin film sample.[205] (f) 3$\omega$ measurement for small molecule organic semiconductors. Dependence of the $\Delta T/P$ for spin-cast TES-ADT films on sapphire substrates; open triangles: bare sapphire and the solid line is a fit to equation similar to *Equation 9*; dashed line: the expected results for 500 nm thick TES-ADT (assuming $\kappa = \kappa_c$ (crystal) = 5 W/m-K); solid symbols: results for vapor annealed films of different thicknesses; open circles and squares: results for non-annealed films.[206]

The thin film's thermal conductivity is calculated utilizing the unknown thermal conductivity of the thin film as a fitting parameter from the best fit between the measured temperature difference and predictions of a heat conduction model. To calculate the thermal conductivity of the thin film, both one-dimensional and two-dimensional heat conduction models were used. The thermal conductivity of boron doped nanocrystalline $Si_{0.8}Ge_{0.2}$ thin film was determined to be 2.9 W/m-K. Using the same technique, the $\kappa_{f,\perp}$ of *n*-type textured $Bi_2Te_{2.7}Se_{0.3}$ thin films with Pt nanoinclusions (247-254 nm-thick) obtained via a PLD method were determined using 2$\omega$ method.[207] In another study, the thin films (0.75 μm-thick) of *n*-type nanocrystalline bismuth-telluride-based (BT) materials were deposited on a glass substrate using a flash evaporation process and their thermal conductivity were examined.[208] The grain size in these thin films was 60 nm. The $\kappa_{f,\perp}$ of the nanocrystalline thin films was measured at RT using a differential $3\omega$ technique. The nanocrystalline thin films had a $\kappa_{f,\perp}$ of 0.8 W/m-K, which was almost half of the value for a sintered bulk sample with an average grain size of 30 µ and nearly the same composition as the nanocrystalline thin films. The $zT$ of the nanocrystalline thin films was estimated to be 0.7, assuming that the $\kappa_{f,\parallel}$ of the nanocrystalline thin films is equivalent to that of the cross-plane direction.

As per the author's information the thermoelectric properties of small molecule organic semiconductors: bis(triisopropylsilylethynyl) pentacene (TIPS-pn), bis(triethylsilylethynyl) anthradithiophene (TES-ADT) and difluoro bis(triethylsilylethynyl) anthradithiophene (diF-TES-ADT) are not reported.[209-211] Because these are organic semiconductors, it is worth reviewing their $3\omega$ thermal property characterization. For each of these materials, several films of different thicknesses were measured to separate the effects of intrinsic thermal conductivity from interface thermal resistance (Figure 18f). For sublimed films of TIPS-pn and diF-TES-ADT (< 100 nm to > 4 µm-thick), the thermal conductivities were similar to that of polymers and over an order of magnitude smaller than that of single crystals, presumably reflecting the large reduction in phonon mean-free path in the films. For thin (≤ 205 nm) crystalline films of TES-ADT (vapor annealed spin-cast films), the thermal resistances were found to be dominated by interface scattering.[206]

### 4.4 $3\omega$ method in combination with suspended ultrathin dielectric membranes

The $3\omega$ method employs harmonic Joule heating along a metal strip placed on a thin film to probe thermal diffusion from the metal strip to the underlying substrate, conveying $\kappa_{f,\perp}$. This method has been extended to the $\kappa_{f,\parallel}$ measurement of thin films, where lateral heat diffusion is tested using metal strips of various widths. Although this method can be applied to a variety of materials in theory, the uncertainty in the computation of $\kappa$ is intrinsically high. In general, it is preferable to evaluate the $\kappa_{f,\parallel}$ of thermally isolated samples from a substrate. To mitigate the effects of parasitic heat conduction from the sample to the substrate, the measurements are conducted with the sample in suspension.

The $3\omega$ technique can also be used in conjunction with suspended ultrathin dielectric membranes to determine the $\kappa_{f,\parallel}$ of the films.[149, 167, 212, 213] One feasible pathway to achieve this is to use the λ-chip configuration from the Volklein method as a measurement platform (Figure 12). In contrast to the conventional $3\omega$ technique, the $3\omega$-Volklein approach utilizes a low frequency (0.4-0.5 Hz) alternating current to generate the heating. During the quasi steady-state conditions, the entire sample volume is penetrated by heat, and the thermal conductivity can be determined directly using Fourier's heat conduction equation without the need for measuring heat capacity and density, based on the measured temperature rise. The following review of such methods can help us understand the details of such methods.

A suspended system for measuring the thermal properties of membranes, where the measurement technique is based on the $3\omega$ dynamic method coupled to a Volklein geometry, is reported.[212] The method was applied to measure the thermal properties of low stress silicon nitride and polycrystalline diamond membranes (100 nm-400 nm-thick). The dimensions of the membranes were 1 mm long and 150 µm wide (Figure *19*c). The measurement method is made up of a heater/thermometer that is centered along the long axis. The methodology involves the generation of sinusoidal joule heating through the application of an alternating current across a transducer (niobium nitride (NbN)) and measure the temperature oscillation at the third harmonic using a Wheatstone bridge set-up (Figure *19*d). The magnitude of the temperature rise and its correlation with the excitation frequency are solely associated with the thermal characteristics of the membrane. The thermal conductivity and specific heat can be inferred by gauging the $V_{3\omega}$ voltage that manifests across the transducer. The transducer is composed of a material exhibiting a notable temperature-dependent resistance. It functions as both a thermometer and a heater. The micro-machined device allowed the measurement of the $\kappa_{f,\parallel}$ of membranes with a sensitivity of less than 10 nW/K (+/−5 × $10^{-3}$ W/m−K at RT) and a very high resolution ($\Delta K/K = 10^{-3}$). For silicon nitride membranes, a $\kappa_{f,\parallel}$ value ~10 W/m−K was observed at RT, in agreement with previously measured values on thin SiN films.[214, 215] Similar to this, thermal conductivity values for diamond membranes between 10 and 100 K have been observed to range from 1 to 10 W/m-K, which is in good accord with published values.[216, 217]

In a similar line, a specific heat measurement technique, where $3\omega$ method coupled to the Volklein geometry was adapted to measure the thin suspended SiN membranes (8K-300K temperature range).[218] The demonstrated device measured the heat capacity of a 70 ng SiN membrane (50 or 100 nm thick), corresponding to a heat capacity of $1.4 \times 10^{-10}$ J/K at 8 K and $5.1 \times 10^{-8}$ J/K at 300 K. Further, the setup for the determination of specific heat measurements additionally enables the evaluation of $\kappa_{f,\parallel}$ of the identical sample at a low frequency, exhibiting exceptional resolution. The Debye temperature was determined by analyzing the temperature-dependent variation of the specific heat of SiN. At low temperatures, there has been a deviation from the Debye $T^3$ law, as other authors have already reported.[152]

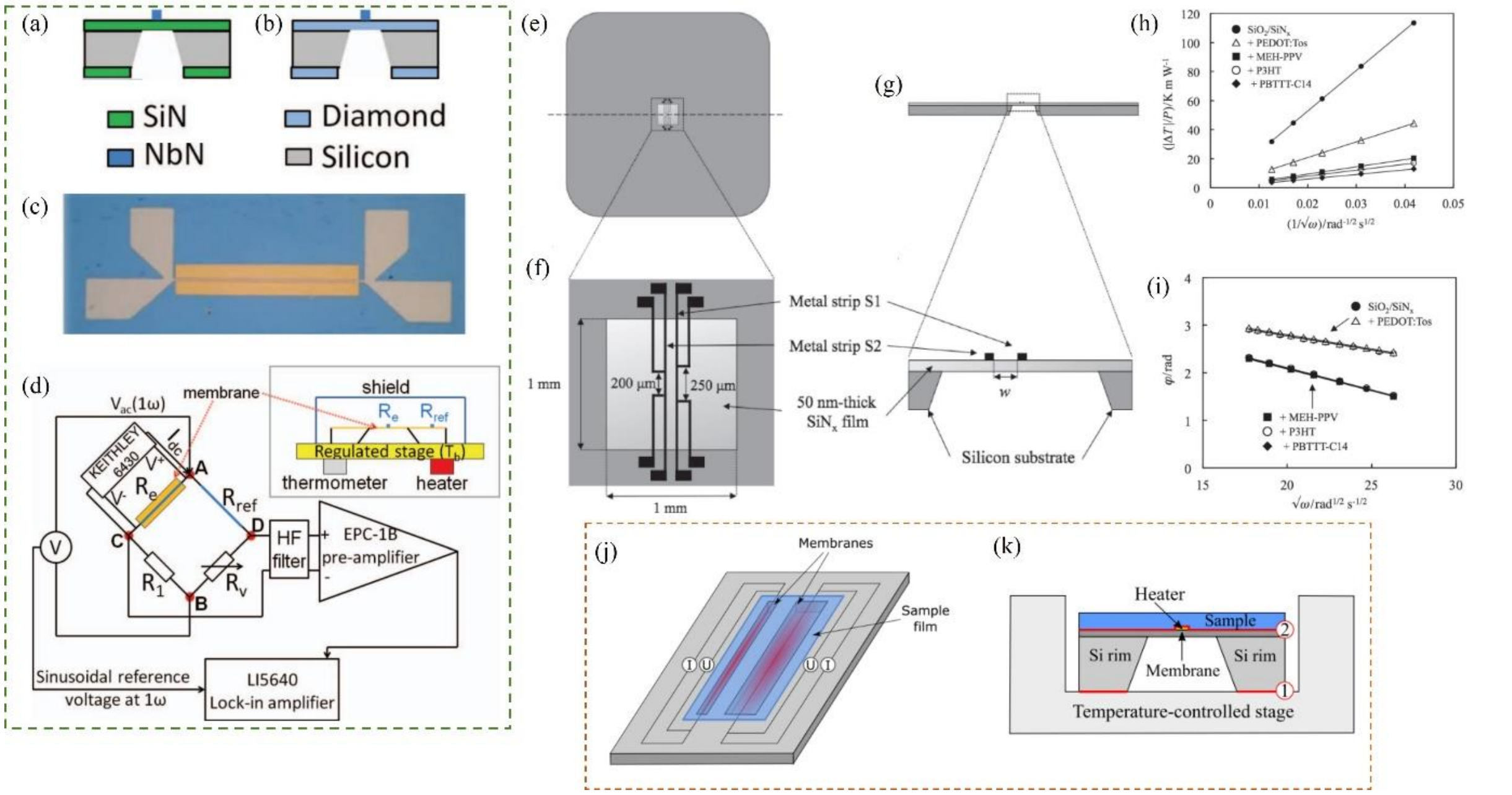

Figure 19: (a-c) 3$\omega$ dynamic method/Volklein geometry. (a) &(b) Schematics of the (a) SiN and (b) diamond membranes. NbN is the transducer; (c) a photograph of the sample. The NbN is grey and the membrane is yellow (1 mm long and 150 µm wide); (d) Electrical measurement system including HF filter, preamplifier, and lock-in amplifier. A, B, C, and D represent the nodes of the Wheatstone bridge. The $V_{3\omega}$ is measured between C and D. The transducer is referred to as $R_e$ and the reference resistance as $R_{ref}$. (d-inset) a schematic of the membrane fixed on the temperature regulated stage covered by the thermal Cu shield. The $V_{3\omega}$ signal is measured between C and D without the 1$\omega$ component.[212](e-i) 3$\omega$ technique with the suspended ultrathin membrane. Schematics of top view (e, f) and cross-section view (g) of the sample; (h-i) Experimental data for $SiO_2/SiN_x$ and $SiO_2/SiN_x$ with polymer layer: $|\Delta T)//P$ vs. $1/\sqrt{\omega}$ (h) and $\phi$ as a function of $\sqrt{\omega}$ (i). The solid lines in each graph represent least-squares fittings of the data with eqs Equation 13 and *Equation 14*.[167](j-k) P(NDI2OD-T2) studies using 3$\omega$-Volklein geometry. (j) measurement chip implementing the Volklein method; (k) Cross-sectional view of the measurement setup with marked interfaces with thermal contact resistances contributing to the measurement uncertainties.[219]

A somewhat different approach (double stripe configuration)[220-222] was taken to combine the 3ω technique with the suspended ultrathin membrane[167] and was applied to determine the $\kappa_{f,\parallel}$ of various π-conjugated polymer TE films (Figure *19*h).[3, 223, 224] The schematic of the sample is depicted in Figure *19*e-g. The study involved the lithographic patterning of metal strips comprising Cr/Au (3 nm/40 nm) on a commercially available silicon nitride ($SiN_x$) membrane. The membrane featured a window size of 1 × 1 mm$^2$. The window's thickness was 50 nm. The two metallic strips, labeled as S1 and S2, had a width of 2.5 µm and lengths of 250 and 200 mm, respectively were patterned. The distance between the two strips was measured to be 47.5 µm. E-beam evaporation was used to deposit a 50-nm-thick $SiO_2$ film on the membrane as an insulating layer. As a first step, thermal properties of $SiO_2/SiN_x$ layer were characterized. A sinusoidal current (angular frequency of 1$\omega$) was supplied to strip S1, causing a temperature oscillation in the membrane with a frequency of 2$\omega$. If the heating frequency is high enough for the thermal penetration length $L_p$ of the layer to be much smaller than the size of the window ($L_p$ <<1 mm), the amplitude of the temperature oscillation $|\Delta T|$ of strip S1 and the phase shift $\varphi$ of the other strip S2, are given, then within the 1$D$ approximation:[222]

$$\frac{|\Delta T|}{P} = \frac{\sqrt{\alpha_{||}}}{2\sqrt{2}\,\kappa_{||} d}\frac{1}{\sqrt{\omega}} \qquad \text{Equation 13}$$

$$\varphi = -\sqrt{\frac{\omega}{\alpha_{||}}}x + \frac{3}{4}\pi \qquad \text{Equation 14}$$

Where, $P$ is the average heating power per length, $\alpha_{\|}$ is the in-plane thermal diffusivity of the suspended layer, $d$ is the thickness of the suspended layer, and $x$ (47.5 μm) is the interval between the two metal strips. The $|\Delta T|$ and $\varphi$ can be detected as a 3$\omega$ voltage of S1 and a 2$\omega$ voltage of S2, respectively. The experiments were carried out at RT in a temperature-controlled vacuum chamber at a pressure of about $5 \times 10^{-3}$ Pa. During the measurement, the dc and ac temperature rises caused by harmonic heating were confined to ~10 and 1 K, respectively. Fitting the measured $|\Delta T|/P$ and $\varphi$ with eqs *Equation 13* and *Equation 14* yielded the $\kappa_{\|}$ and $\alpha_{\|}$ of the $SiO_2/SiN_x$ layer. Figure *19*h and i show the experimental data, in which $|\Delta T|/P$ and $\varphi$ as a function of $1/\sqrt{\omega}$ and $\sqrt{\omega}$, respectively are plotted. From these plots, the thermal conductivity of $SiO_2/SiN_x$ was estimated to be 1.0 W/m-K. Similar measurement was then performed on a polymer layer coated on the membrane. Spin-coating was used to produce P3HT, MEH-PPV, and PBTTTC14 films (1 μm-thick for all samples), and chemical polymerization was used to fabricate PEDOT:Tos films (200 nm-thick). These thicknesses were sufficient to ensure that the thermal conductance ($\kappa \times d$) of the polymer layers is comparable to or greater than that of the $SiO_2/SiN_x$ layer, allowing sufficient sensitivity to estimate the thermal conductivity of the polymer films. The polymer samples' experimental data were fitted with equations *Equation 13* and Equation 14 (Figure *19*h and i). By subtracting the thermal conductance of the $SiO_2/SiN_x$ layer, the $\kappa_{\|}$ values of the P3HT, MEH-PPV, PBTTT-C14, and PEDOT:Tos films were estimated to be 0.38, 0.37, 0.39, and 0.86 W/m-K, respectively. The relative uncertainties related to $\kappa_{\|}$ were assessed to be below 15%.

The effect of annealing on the $\kappa$ of the conjugated polymer P(NDI2OD-T2) was investigated using 3$\omega$-Volklein method.[219] This study is quite similar to the SiN and polycrystalline diamond studies discussed above.[212] In this work, a sample of interest is spin-coated on top of a Si-based device including two free-standing high aspect ratio $Si_3N_4$ membranes of different areas (Figure *19*j-k). Each membrane has a resistive thermometer and a microfabricated wire aligned with the longitudinal axis that serves as a heater, providing a temperature differential across the membrane. The two measuring zones of the sample film on top of the membranes are isolated by effective heat sinking over the Si rim. Because of the large aspect ratio of the membranes (and hence the effective sample areas), the heat flux is predominantly 1D in the plane of the membranes and perpendicular to the heater wires. As a result, the measurement investigates the sample's $\kappa_{f,\parallel}$, which is calculated using a differential approach.

3$\omega$ approach/suspended ultrathin membrane method enables easy sample deposition via PVD (thermal evaporation and magnetron sputtering) and solution-based techniques (spin coating, drop casting, or ink-jet printing). To guarantee adequate measurement sensitivity, however, the $\kappa$ of the test film should be as large as feasible when compared to that of the bare membrane. This technique should be carried out in a vacuum.

### 4.5 SThM-3$\omega$ Method

The Scanning Thermal Microscopy (SThM)-3$\omega$ method is a microprobe-based approach in which the microprobe serves as a thermometer and heater while in contact with or near the sample surface.[225] Through the localized delivery of heat to the sample and subsequent measurement of its thermal responses, it becomes feasible to investigate the $\kappa$ of the sample with a significant degree of lateral resolution. The probe may consist of either a TC junction[226] or a resistive metallic probe.[227] In the latter scenario, the detection of temperature variation is accomplished through the measurement of resistance changes of the probe. If an angular frequency ($\omega$) ac current is used to heat the resistive metallic probe, the 2$\omega$ temperature oscillation that results should be proportional to the voltage rise in the third harmonic (i.e.,$V_{3\omega}$) that can be monitored by a lock-in amplifier.

An experimental setup to perform the SThM-3$\omega$ measurement is schematically shown in Figure 20.[227] Several analytical and numerical models have been suggested to establish a correlation between the measurable quantity $V_{3\omega}$ and thermal conductivity. These models consider several factors, including the geometrical and dimensional parameters, material properties of the probe, effective thermal resistance between the sample and probe, and the effective coefficient of heat losses by the entire probe surface to its surroundings.[227, 228] This methodology enables expedited characterization of diverse sample types (bulk materials, films, or nanostructures),[225, 229] with minimal sample preparation. This technique offers a superior lateral spatial resolution[225] that is unachievable by alternative methodologies. The SThM-3$\omega$ technique is widely regarded as the optimal method for examining heat transfer in nanoscale contacts and interfaces,[230] due to its ability to achieve sub-nW heat flux and sub-10 nm thermal spatial resolution.

SThM-$\omega$ method has been used to measure the $\kappa$ of several inorganic nanostructures: nanowires of Si,[229, 231] SiGe, and $Bi_2Te_3$,[232]. Further, SThM-3$\omega$ technique has been employed to examine the heat transfer phenomena in supported thin films[233, 234] and in 2D materials[235-238]. In recent years, novel SThM-3$\omega$ configurations have been devised to investigate thermal and TE transport at the nanoscale.[239] These configurations have uncovered significant phenomena, including local Joule heating, as well as Seebeck and Peltier effects in graphene and nanowire heterostructures. The aforementioned measurements provided additional understanding regarding phonon transport in the nanoscale regime,[240, 241] and demonstrated the significant benefits of utilizing thermal characterization instruments that possess both thermal and topographic mapping functionalities. Further, the SThM-3$\omega$ technique has been utilized to examine the heat transfer phenomenon in supported thin films and two-dimensional materials. Because SThM-3$\omega$ measurements are often performed in air, the method is in theory useful to studies of the humidity dependent thermal conductivity of organic hybrid thermoelectric materials.

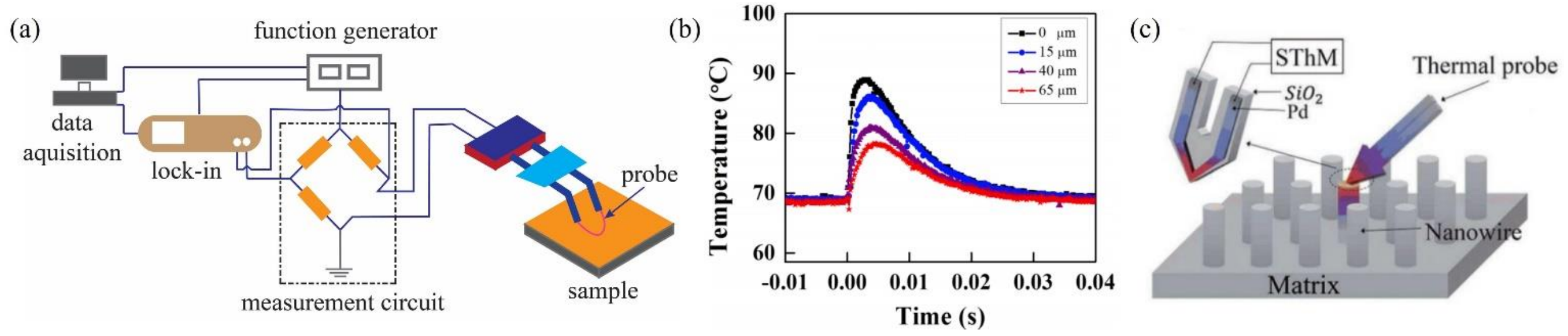

Figure 20: (a) The experimental setup for the SThM-3$\omega$ measuring technique; (b)Thermal diffusivity of $Bi_2Te_3$. Temperature-time profiles for pristine $Bi_2Te_3$ film obtained at varying distances from the heater arm.[242](c) Schematic of SThM-3$\omega$ method for NW thermal conductivity measurement.[156]

The thermal diffusivity of pristine $Bi_2Te_3$ and $Bi_2Te_3$:Si composite films was measured using a SThM-3$\omega$ combined with Parker's flash method (Figure 20b, only $Bi_2Te_3$ is shown).[242] SThM-3$\omega$ has also been employed in characterizing the $\kappa$ of nanostructures. The $\kappa$ of $Bi_2Te_3$ nanowires (300 nm in diameter) was determined using the SThM-3$\omega$ method and high vacuum thermal scanning wave microscopy.[232] Similarly, the $\kappa$ of $Si_{0.8}Ge_{0.2}$ nanomeshes and $Ag_2Se$ film was investigated using SThM-3$\omega$, yielding 0.55 W/m-K and 0.64 W/m-K respectively at RT.[243] Another $\kappa$ measurement using SThM-3$\omega$ shows that Si NWs (50 nm diameter, Figure 20c) exhibit reduced $\kappa$ in comparison with Si NWs of 200 nm in diameter and the $\kappa$ of SiGe NWs is significantly lower than that of SiGe bulk, even at larger diameters than Si NWs.[229]

The primary challenge encountered in conducting thermal conductance measurements through the SThM-$\omega$ technique is to reduce the fluctuations in the contact area between the tip and sample to prevent any alterations in resistance due to topographical factors. Therefore, a meticulous examination of the correlation between topographical and thermal resistance data is necessary. The challenge further adds complexity to the direct comparison of thermal transport information across various SThM-3$\omega$ configurations and thermal sensors. Unlike the previous approaches, SThM-3$\omega$ does not allow direct access to the $\kappa$ of the examined sample. The $\kappa$ measurement requires additional modeling, vital assumptions, and many calibration processes on reference samples.

### 4.6 Frequency-domain thermoreflectance (FDTR) technique

The FDTR technique is a prompt and non-invasive approach used for the determination of $\kappa$ in sub-micrometer thin films and bulk materials. Similar to the TDTR technique, the FDTR approach employs a modulated pump laser to periodically generate thermal energy on the sample surface and subsequently detects the thermoreflectance change using the probe laser beam.[244, 245] In contrast to the TDTR technique, which involves measuring the thermoreflectance change with respect to the delay time between the pump beam and the probe beam, the FDTR approach records the output signals as a function of modulated frequency. Both ultrafast pulsed lasers and continuous wave lasers can be employed in this technique, and two major experimental setups have been proposed for measurements:[244] (a) Modifying the TDTR system (Figure 15a-c) by setting the position of the mechanical delay stage and changing the modulation frequency; (b) The CW lasers are utilized as the pump and probe beams (Figure *21*a). In this instance (b), the pump laser has been externally modulated, and the probe beam is aimed towards the sample surface without going through the mechanical delay stage. In comparison to the FDTR arrangement based on the ultrafast laser system and the mechanical delay stage, the setup is therefore less complex. The key component of the technique is the phase lag between the pump heat wave created by the pump laser and the harmonic response of the sample as measured by the probe laser employing a lock-in amplifier (i.e. $\varphi_{lag} = \varphi_{pump} - \varphi_{probe}$).

A comparison was made between the measurement sensitivity of the FDTR method and that of the TDTR method. It was noticed that both methods are suitable for analyzing small samples with dimensions on the order of 100 × 100 × 100 $\mu m^3$.[246] The FDTR method is considered an effective method for investigating the $\kappa_{f,\perp}$ of nanoscale thin films due to its thermal penetration depth, which typically ranges from 200 nm to several micrometers at a modulation frequency of 0.1-20 MHz.[247] The $\kappa_{f,\perp}$ of *n*-type niobium-doped titanium dioxide ($TiO_2$:1.5 at.%Nb) thin films produced by reactive magnetron sputtering was measured via FDTR and reported to be 1.3 W/m-K and 2.3 W/m-K, for 150 nm and 750 nm thick, respectively.[248] In this experiment, on the surface of the specimens, a 60 nm thick Au transducer was evaporated. Using FDTR, the experimental thermal conductivity values of ZnSe and ZnTe are reported to be ~17 W/m-K and 14 W/m-K respectively, which compared well with a computed results of around 23.2 W/m-K and 13.72 W/m-K for ZnSe and ZnTe respectively at 300K obtained via first-principle calculations. Au film of ~80-100 nm was used as a transducer.[249]

Based on the FDTR technique under periodic heating, Wong-Ng et al. have developed a scanning thermal effusivity (*e*) measurement system.[250] This method can measure the *e* of ceramics, metals, glass, and plastics by employing a metal as the film layer. A lock-in amplifier determines the amplitude and phase difference between the pump signal and the reference signal from the pulse generator (phase lag, $\varphi_{lag}$). Since *e* is related to thermal conductivity $\kappa$, via $e = (\kappa c_p \rho)^{1/2}$, where $c_p$ is the specific heat capacity, and $\rho$ is the density, once *e* is obtained experimentally, then $\kappa$ can be calculated. In this study, the $\varphi_{lag}$ of five bulk samples of whose *e* values are known: $SiO_2$, $SrTiO_3$, $LaAlO_3$, $Al_2O_3$, and Si were measured. The dependence of the $\varphi_{lag}$ on *e* of these five compounds was determined to be linear, i.e. $e = -402.34\ \varphi_{lag} + 26352$. Therefore, once the $\varphi_{lag}$ value for an unknown material is measured experimentally, one can estimate *e*. To test $\kappa$ screening tool, 800-nm-thick conventional $Ba_2YCu_3O_{6+x}$ film (on a $SrTiO_3$ substrate), followed by a deposition of 100-nm-thick Mo metallic film was prepared and its $\varphi_{lag}$ was measured. Using the equation, $e = -402.34\ \varphi_{lag} + 26352$, the *e* was estimated to be 1370 $Js^{1/2}M^{-2}\ K^{-1}$ (measured phase lag value, $\varphi_{lag} = 62.1^o$). The $\kappa$ of the film was then determined to be 12.0 $Js^{-1}M^{-1} K^{-1}$, which agrees well (within 10%) with the reported value of 12.9 $Js^{-1}M^{-1} K^{-1}$.[251] To assess the application of 2D screening tool on Si, the $\varphi_{lag}$ values of a Si sample in the yz-direction were determined and using the measured $\varphi_{lag}$ values and other parameters, the $e$ and $\kappa$ values of the Si sample in an area of approximately 1 mm × 1 mm in increments of 100 µm were determined. Figure 21b shows the 2D 10 × 10 data net of the $\kappa$ values. The average $\kappa$ was 1.65 ± 0.148 W/cm-K, which is within 8% of the literature reported value of 1.56 W/cm-K. Ong et al. have used FDTR to measure the $\kappa$ of a series of materials: ligand-stabilized 3D inorganic nanocrystal arrays (NCAs) layer (30-150 nm thick)[252] and superatomic crystals (SACs).[246] The study revealed that the $\kappa$ values of NCAs and SACs fall within the range of 0.1-0.3 W/m-K at RT. The RT $\kappa$ of amorphous $SiO_2$, bulk $Sr_{0.03}La_{0.97}TiO_3$, and Si single-crystal coated with 100 nm thick Nb films and P3HT (1.6 µm thick) and $SiO_2$ films have been measured via a fiber-aligned FDTR (FAFDTR) experiment (Figure *21*).[253] For P3HT, the measured value was ~0.17 W/m-K, which was close that obtained by the conventional 3$\omega$[167] or TDTR[193] methods. This FAFDTR (Figure *21*) employs fiber-coupled diode lasers to generate pump (red) and probe (blue) beams, which are then combined into a single fiber, resulting in perfect pump-probe alignment at the sample surface. Periodic adjustments to the pump power (dashed line) lead to periodic fluctuations in temperature on the surface of the sample. When the probe beam is reflected, it gets modulated due to the temperature dependent reflectivity of the material. The variation in phase lag between the periodic signals of the pump beam and the reflected probe beam is dependent on the modulation frequency. This phenomenon is utilized to ascertain the thermal conductivity of the sample.

The implementation of the FDTR technique necessitates a polished sample surface that is free of any irregularities, in addition to the deposition of a metallic coating. Although this technique was modified to measure the $\kappa_{f,\parallel}$ of anisotropic materials using the beam-offset setup, a significant uncertainty in detecting the $\kappa_{f,\parallel}$ below 5 W/m-K was noticed,[254] bringing difficulties to quantifying the $\kappa_{f,\parallel}$ of organic materials, which is typically < 5 W/m-K.[3]

Table *4* summarizes the key aspects of several techniques for characterizing $\kappa$, as well as their advantages and disadvantages, as discussed in this review.

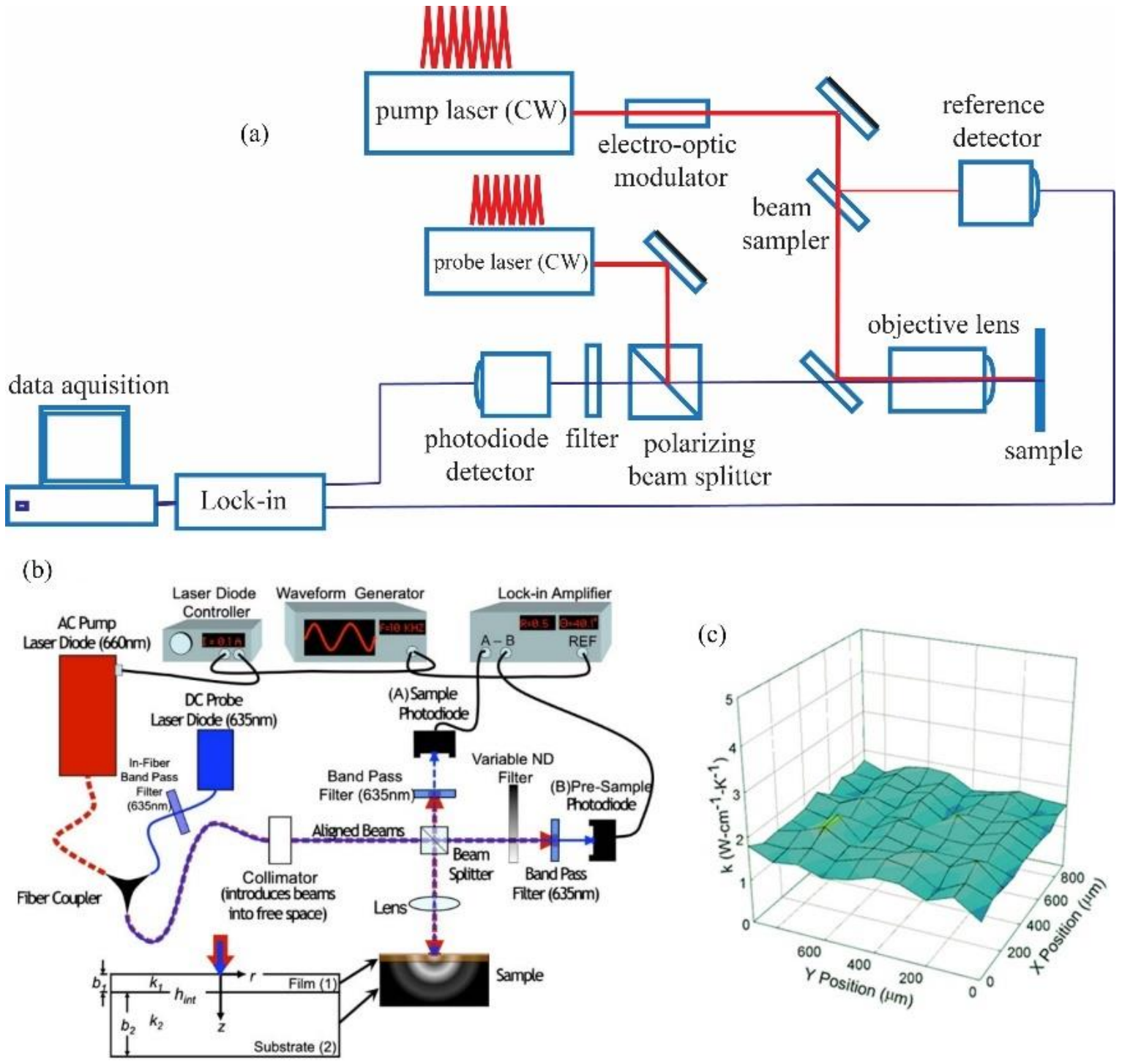

Figure 21: (a) FDTR system based on CW lasers; (b) Schematic figure of FAFDTR;[253] (c) $\kappa$ data of the Si sample in an area of ~1 mm × 1 mm in increments of 100 µm in both $y$ and $z$ directions.[250]

Table 4: Key aspects of steady state methods, time domain methods, and frequency domain methods for measuring $\kappa_{f,\perp}$, $\kappa_{f,\parallel}$

| Method | $\kappa_{f,\perp}$, $\kappa_{f,\parallel}$ measurement possibilities | Advantages | Disadvantages |
|---|---|---|---|
| Volklein method | $\kappa_{f,\parallel}$; inadequate for measuring the $\kappa_{f,\perp}$ | Suitable for films obtained via different deposition processes; $\kappa$ directly obtainable; substrate effect elimination; use of prefabricated metal stripes and insulating layer is possible; possible to measure other TE parameters on a single chip; allows films of sub-micrometer thick | The sample needs to have a high heat conductance compared to the bare membrane; deposition of insulating coating is needed on the metal strip; measurements must be done in vacuum |
| Microfabricated suspended device | $\kappa$ of nanowires or nanotubes along their axial direction; $\kappa_{f,\parallel}$ for thin film stripes | Applicable to nanoscale or microscale samples; Substrate effect elimination; $\kappa$ directly obtainable; possible to measure $S$ and $\sigma$; also applicable to polymers and their composites | Demands substantial microfabrication work; transferring as-prepared samples to microdevices may be difficult; thermal contact resistance and measurement sensitivity are crucial factors that must be carefully evaluated |
| Steady-state infrared thermography | $\kappa_{f,\parallel}$ | Noncontact method; $\kappa$ directly obtainable; applicable to sub-micrometer thick highly porous SiGe films, composite material film, and organic/inorganic hybrid materials, organic films | Sample must have optical absorptance and large emissivity; rigorous calibration of light source and IR camera is needed |
| TDTR method | Mainly for $\kappa_{f,\perp}$,of films but can be adapted to measure $\kappa_{f,\parallel}$ | Noncontact method; $\kappa$ can be obtained without knowing $C_v$ information; high level of adaptability for samples ranging from sub-micrometer-thick films, SLs, and bulks; does not require intricate sample preparation; works for various inorganic materials and organic/hybrid TE materials | Lack of commercial tools; high precision optics required; surface coating is needed; surface of sample must be optically smooth; large uncertainty for $\kappa_{f,\parallel}$ when $\kappa_{f,\parallel} < 5$ W/m-K especially for spin-coated or cast organic material films; robust data reduction procedure is needed; cannot be used to measure the $\kappa$ of monolayer or few-layer 2D materials; modulation frequencies and laser spot sizes limit $\kappa$ measurement |
| Conventional 3$\omega$ method | Mainly for $\kappa_{f,\perp}$,of films but can be used to characterize $\kappa_{f,\parallel}$; of bulk materials | Suited for sub-micrometer thick films; insensitive to radiation and convective heat losses; $\kappa$ directly obtainable; high signal-to-noise ratio | Microfabrication procedures are needed for metallic heater/sensor; in some cases (conducting and semiconducting materials) electrical isolation of samples from metallic heater/sensor using an additional insulating layer is needed which may reduce the accuracy |
| 3$\omega$ approach/suspended ultrathin membrane method | $\kappa_{f,\parallel}$ and $\kappa_{f,\perp}$ | Applicable to variety of materials; elimination of substrate effect; can obtain $\kappa$ directly; prefabricated metal stripes and insulation layers can be used; method enables easy sample deposition via PVD and spin coating, drop casting, or ink-jet printing | $\kappa$ of test film should be as large when compared to that of bare membrane |
| SThM-3$\omega$ method | Nanoscale temperature and $\kappa$ mapping | $\kappa$ of bulks, thin films, or nanostructures; can obtain $\kappa$ directly; quantitative $\kappa$ measurements with a resolution of < 100 nm; minimum sample preparations; high spatial resolution | Fluctuations in the contact area between tip and sample must be reduced to prevent any alterations in resistance due to topographical factors; complex models and thorough calibrations are required |
| FDTR method | Mainly for $\kappa_{f,\perp}$,of films but can be adapted to characterize $\kappa_{f,\parallel}$ | Fast and non-contact method; suitable for both sub-micrometer thick films and bulk materials; simpler than TDTR; also applicable to polymers films | Requires deposition of a metallic coating on the sample; large uncertainty in detecting the $\kappa_{f,\parallel}$ of < 5 W/m-K; requires polished sample surface; efficient data reduction procedure is needed; specific heat information is needed |

## 5 Summary and outlook

This review provided a concise overview of the prevalent methodologies employed for the characterization of the $S$, $\sigma$, and $\kappa$ of TE films. Material growth and thin film fabrication methods have advanced, allowing the development of a thriving research area in electrical and thermal transport at the nanoscale. Recent scientific discoveries in nano-structured thermoelectric material research would not have been achieved without the development of innovative thin film characterization techniques. Thin film thermoelectric characterization techniques can be extremely valuable as the need for thermoelectric energy conversion, cooling, and heat pumping grows in a wide range of scientific disciplines and applications. Further, a significant proportion of contemporary electronic, opto-electronic, and solar energy components are composed of materials featuring numerous interfaces and nanoscale contacts. Consequently, it is vital to investigate the interfacial thermal energy transfer and heat transport at contacts with dimensions in the nanometer range. The advancement of cutting-edge methodologies for comprehending complex phenomena occurring at the nanoscale is of utmost importance.

The determination of cross-plane TE transport properties presents a considerable challenge as a result of the presence of substantial parasitic resistances, encompassing both thermal and electrical resistances, originating from different interfaces and extra layers: buffer layer and substrate, through the transport pathways. The assessment of thin films in-plane is comparatively straightforward, given that the dimensions of the sample are typically sizable. The determination of in-plane $S$ involves the measurement of the open-circuit voltage while subjecting the material to a temperature gradient. Accurate measurement of temperature and voltage can mitigate the effects of three-dimensional heat flux in the measurement system. The inclusion of additional layers in the conducting sample may potentially exert a significant influence and take precedence over the outcomes of in-plane measurements. The aforementioned errors may be rectified by transferring the thin film onto an insulating substrate and subsequently eliminating the problematic layers via material processing methodologies. Complications emerge when attempting to characterize the $S\perp$ and $S_{\|}$ because most thin films are formed on semiconductor substrates whose TE effects may outweigh those of the test films. Several ways have been explored to overcome these challenges, such as removing the substrate or growing the film atop insulating layers.

The $S_{\perp}$ of a thin film can be measured using a DC technique or a $3\omega$ method. The temperature oscillation resulting from the application of a sinusoidal current to a micro-heater deposited on a thin film is determined by measuring the third harmonic component of the voltage response in the heater line. By locking on to the signal at $2\omega$, the Seebeck voltage caused by the temperature oscillation is measured separately. To remove the contributions of the substrate and interfaces, a reference sample (without film) must be measured. However, because the interfaces in the sample device are not similar to those in the reference device, this method ignores the effect of interfacial thermal resistance.

Screening techniques for the $S$ are utilized to investigate chemical and functional heterogeneities that may emerge from a range of factors, including synthesis methodologies, defects, and combinatorial thermoelectric materials. The utilization of particular production techniques can give rise to non-homogeneity in the material owing to differences in composition and doping throughout the sample, resulting in local variations in thermoelectric properties. Therefore, it is essential to perform a methodical analysis of local $S$ variations for both bulk and film samples.

The four-probe approach, the bridge method, and the vdP method are the most often used methods for measuring resistivity. The vdP approach is best suited to thin, arbitrarily shaped samples with contacts positioned anywhere on the periphery. The bridge approach is quite precise. However, its implementation is time-consuming.

Among the three TE properties to determine the $zT$ value, the assessment of $\kappa$ often presents a significant difficulty. The primary reason for this is the presence of undesired heat losses, such as those resulting from blackbody radiation, and thermal contact resistance, which arises from the interface between the sample and the temperature sensor. These factors can introduce additional inaccuracies to the ultimate outcomes. It is imperative to implement measures aimed at mitigating or quantifying these influences. The investigation of anisotropic $\kappa$ has gained significance in light of the emergence of novel TE materials. The utilization of appropriate methodologies, including but not limited to $3\omega$ techniques and thermoreflectance techniques, could potentially aid in addressing the aforementioned problem. The $3\omega$ method is a frequently employed technique that utilizes an oscillatory heating current applied to a metal strip that has been deposited onto the thin film under investigation. Although this approach has demonstrated high reliability, it is not directly applicable to thin films made of metals or semiconductors.

Because of the excellent accuracy of the Pt thermometers and direct temperature calibration, the thermal bridge approach provides high-temperature resolution of 0.05 K in a temperature range of 4 to 400 K. As opposed to the mesoscopic testing method, this method enables faster thermal stabilization after each temperature change. In this system, the cumulative thermal resistance of the entire system ($R_t$) comprises the thermal resistance of the suspended sample, the thermal resistance contribution from the part of the sample connected with the membranes, the internal thermal resistances of the two membranes, and the additional thermal resistance contribution from the part of the membranes connected with the heaters or thermometers. The key challenge is estimating the thermal contact resistance components that would undoubtedly contribute to the measured $R_t$.[239] The first is the thermal contact resistance between the suspended sample's two ends and the $SiN_x$ membranes.[155]The estimation of this resistance requires the use of a fin resistance model[160] Another part of $R_t$ is the thermal contact resistance between the sample-membrane interface and the thermometer ($R_{smt}$). This comes from the non-uniform temperature distribution on the heating membrane. This can be ignored only when a uniform temperature distribution in the membrane can be assumed, i.e., when the thermal resistance of the suspended sample is larger than the internal thermal resistance of the membrane. However, this is not the case for materials with high $\kappa$. To address these challenges, numerical simulations of heat transfer have been performed to determine the precise rise in temperature occurring at the interface of the sample and the heated membrane.[159, 255] Other techniques, such as adding high $\kappa$ materials to the membranes, have been proposed to minimize $R_{smt}$ and enhance temperature uniformity. Recent research has demonstrated that the utilization of an integrated device composed of the identical device layer as the membrane can effectively reduce the thermal contact resistance between the sample and the membrane.[256] Additional challenges associated with this methodology pertain to the fabrication of the apparatus and the installation of the specimen or deposition of the film atop thin and low $\kappa$ supporting films, both of which are intricate, time-consuming, and not appropriate in many circumstances.

Steady-state infrared thermography, a noncontact method for $\kappa_{f,\parallel}$ measurement is ideal for sub-micrometer-thick composite material films, organic/inorganic hybrid materials, and organic materials. However, the samples must have high optical absorptance and emissivity, which many materials do not have.

As evidenced by Table 1, Table *2*, and Table S1 (Supplementary information), over the course of the last twenty years, the TDTR method has undergone significant advancements and has emerged as a potent and adaptable instrument for gauging the thermal characteristics of TE thin films. The manipulation of laser spot sizes and modulation frequencies during TDTR experiments can regulate the heat transfer regime and enhance the sensitivity of TDTR signals to diverse parameters. This approach facilitates the measurement of $\kappa_{f,\perp}$, $\kappa_{f,\parallel}$, and the thermal boundary conductance of interfaces. The utilization of TDTR method has demonstrated its efficacy in the assessment of $\kappa_{f,\perp}$ of diamond, which exhibits a remarkably high $\kappa$ of 2000 W/m-K, as well as in the evaluation of the ultralow $\kappa$ of disordered $WSe_2$ thin films (~0.03 W/m-K). The TDTR technique, which relies on optical pumping and probing, necessitates that the surface of the sample be optically smooth in order for the detector to receive the probe beam via specular reflection. It is thought that the modulation of the diffusely scattered probe light is primarily attributed to thermoelastic effects. If the detector receives this modulated signal, it may lead to an erroneous contribution. The measurable range of $\kappa$ through TDTR is constrained by the feasible modulation frequencies and laser spot sizes. Typically, the modulation frequency utilized in TDTR falls within the 0.2-20 MHz range. Measurements using TDTR at frequencies ($f$) below 0.2 MHz are often challenging due to the low signal-to-noise ratio resulting from $1/f$ noise and the significant ambiguity in phase determination due to pronounced pulse accumulation. Measurements of TDTR at frequencies above 20 MHz pose significant challenges due to the low amplitude of the out-of-phase signals and the susceptibility of the detector and signal cables to high levels of radio-frequency noise. The present TDTR technique is inadequate for the determination of $\kappa$ values of 2D materials that are composed of a single layer or a small number of layers. Hence, it is imperative to make noteworthy advancements in the determination of $\kappa$ for monolayer and few-layer 2D materials. The presence of the transducer would result in in-plane heat spreading during TDTR measurement, decreasing the measurement's sensitivity to $\kappa_{f,\parallel}$. The transducer film also makes data processing more difficult since it necessitates rigorous calibration or fitting to establish additional parameters including the film's conductivity and thickness as well as the interface conductance between the transducer and the sample. Therefore, there is a need for further research into potential solutions.

The SThM-3$\omega$ has been demonstrated to be an effective tool for gaining thermal insights into various kinds of nanomaterials. SThM-3$\omega$ can be used to characterize nanoscale TE materials by acquiring qualitative $\kappa$ contrast images or quantitative $\kappa$ measurements with a resolution of less than 100 nm. More development in this field is necessary to attain improved temperature reading accuracy, resolution, temporal response, and/or sensitivity, especially when the size of the material gets smaller and smaller. Furthermore, it is demonstrated that the SThM-3$\omega$ could function in a variety of conditions: air, liquids, and rarefied gases.

As reported by many studies that have demonstrated the feasibility of concomitantly determining the $\kappa$, $S$, and $\sigma$ on a single sample, it is both time- and resource-saving. [22,61,157] The inclusion of this feature is of great importance as the computation of the $zT$ value based on the TE properties obtained from distinct samples may result in inaccuracies. This phenomenon can be attributed to the inherent inhomogeneity of samples, which persists even when they are selected from a single batch. Consequently, an increasing number of researchers are directing their attention towards the precise and integrated quantification of TE variables. Many different material properties can be easily improved through combinatorial synthesis before scaling up to massive sizes for practical energy conversion applications. However, as the research gains momentum, much needs to be learned about how these characterization tools can be modified, how systematic/random errors can be detected, and how reproducibility can be improved. It is important to note that testing platforms utilizing resistive thermometry, integrated heaters/coolers have the capability to measure both thermal and electrical properties within a single device. This approach is considered to be one of the most precise methods for determining $zT$ in the in-plane direction of thin films.

Each specific technique discussed above possesses distinct advantages and drawbacks. The selection of an appropriate methodology is primarily determined by various factors: the dimensionality, geometry and size, optical, electrical, and thermophysical properties (specific range) of the subject under investigation, the temperature range of the measurement, and the desired level of measurement accuracy. Once a particular methodology has been chosen, it is of utmost importance to comprehend potential sources of error. The implementation of precise sample preparation techniques and meticulous calibration of instrumentation are fundamental components in the execution of dependable measurements.

It has been reported that the multiple layered PbTe/PbSe nanolaminate structures (synthesized by ALD) on microporous Si substrate template exhibit significantly enhanced $S$ in both horizontal and vertical directions, in stark contrast to the case when the same ALD thermoelectric nanolaminates were grown on regular planar bulk Si substrates.[257] More research is needed to determine how porous Si substrates might reduce substrate effects and simplify sample preparation while measuring TE parameters. Furthermore, research into how such platforms might be made CMOS-compatible is required.

It is fairly simple to produce a small-area TE film in a research laboratory using an appropriate fabrication technology and characterize the film for TE properties using a suitable technique. However, because the microstructures in various samples may differ, this process becomes difficult when numerous devices are built from a single film (particularly composite films). It is time-consuming and inefficient to characterize all of the films from a single batch. Large-area film deposition processes have recently become a reality. In such cases, scaling up present measurement techniques to large-area films may become troublesome and necessitate additional research. In terms of simplicity, accuracy, and time consumption, using the same platform from fabrication to testing is preferable to film transfer procedures (which may involve contamination and mechanical damage).

Recently, polymer TE materials and inorganic/organic TE films have gained more attention for the application of wearable devices. Such films are usually deposited on a flexible substrate such as polyimide (PI) or Kapton. However, these substrates do not withstand high temperatures. Therefore, caution must be exercised while preparing the testing samples and choosing the characterization tools.

While current techniques continue to advance in terms of precision, simplicity of measurement, and versatility, they continue to contribute to the resolution of complex problems and the discovery of complex processes at the nanoscale level. TE instrumentation is more of a technical field. The bulk of thermoelectrics researchers may not have an engineering background. In such instances, a collaborative approach to exchanging instrumentation know-how is strongly suggested.

**Conflicts of interest:** No competing interests.

**Acknowledgement:** The author would like to pay tribute to his supervisor, the late Prof. Vitalij Pecharsky, whose enduring influence and commitment to group members serve as a source of inspiration. Some of this research was performed during an appointment at the Ames Laboratory, which is operated for the U.S. DOE by Iowa State University under contract # DE-AC02-07CH11358.